\documentclass[traditabstract]{aa}
\usepackage{tabularx}
\usepackage{amssymb}
\usepackage{amsmath} 
\usepackage{txfonts}
\usepackage{balance}
\usepackage{natbib}
\usepackage{threeparttable}
\bibpunct{(}{)}{;}{a}{}{,} % to follow the A&A style
\usepackage[usenames, dvipsnames]{color}
\usepackage{xspace}
\usepackage[normalem]{ulem} % to strike through with \sout{}
\usepackage{multirow}
\usepackage{rotating}
\usepackage{cancel}
\usepackage{graphicx}
\usepackage{caption}
\usepackage{ifthen}
\usepackage{booktabs}
\usepackage{tabularx}
\usepackage{float}
\usepackage{gensymb}
\usepackage{subcaption} 
\usepackage{booktabs}
\usepackage{dblfloatfix}
\usepackage{placeins}

\begin{document}

\title{Galaxy populations in the most distant SPT-SZ clusters\\  II. Galaxy structural properties in massive clusters  at $1.4\lesssim z\lesssim1.7$}
\titlerunning{Galaxy structural properties in massive clusters at $1.4\lesssim z\lesssim1.7$}
\authorrunning{Strazzullo et al.}

\author{V.~Strazzullo\thanks{veronica.strazzullo@inaf.it}\inst{\ref{inafTS}}$^,$\inst{\ref{units}}$^,$\inst{\ref{ifpu}}$^,$\inst{\ref{LMU}}
  \and M.~Pannella\inst{\ref{units}}$^,$\inst{\ref{inafTS}}$^,$\inst{\ref{ifpu}}$^,$\inst{\ref{LMU}} 
   \and J.~J.~Mohr\inst{\ref{LMU}}$^,$\inst{\ref{MPE}}$^,$\inst{\ref{ECUniverse}}
  \and A.~Saro\inst{\ref{units}}$^,$\inst{\ref{ifpu}}$^,$\inst{\ref{inafTS}}$^,$\inst{\ref{infn}}
  \and M.~L.~N.~Ashby\inst{\ref{cfa}}
  \and ~\\ M.~B.~Bayliss\inst{\ref{UCincinnati}}
  \and R.~E.~A.~Canning\inst{\ref{UPortsmouth}}
    \and B.~Floyd\inst{\ref{UMKC}}
    \and A.~H.~Gonzalez\inst{\ref{UFlorida}}
 \and G.~Khullar\inst{\ref{UPittsburgh}}
 \and K.~J.~Kim\inst{\ref{UCincinnati}}
 \and  ~\\M.~McDonald\inst{\ref{MIT}}
 \and C.~L.~Reichardt\inst{\ref{melbourne}}
\and K.~Sharon\inst{\ref{Umich}}
\and T.~Somboonpanyakul\inst{\ref{KavliStanford}}
}

\institute{INAF — Osservatorio Astronomico di Trieste, via Tiepolo 11, I-34131, Trieste, Italy  \label{inafTS}
  \and Astronomy Unit, Department of Physics, University of Trieste, via Tiepolo 11, I-34131 Trieste, Italy \label{units}
  \and IFPU - Institute for Fundamental Physics of the Universe, Via Beirut 2, 34014 Trieste, Italy \label{ifpu}
  \and Faculty of Physics, Ludwig-Maximilians-Universit\"{a}t, Scheinerstr.\ 1, 81679 Munich, Germany\label{LMU}
\and Max Planck Institute for Extraterrestrial Physics, Giessenbachstr. 1, 85748 Garching, Germany\label{MPE}
\and Excellence Cluster Universe, Boltzmannstr.\ 2, 85748 Garching, Germany \label{ECUniverse}
\and INFN - National Institute for Nuclear Physics, Via Valerio 2, I-34127 Trieste, Italy \label{infn}
\and Center for Astrophysics $|$ Harvard \& Smithsonian, 60 Garden St, Cambridge MA, 02138, USA \label{cfa}
\and Department of Physics, University of Cincinnati, Cincinnati, OH 45221, USA \label{UCincinnati}
\and Institute of Cosmology and Gravitation, University of Portsmouth, Burnaby Road, Portsmouth PO1 3FX, UK \label{UPortsmouth}
\and Department of Physics and Astronomy, University of Missouri, Kansas City, MO 64110, USA  \label{UMKC}
\and Department of Astronomy, University of Florida, Gainesville, FL 32611 \label{UFlorida}
\and Department of Physics and Astronomy and PITT PACC, University of Pittsburgh, Pittsburgh, PA 15260, USA \label{UPittsburgh}
\and Kavli Institute for Astrophysics and Space Research, Massachusetts Institute of Technology, 77 Massachusetts Avenue, Cambridge, MA 02139 \label{MIT}
\and{School of Physics, University of Melbourne, Parkville, VIC 3010, Australia}\label{melbourne}
\and Department  of  Physics,  University  of  Michigan,  450  Church  Street,  Ann  Arbor,  MI,  48109 \label{Umich}
\and Kavli Institute for Particle Astrophysics \& Cosmology, P.O. Box 2450, Stanford University, Stanford, CA 94305, USA \label{KavliStanford}
}

\date{ }

\abstract {We investigate structural properties of massive galaxy
  populations in the central regions ($<0.7r_{500}$) of five very
  massive (M$_{200}>4 \times 10^{14}$~M$_{\odot}$), high-redshift
  ($1.4 \lesssim z \lesssim 1.7$) galaxy clusters from the 2500
  deg$^2$ South Pole Telescope Sunyaev Zel’dovich effect (SPT-SZ)
  survey. We probe the connection between galaxy structure and broad
  stellar population properties at stellar masses of
  log(M/M$_{\odot}$)$>$10.85. We find that quiescent and star-forming
  cluster galaxy populations are largely dominated by bulge- and
  disk-dominated sources, respectively, with relative contributions
  being fully consistent with those of field counterparts. At the same
  time, the enhanced quiescent galaxy fraction observed in these
  clusters with respect to the coeval field is reflected in a
  significant morphology-density relation, with bulge-dominated
  galaxies already clearly dominating the massive galaxy population in
  these clusters at $z\sim1.5$.  At face value, these observations
  show no significant environmental signatures in the correlation
  between broad structural and stellar population properties. In
  particular, the Sersic index and axis ratio distribution of massive,
  quiescent sources are consistent with field counterparts, in spite
  of the enhanced quiescent galaxy fraction in clusters. This
  consistency suggests a tight connection between quenching and
  structural evolution towards a bulge-dominated morphology, at least
  in the probed cluster regions and galaxy stellar mass range,
  irrespective of environment-related processes affecting star
  formation in cluster galaxies.  We also probe the stellar mass -
  size relation of cluster galaxies, and find that star-forming and
  quiescent sources populate the mass-size plane in a manner largely
  similar to their field counterparts, with no evidence of a
  significant size difference for any probed sub-population. In
  particular, both quiescent and bulge-dominated cluster galaxies have
  average sizes at fixed stellar mass consistent with their
  counterparts in the field. }
 
\maketitle

% ----------------------------------------------------
\section{Introduction}
\label{sec:intro}

Quiescent, early-type galaxies dominating the massive galaxy
population are considered to be a defining signature of cluster
environments up to $z\sim1$, marking the extremes in the long-known
relations between environment and galaxy properties both in terms of
galaxy morphology \citep[morphology-density relation,
  e.g.][]{dressler1980,goto2003,vanderwel2008,cappellari2011} and of
star formation \citep[colour-density and star formation rate--density
  relations,
  e.g.][]{carlberg2001,gomez2003,balogh2004,kauffmann2004,baldry2006,peng2010,boselli2016}. Fossil
record results up to $z\sim1$ \citep[e.g.][]{gobat2008,
  mei2009,mancone2010,jorgensen2017, webb2020, khullar2022} and the direct
observation of proto-clusters at $z>2$
\citep[e.g.][]{wang2016,oteo2018,miller2018} point towards these
sources being formed in massive star formation events at high
redshift, followed by efficient suppression of star formation leading
to the emergence of the red sequence by $z\sim2$
\citep[e.g.][]{kodama2007,zirm2008,strazzullo2016,willis2020,trudeau2022}.

A tight correlation between suppressed star formation and an early-type
morphology has long been observed in the nearby Universe
\citep[e.g.][]{humason1936,morganemayall1957,devaucouleurs1961,fioceroccavolmerange1999,strateva2001}. The
evolution of this correlation at higher redshifts is necessarily
intertwined with the nature and timescales of the most relevant
processes in galaxy evolution, and it has thus been extensively
investigated.  Many studies have found that galaxy morphology and
stellar population properties are already tightly coupled at high
redshift up to at least $z\sim2$, with quiescent sources typically
being characterised by a centrally denser, more bulge-dominated structure
\citep[e.g.][]{wuyts2011,bell2012,lee2013,lang2014,kawinwanichakij2017,kim2018,mowla2019,matharu2019,osborne2020,suess2021,lustig2021}
-- though this conclusion is not necessarily uniform across the
literature \citep[e.g.
][]{stockton2008,vanderwel2011,mclure2013,hill2019,stockmann2020}. Probing
the environmental dependence of this correlation constrains the
physical link between suppression of star formation and structural
evolution in different environments.

Constraining such a link is especially relevant at $1.5\lesssim z
\lesssim 2$, which is often considered as the epoch when massive
galaxies in the central regions of cluster progenitor environments
transition from highly star-forming $z\gtrsim2$ proto-clusters, to
established $z\lesssim 1$ clusters where star formation is already
efficiently suppressed. Indeed, a wealth of literature over the last
decade has discussed this critical phase, revealing significant star
formation activity even in massive galaxies in clusters at $z>1.3$,
and sometimes even in cluster core regions \citep[e.g.][]{tran2010,
  tran2015, hayashi2010, brodwin2013,santos2015, alberts2016,
  nantais2017, alberts2021}. At the same time, other studies have
highlighted that environmental quenching is already active at this
redshift with massive, quiescent galaxies already characterising
(proto-)cluster cores even up to $z\sim2$ (e.g.
\citealt{papovich2010,strazzullo2010b,strazzullo2018,strazzullo2019,spitler2012,tanaka2013,andreon2014,cooke2016,lemaux2019,
  zavala2019,willis2020}, and rare examples even beyond, e.g.
\citealt{mcconachie2022}), although with a likely significant
cluster-to-cluster variation.

In this context, the evolution of the morphology-density relation
together with the star formation rate--density relation has been an important
subject of study. Indeed, the environmental dependence of structural
properties of galaxies at high redshift, and particularly in cluster
environments, may provide fundamental insight into the
environment-related processes affecting the evolution of galaxies in
the critical proto-cluster to cluster transition phase.

Several investigations have shown that, at least for massive galaxies,
the morphology-density relation is already in place by $z\sim1$,
possibly with some evolution with cosmic time at lower masses and
depending on the sample selection
\citep[e.g.][]{postman2005,capak2007b,vanderwel2007,holden2007,mei2012,paulinoafonso2019}.
We focus here on the comparison between galaxies in the `average
field' (i.e. a representative galaxy population averaging environments
across large volumes) and in the most extreme, massive cluster
environments at $1.5 \lesssim z \lesssim 2$. Very massive clusters in
this transition epoch are possibly affected by the strongest star
formation suppression.  Therefore, an environmental dependence of
structural properties (e.g. galaxy size, surface brightness profile,
and radial gradients in stellar population properties), in particular
for quiescent galaxies, would reflect different evolutionary paths in
cluster versus field environments, especially in connection with their
quenching. More specifically, at this redshift one might expect that a
putative widespread quenching event across the (massive) cluster
galaxy population might leave a signature on the galaxy structure, and
notably an excess of quiescent galaxies with structural properties
differing from those of reference field analogues.

Studies investigating such aspects in clusters at this redshift have
highlighted the presence of a morphology-density relation, at least in
some systems up to $z\sim2$, with an excess of early-type galaxies
with respect to field environments, albeit with a potentially
significant cluster-to-cluster variation possibly also partly
depending on cluster properties and the local galaxy density
\citep[e.g.][]{newman2014,sazonova2020,mei2022}. A variety of findings
have been presented on the environmental dependence of quiescent
galaxy sizes, as well as of morphology proxies such as the
\citet{sersic1968} index, axis ratio, and concentration. Lower Sersic
indices and axis ratios in cluster versus\ field environments, at
least in some range of stellar mass and/or cluster-centric
distance \citep[e.g.][]{papovich2012,bassett2013,chan2021}, are
suggestive of a signature of environmental quenching mechanisms
affecting the galaxy structure in a different way with respect to
dominant quenching processes in the field. Conversely, other studies
suggested a higher fraction of bulge-dominated quiescent galaxies in
cluster versus\ field environments
\citep[e.g.][]{newman2014,noordeh2021}, or yet a negligible
environmental dependence of quiescent galaxy structural properties
\citep[e.g. ][]{saracco2017,kawinwanichakij2017} in line with results
in the same stellar mass range at lower redshift \citep[e.g.][]
{wijesinghe2012,carollo2016,paulinoafonso2019} highlighting the tight
connection between the morphology-density relation and the high
quiescent fraction in dense environments.

Similarly, a potential environmental dependence of galaxy sizes has
been widely investigated in the last decade and is still debated, with
different studies finding quiescent cluster galaxies to be larger
\citep[e.g.][and references
  therein]{papovich2012,lani2013,strazzullo2013,delaye2014,chan2018,andreon2018,noordeh2021,afanasiev2022},
similarly sized \citep[e.g.][]{newman2014,saracco2017}, or smaller
\citep{raichoor2012b,matharu2019} than field analogues. Each of these
different results suggests a different evolutionary path -- and thus
more specifically different environment-related mechanisms affecting
such a path -- for the quiescent population \citep[e.g. see discussion
  in ][and references therein]{andreon2018,matharu2019}. Such
potentially conflicting results -- at least at face value -- are likely
at least partly driven by the considerable measurement uncertainties,
often small sample sizes, a different galaxy sample selection, and
intrinsic differences in the analysis approach.

In this work, we investigate structural properties of galaxies in a
sample of five very massive galaxy clusters at $1.4 \lesssim z
\lesssim 1.7$ selected through the detection of the Sunyaev Zel'dovich
effect \citep[SZE, ][]{sunyaevzeldovich1972} signature, as detailed in
Sect.~\ref{sec:clustersample}. As discussed more specifically in
Sect.~\ref{sec:alldata}, this sample is particularly well suited for
cluster galaxy evolution studies at this redshift, because of two
important aspects: 1) the cluster selection does not rely on galaxy
population properties, and thus does not bias the investigation of
cluster galaxies; and 2) the high cluster masses imply a likely
significant impact of environmental effects, as well as large numbers
of cluster members, improving the statistics of galaxy studies. We
have already investigated galaxy populations in these clusters with
respect to environmental quenching and broad stellar population
properties in \citet[][hereafter S19]{strazzullo2019}, suggesting that
the suppression of star formation is already efficient in the cores of
these systems. In this work, we combine these results with the related
investigation of structural properties of the same populations of
cluster galaxies, by comparison with coeval field counterparts, to
investigate environmental signatures.

In the following we adopt a flat $\Lambda$CDM cosmological model with
$\Omega_{\textrm{M}}$=0.3 and $H_{0}$=70 km s$^{-1}$
Mpc$^{-1}$. Magnitudes are quoted in the AB system and a
\citet{salpeter1955} initial mass function is assumed throughout.

\section{Data and measurements}
\label{sec:alldata}

\subsection{Cluster sample}
\label{sec:clustersample}

The cluster sample studied in this work was selected from the
\citet{bleem2015} catalogue of galaxy clusters identified through
Sunyaev-Zel'dovich effect in the 2500 deg$^2$ South Pole Telescope
\citep[SPT, ][]{carlstrom2011} SPT-SZ survey. From the whole SPT-SZ
sample of confirmed clusters associated with SZE detections with a
signal-to-noise ratio S/N$>$5 as presented in \citet{bleem2015}, we
selected all systems with photometric or spectroscopic redshift
$z>1.4$ for a dedicated follow-up with the {\it Hubble} Space Telescope
({\it HST}) and {\it Spitzer} Space Telescope (see
Sect.~\ref{sec:dataphot}). These are five clusters at $1.4\lesssim z
\lesssim 1.7$ with estimated virial masses M$_{200}\sim 5 \times
10^{14}$~M$_{\odot}$ (see Table~\ref{tab:sample} and S19; cluster
names are shortened to SPT-CLJxxxx hereafter). Irrespective of
selection, the clusters in this sample are among the few known
examples of the first, rarest massive clusters emerging at this
redshift
\citep[see][]{mullis2005,rosati2009,andreon2009,andreon2014,stanford2012,brodwin2012,bayliss2014,newman2014,tozzi2015,finner2020,dicker2020,dimascolo2020}.

\begin{table*}[htbp!]
\caption{Cluster sample. Columns 1 to 5 list main cluster
  properties: name, coordinates, redshift, mass M$_{500,c}$ and scale
  radius r$_{500}$, as indicated. Redshifts marked with a $^\dagger$
  are spectroscopic determinations from \citet{bayliss2014},
  \citet{khullar2019}, \citet{mantz2020}; the remaining photometric
  redshifts are from S19. Columns 6 and 7 list, for each cluster, the
  limiting m140 magnitude and stellar mass completeness limit adopted
  in this work (from S19). Column 8 lists the quiescent galaxy
  fraction as estimated in S19 in the $r<0.7r_{500}$ region and for
  stellar masses above log(M/M$_{\odot}$)$>$10.85 (left), and the
  quiescent fraction for the corresponding reference field sample
  (right).
\label{tab:sample}}
\vspace{0.2cm}
  \begin{tabular}{c c c c c|c c c}
     \toprule
cluster & coordinates  &  redshift  & M$_{500,c}$  & r$_{500}$ & m140$_{\textrm{lim}}$ & log(M$_{*,compl}$/M$_{\odot}$) & $f_{q}$ vs\ $f_{q,field}$ \\
 &   &    &\small{[10$^{14} $M$_{\odot}$]}&\small{[Mpc]}&\small{[mag]} & & [\%] \\
\hline
\small{SPT-CLJ0421-4845} & \small{$04^h21^m16.9^s,  -48\degree 45'40''$  } & \small{$1.38^{+0.02}_{-0.02}$} & \small{$2.90^{+0.65}_{-0.72}$}&\small{ 0.60$\pm$0.04} &\small{ 23.5}&\small{ 10.54}& \small{$80^{+6}_{-12}$~~$56^{+11}_{-12}$}\\
\small{SPT-CLJ0607-4448} & \small{$06^h07^m35.6^s,  -44\degree 48'12''$  } & \small{$1.401^\dagger$       }&\small{$3.28^{+0.76}_{-0.75}$ }&\small{0.62$\pm$0.04}  & \small{ 23.5}&\small{ 10.56}& \small{$64^{+11}_{-16}$~~$59^{+10}_{-12}$}\\
\small{SPT-CLJ2040-4451} & \small{$20^h40^m59.6^s,  -44\degree 51'37''$  } & \small{$1.478^\dagger$       }&\small{$3.44^{+0.75}_{-0.80}$ }&\small{0.61$\pm$0.04}&   \small{ 23.5}&\small{ 10.60}& \small{$88^{+4}_{-9}$~~$53^{+11}_{-12}$}\\
\small{SPT-CLJ0446-4606} & \small{$04^h46^m 55.8^s, -46\degree 06'04''$  } & \small{$1.52^{+0.13}_{-0.02}$}  &\small{$2.74^{+0.65}_{-0.69}$}&\small{ 0.56$\pm$0.04}& \small{ 23.4}&\small{ 10.65} & \small{$85^{+5}_{-16}$~~$56^{+11}_{-12}$}\\
\small{SPT-CLJ0459-4947} & \small{$04^h59^m42.5^s   -49\degree 47'14''$  } & \small{$1.71^\dagger$        }&\small{$2.85^{+0.64}_{-0.68}$ }&\small{0.53$\pm$0.04}&   \small{ 23.2}&\small{ 10.85}& \small{$83^{+6}_{-12}$~~$49^{+8}_{-8}$}\\
\bottomrule
  \end{tabular}
%  }
\end{table*}

The spectroscopic follow-up of the high-redshift tail of the SPT-SZ
sample, carried out in large part after the selection of the sample
studied here (see S19), confirms that this sample includes five of the
six most distant SZE detections with $S/N>5$ in the SPT-SZ survey at
$z\gtrsim 1.4$
\citep[e.g.][]{khullar2019,strazzullo2019,mantz2020}. Indeed, as
discussed in S19, because of photometric redshift uncertainties one
cluster with photo-z$<$1.4 in the \citet{bleem2015} catalogue used for
our cluster sample selection, turned out to be at $z>1.4$ from later
spectroscopic follow-up. On the other hand, based on such follow-up
\citep{stalder2013,khullar2019} the possibility that we are missing
further $z > 1.4$ clusters due to photo-z uncertainties is rather
small, and we thus consider the sample studied here as representative
of the SPT-SZ cluster population above $z\sim1.4$.

As discussed in detail in \citet{andersson2011},
\citet{bocquet2015,bocquet2019} and \citet{dehaan2016}, the $S/N$ of a
cluster SZE detection is a proxy of cluster mass that is largely
redshift independent and has a relatively low scatter ($\sim$20\%).
As described in S19, even after accounting for potentially significant
contamination of the SZE measurement by mm-wave emission from
star-forming cluster galaxies, this sample is deemed to be
representative of the cluster population in the probed mass and
redshift range, independent of the properties of cluster galaxies.
This cluster sample is therefore especially advantageous for
investigations of galaxy evolution in early massive cluster
environments.

S19 investigated broad galaxy population properties in this
sample with respect to environmental quenching, finding already
established red-sequence populations and typically enhanced quiescent
galaxy fractions among massive galaxies in the inner cluster region
($<r_{500}$\footnote{Overdensity radii $r_{500}$ and $r_{200}$ quoted
  in the following are the cluster-centric radii within which the mean
  density is 500 and 200 times, respectively, the critical density of
  the Universe at the cluster redshift. The cluster masses $M_{500}$
  and $M_{200}$ refer to the mass enclosed within these radii.}) with
respect to coeval field analogues (see Table~\ref{tab:sample}
and S19). This supports an efficient
suppression of star formation in massive cluster environments already
at this redshift, at least in the probed cluster central regions
($r/r_{500}<0.7$) and stellar mass range (log(M/M$_{\odot}$)$>$10.85),
with environmental quenching efficiencies typically in the range
$\sim$0.5-0.8 (or possibly higher, see S19 for full details).

\subsection{HST and Spitzer observations, photometric measurements and derived properties}
\label{sec:dataphot}

All five clusters were observed with {\it HST} with the Advanced
Camera for Surveys (ACS) in the F814W band ($\sim$4800 s per cluster),
and with the Wide Field Camera 3 (WFC3) in the F140W band ($\sim$2400
s per cluster; all data from GO-14252, PI: Strazzullo, with the
exception of F140W band imaging of cluster SPT-CLJ2040, for which we
used $\sim$9200~s archival observations from the ``See Change''
program GO-14327, PI: Perlmutter). Data reduction and first analysis
of these observations are described in S19. For the surface brightness
modelling discussed in Sect.~\ref{sec:datamorph}, we exploit the 6
dithers available in F140W band for each cluster field to produce
mosaics with a pixel scale of 0.03".

Similarly, all five clusters were observed to homogeneous depth with
the Infrared Array Camera (IRAC) on board the {\it Spitzer} Space
Telescope (PID 12030; PI Strazzullo), with each cluster observed for
5500~s in both the 3.6 and 4.5$\mu$m bands, with the exception of
SPT-CLJ2040 that, lying in a region of higher background, was observed
for 7500~s in each band to ensure similar point-source sensitivity. We again refer the
reader to S19 for full details on data reduction and characteristics.

\subsubsection{Galaxy samples in cluster fields: definition and properties of candidate cluster members}
\label{sec:galsamples}

The available spectroscopic coverage in these clusters is not
sufficient for a galaxy population study based on spectroscopically
confirmed members. This work thus relies (as for S19) on
photometrically-selected candidate cluster member samples.  More
specifically, this analysis is based on the same samples of candidate
cluster members used in S19. We refer the reader to Sects.~2 and 3 of
S19 for a detailed description of the definition and characterisation
of these samples, and only briefly summarise here the most important
aspects relevant to this work.

Candidate cluster member samples are selected purely based on
photometry in the {\it HST} F814W and F140W and {\it Spitzer}
3.6~$\mu$m and 4.5~$\mu$m bands (in the following, magnitudes in these
bands are denoted by m814, m140, [3.6], [4.5], respectively). We focus
on the central cluster regions where homogeneous {\it HST} imaging is
available, limiting the analysis to cluster-centric
distances\footnote{Following S19, the adopted definition of cluster
centre is the position of the brightest red galaxy within 100~kpc
(proper) of the projected number density peak of candidate cluster
members. The adopted cluster centre thus corresponds to the most
massive galaxy lying at the centre of the red galaxy concentration
associated with the cluster. See Sect.~3.4 in S19 for full details.}
$r<0.7 r_{500}$. We further focus on the m140-selected
galaxy\footnote{To remove stars from our samples, point-like sources
were identified based on the SExtractor’s MAG\_AUTO (``total
magnitude'') vs FLUX\_RADIUS sequence, down to a F140W band magnitude
m140=22 mag. This is a purely morphological criterion and thus
unresolved non-stellar sources might be misclassified, but as
discussed in Sect.~2.1 of S19, the contamination from non-stellar
objects of our sample of 120 removed point-like sources across all
five clusters is at most at the few percent level. } population down
to m140$\sim$23.2-23.5 mag, depending on the cluster redshift, as
reported in Table~\ref{tab:sample}. The F140W imaging used for source
detection is much deeper than this limit, with F140W-band source
catalogues estimated to be $>$95\% complete at m140$<$24 mag for all
clusters.  The adopted m140 limits are driven by: 1) the F814W imaging
depth, in order to secure a S/N$>$5 on the colour measurements used in
the following also for red sources (see Sect.~3 in S19 for full
details); and 2) specifically relevant to this work, the S/N in the
F140W band imaging required to carry out the surface brightness
analysis (see Sect.~\ref{sec:datamorph}).

Within the m140-selected galaxy sample, candidate cluster members are
colour-selected mainly based on their [3.6]-[4.5] colour, with the
addition of optical/NIR colour constraints to remove low-redshift
interlopers.  The obtained cluster candidate member sample is deemed
to be highly complete (Sect. 3.3 in S19) but affected by residual
contamination from interlopers. Although such contamination is
estimated to be rather small ($\sim$10\%) for red galaxies in the
cluster region of interest, it may increase up to even $>$50\% as a
function of galaxy colour for the blue population (Fig.~1 in S19). This
residual foreground and background contamination is thus dealt with
statistically, by means of a suitable control field sample (full
details in Sect. 3.2 in S19). As a brief summary, we assign to each
galaxy a weight, which is based on the excess number density of
colour-selected (as mentioned above) galaxies in the cluster
versus control field at the given galaxy location in the m140 -
(m814-m140) - (m140-[3.6]) magnitude-colour-colour space. Such weight
thus corresponds, for each galaxy, to the statistical excess of the
candidate member sample over the control field density at the
magnitude and colours of the given galaxy. We then use these weights in
the analysis to account for the interloper contamination of the
candidate cluster member sample.

\begin{figure*}[ht!]
 \includegraphics[width=0.36\textwidth,viewport= 53 411 395 706, clip]{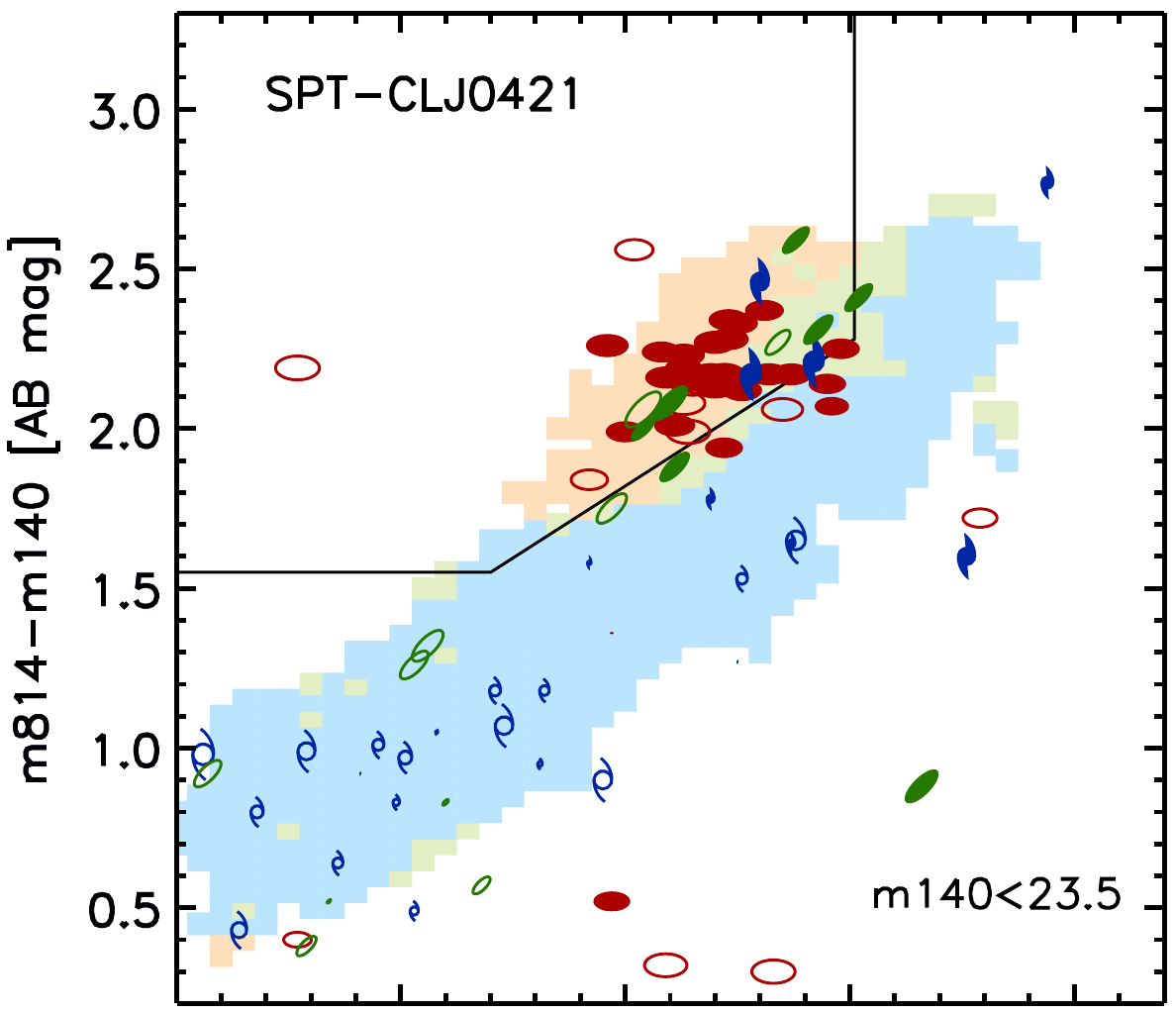}%
 \includegraphics[width=0.31\textwidth,viewport= 101 411 395 706, clip]{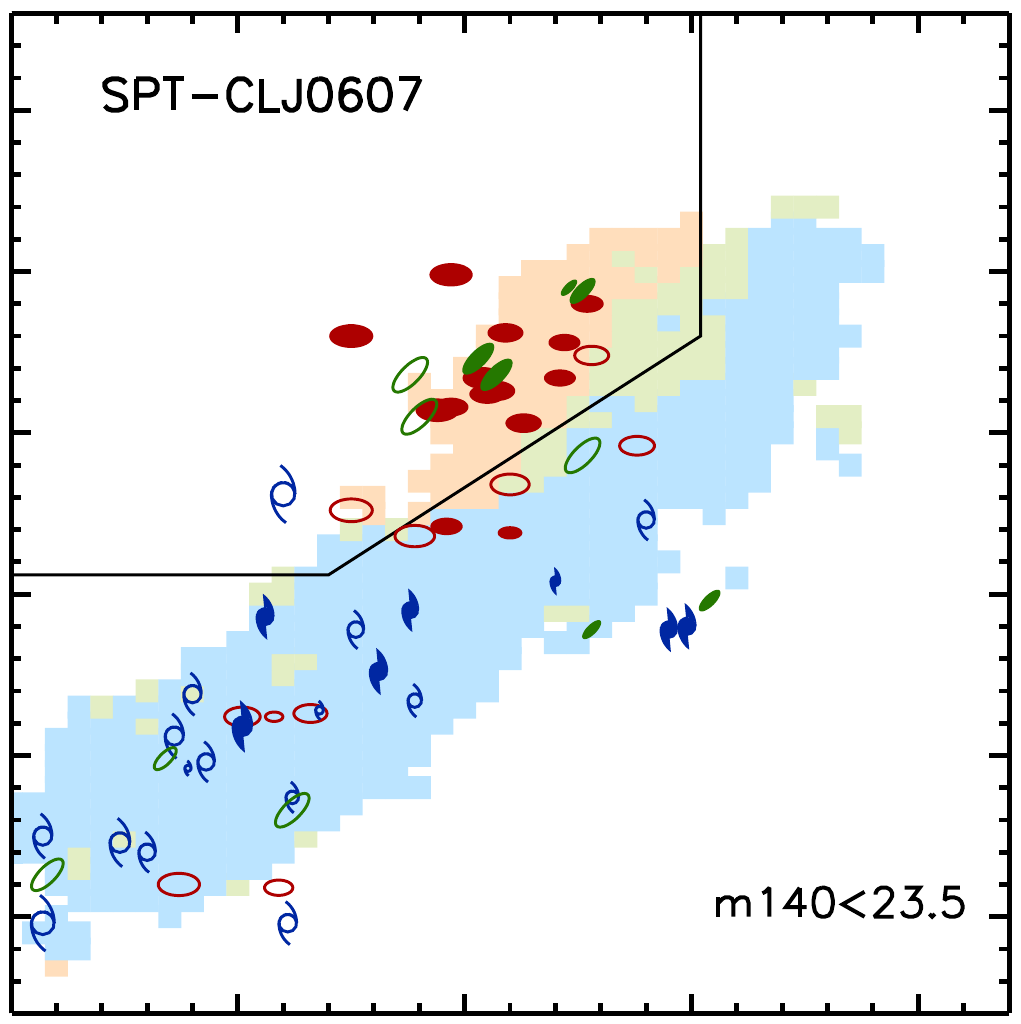}%
  \includegraphics[width=0.3\textwidth,viewport= 131 505 350 678, clip]{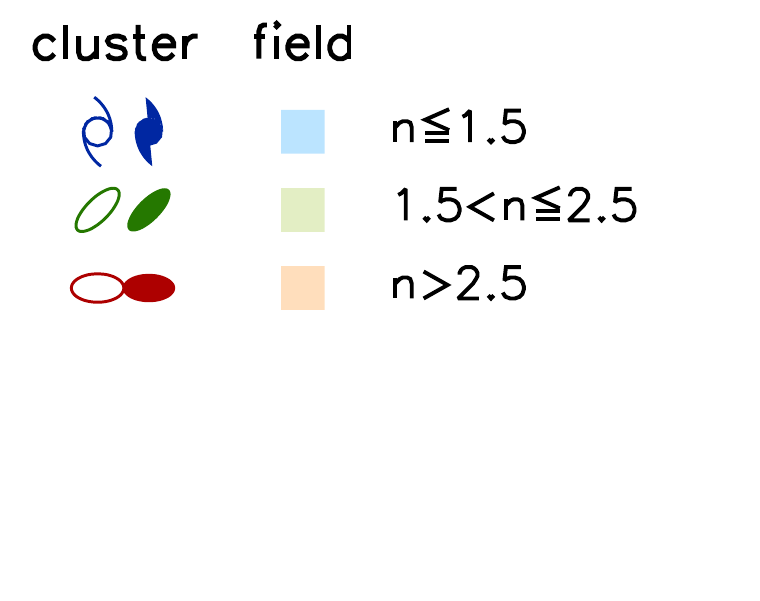}
 \includegraphics[width=0.36\textwidth,viewport= 53 371 395 706, clip]{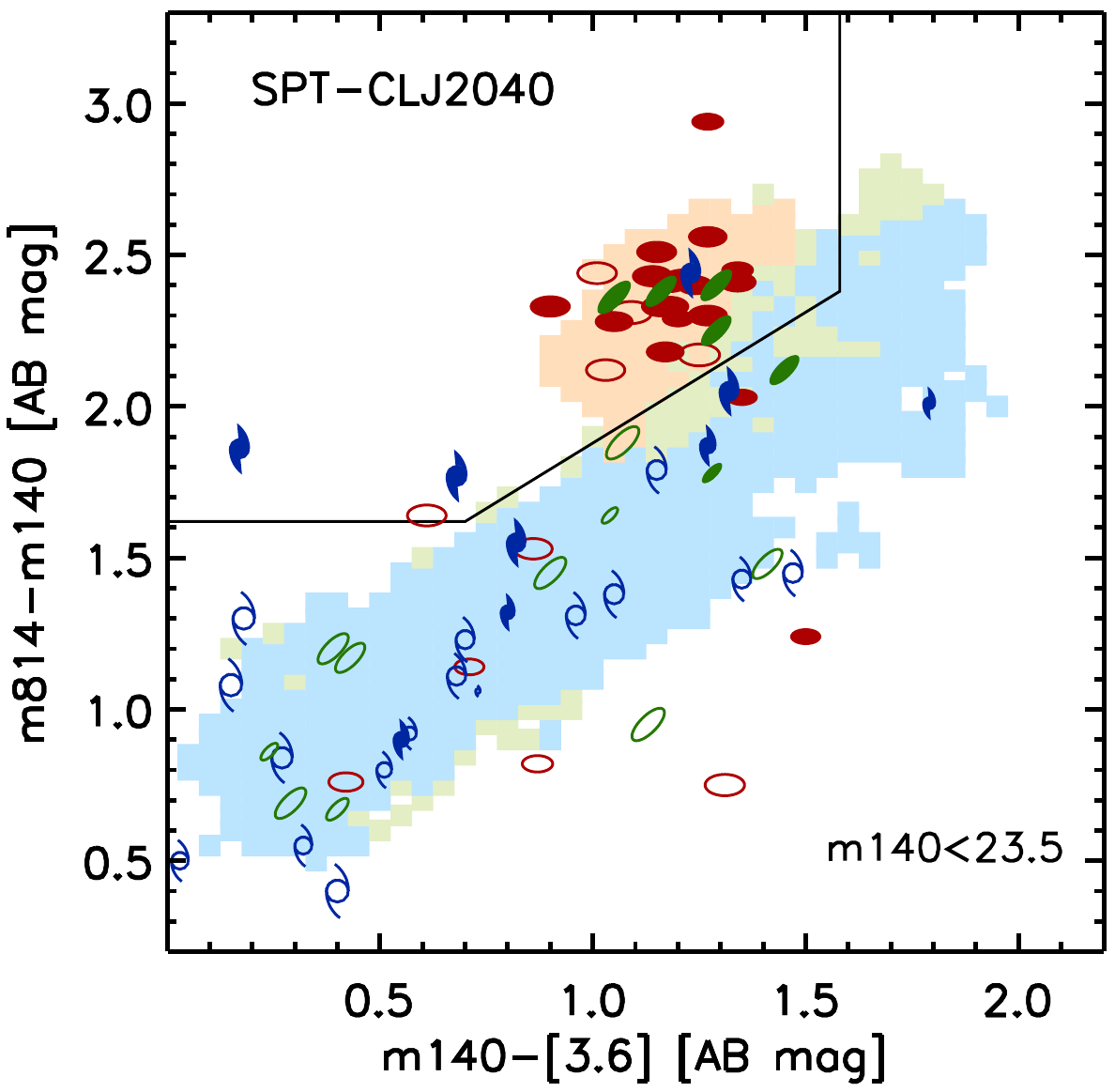}%
 \includegraphics[width=0.31\textwidth,viewport= 101 371 395 706, clip]{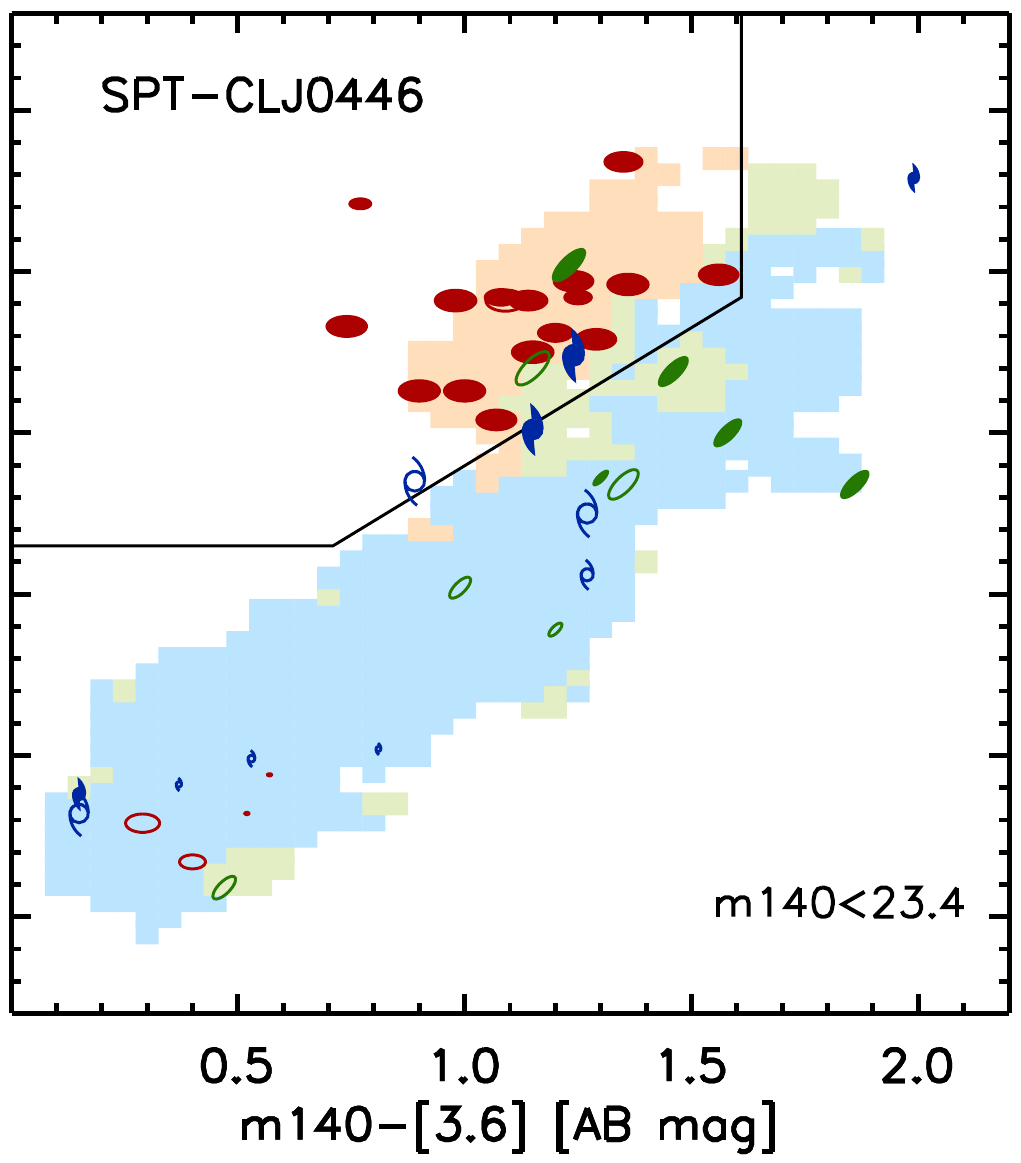}%
 \includegraphics[width=0.31\textwidth,viewport= 101 371 395 706, clip]{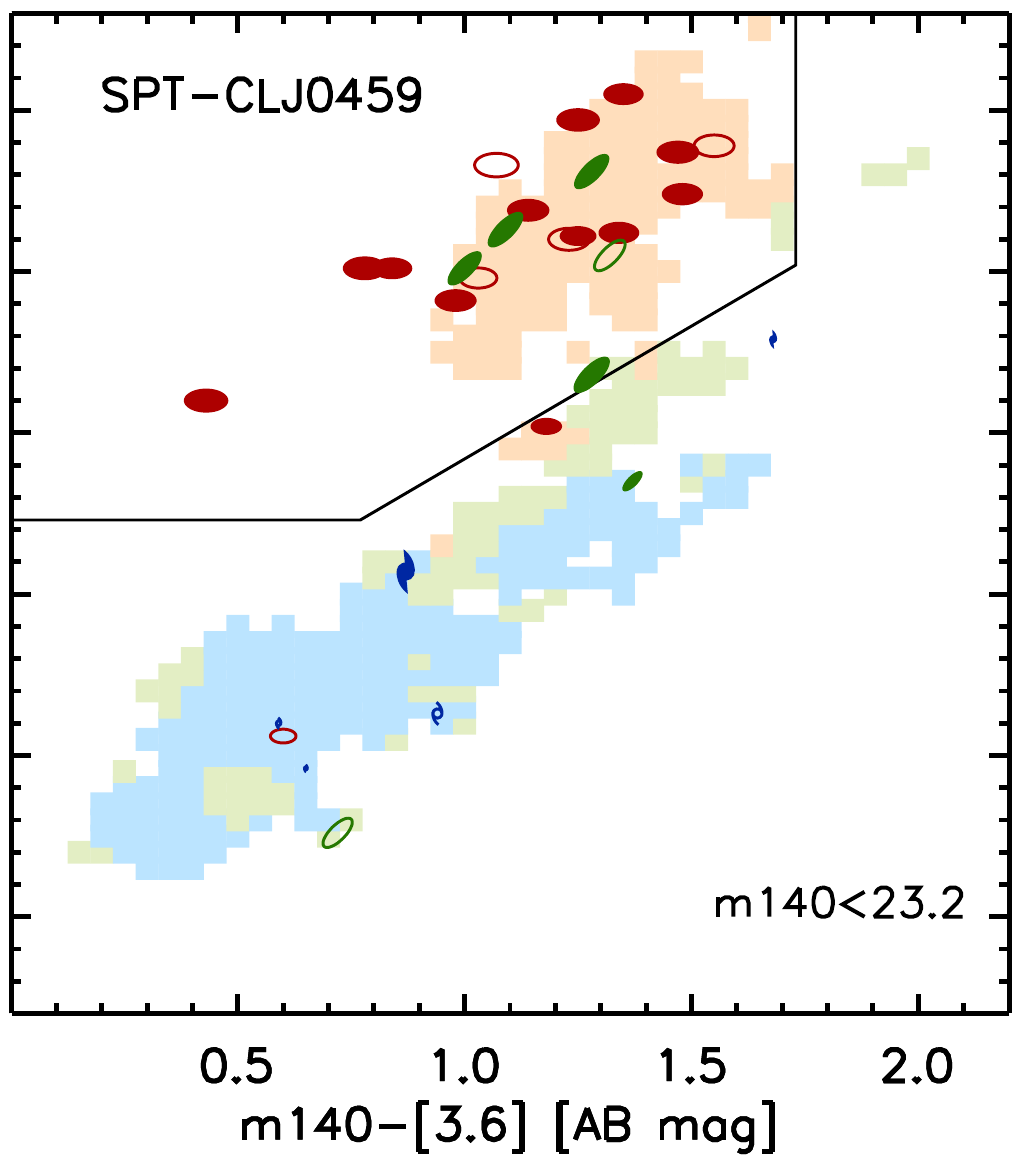}

 \caption{Observed m814-m140 vs\ m140-[3.6] colour-colour diagram
   of candidate cluster members at $r/r_{500}<0.7$ for all clusters,
   down to the m140 limit indicated in each panel, coded by best-fit
   Sersic index as indicated in the legend (dark red, green and blue
   symbols).  Filled and empty symbols show, respectively, galaxies
   above and below the stellar mass completeness limit for the
   individual cluster (Table~\ref{tab:sample}). Symbol size scales
   with the statistical background-subtraction weight
   (Sect.~\ref{sec:galsamples}; smaller symbols are overall more
   likely to be interlopers).  In each panel, the black line marks the
   adopted separation between quiescent and star-forming galaxies,
   following S19. The background colour scale traces the median Sersic
   index (see legend) across the colour diagram (within a colour
   distance of 0.1 mag) of galaxies in the reference field sample with
   photo-zs within $\pm$0.1 from the cluster's redshift, and same m140
   magnitude limit (only colour bins with at least five galaxies are
   shown).
  \label{fig:UVJncoded}}
\end{figure*}

Exactly as in S19, candidate cluster members are classified as
star-forming or quiescent with a photometric criterion similar to the
routinely used UVJ selection \citep[][]{labbe2005,williams2009}.  In
the probed redshift range $z\sim1.4-1.7$, the F814W, F140W and
3.6~$\mu$m bands probe, respectively, the rest-frame ranges
$\sim$2900-3400$\AA$, $\sim$5000-6000$\AA$ and
$\sim$13000-15000$\AA$. Indeed, these passbands were explicitly
selected to approximate the restframe UVJ colour diagram when designing
the imaging follow-up of these clusters. As explained in full detail
in S19 (their Sect.~5.2 and Fig.~6) we thus use the m814–m140
vs m140–[3.6] colour diagram to disentangle quiescent galaxies from
star-forming (including dusty) sources with the colour
selection\footnote{The colour selection in the m814–m140 vs m140–[3.6]
  diagram is different from cluster to cluster because, being based on
  observed colours, it is redshift dependent. The actual selection
  adopted for each cluster is shown in Fig.~\ref{fig:UVJncoded}.} shown in
Fig.~\ref{fig:UVJncoded}, which was determined based on \citet{BC03}
stellar population models and empirically confirmed to be analogous to
the standard UVJ classification by comparison with UVJ-selected
galaxies at the clusters' redshifts from the GOODS-S field (see
Sect.~5.2 in S19). To limit the impact of the boundary region in the
colour diagram along the dividing line between quiescent and
star-forming populations, some results in the following are also
quoted excluding from the analysis a band of $\pm$0.1 mag around the
dividing line.

Finally, for these galaxy samples we use the same empirical stellar
mass estimates as in S19. These are derived by scaling the 3.6~$\mu$m
flux\footnote{In a minority of cases where a reliable measurement of
  the 3.6~$\mu$m flux is not available, stellar masses were estimated
  by scaling the F140W band flux with a M/L ratio calibrated on the
  m814-m140 colour. The uncertainty on such estimates deriving from the
  empirical calibration is $\sim$30-40\%. These F140W-flux based
  masses are only used for a very small fraction of sources in the
  mass complete samples used in the following, and any small related
  systematic with respect to the 3.6~$\mu$m-flux based masses does not
  appreciably affect our results, as discussed in Sect. 2.3.1 of S19.}
to stellar mass with a mass-to-light ratio (M/L) based on the
m814–m140 colour, as calibrated on stellar masses derived by parametric
SED modelling for a reference sample of galaxies at each cluster redshift
from the GOODS-S field (full details in Sect.~2.3.1 in S19). The
estimated uncertainty on the individual stellar mass estimates
deriving from the scatter in this empirical calibration is of order
20-30\% - as stressed in S19 this relies on the assumptions adopted to
estimate stellar masses for the reference sample, and thus this
uncertainty does not correspond to the stellar mass uncertainty on an
absolute scale.

In contrast with stellar mass estimates adopted in S19 that were
scaled to a ``total flux'' as estimated by SExtractor
\citep{sextractor} FLUX\_AUTO, mass estimates used in the following
are scaled to total fluxes as measured by T\_PHOT in the attempt to
more closely recover the whole source flux as also done for the field
sample (see Sect.~\ref{sec:datafield}). Differences on the total flux
are $\sim$10\%, implying that stellar mass estimates used in the
following are larger than those in S19 by the same amount.

In this respect, we further note that for several aspects of the
analyses presented here we compare the cluster galaxy samples with
reference field samples in four fields from the CANDELS survey, as
discussed in Sect.~\ref{sec:datafield}. We stress that the stellar
masses measured for the cluster galaxy sample as discussed above are
not derived in the same way as those for the field reference sample
(see Sect.~\ref{sec:datafield}). While this was also the case for our
analysis in S19, the S19 mass estimates were directly calibrated on
galaxies from the very same galaxy catalogue in GOODS-S that was also
adopted as a control field sample, where we could very closely match
the photometry in the relevant bands between cluster and field
samples. Instead, in this work we use independent stellar mass and
photometric redshift catalogues for the four adopted reference fields
from \citet{skelton2014}, as detailed in Sect.~\ref{sec:datafield}.

The proper comparison of results for cluster galaxies and reference
field counterparts relies on the proper comparison of their stellar
mass estimates. As there is no overlap between the cluster and field
sky areas, we cannot directly compare mass estimates for individual
sources. We thus attempt a quantification of the impact of potential
systematics in the mass estimates of cluster versus field samples coming
from their different measurement procedure (i.e. \citet{skelton2014}
SED-fitting based masses versus empirical estimates adopted for cluster
galaxies) and possible systematic offsets in photometric measurements
driving stellar mass estimates. We evaluate the impact of the former
by computing empirical stellar mass estimates for the
\citet{skelton2014} sources with the same procedure adopted for
cluster galaxies, separately for quiescent and star-forming sources
and as a function of redshift across the range of interest
$1.3<z<1.8$. This evaluation suggests stellar mass offsets of order
$\lesssim$10\% for galaxies at $z<1.6$, and of order $\sim$10-15\% at
$z>1.6$. In the following, we correct by default the stellar masses of
cluster galaxies by the estimated systematics\footnote{The
  corrections are estimated as a function of redshift, and for
  quiescent and star-forming sources, separately. The applied
  corrections turn out to be as follows: at $z<1.45$ (i.e. for
  SPT-CLJ0421 and SPT-CLJ0607), all mass estimates are increased by
  0.01 dex; at $1.45 < z < 1.6$ (i.e. for SPT-CLJ2040 and SPT-CLJ0446)
  mass estimates are decreased by 0.04 and 0.06 dex for quiescent and
  star-forming galaxies, respectively; at $z>1.6$ (i.e. for
  SPT-CLJ0459), mass estimates are decreased by 0.04 dex and increased
  by 0.07 dex for quiescent and star-forming galaxies, respectively. }
to bring them on the same scale of the reference field catalogue, and
minimise the impact on the cluster versus field galaxy comparison.

Concerning systematics in the photometry driving stellar mass
estimates, although galaxies in both cluster and field samples used in
the following analyses are all observed with sufficiently high S/N to
limit concerns on the uncertainty of photometric measurements, some
differences may occur due to the definition of the adopted total
fluxes and/or their actual measurement (also depending on specific
characteristics of the images). In the S/N range of interest, we
consider such potential systematics very unlikely to exceed a factor
$\sim$10\%, and given our adopted empirical estimation of stellar
masses discussed above we thus consider the impact of a 0.1 mag
systematic on the 3.6~$\mu$m flux and/or on the m814-m140 colour, which
result in systematics on the estimated stellar mass of about 10\% and
$<$5\%, respectively.

The sources of systematics on stellar mass estimates discussed above
are largely independent from each other, and we thus estimate that any
residual potential systematics affecting stellar masses in the cluster
versus field comparisons carried out in the following will stay below
the $<$20\% level overall. We comment in the following on the impact
of such potential systematics on our results, as relevant.

For the analyses presented in the following, we consider cluster
galaxy samples complete down to a stellar mass limit defined as in
S19. A mass completeness limit is computed, for each cluster, as the
stellar mass of an unattenuated \citet{BC03} simple stellar population
(SSP) of solar metallicity, formed at $z_{f} \sim 8$, having at the
cluster redshift a F140W-band magnitude equal to the m140 limit
adopted for the given cluster (see Table~\ref{tab:sample}). These mass
completeness limits for the individual clusters range between
log(M/M$_{\odot}$)=10.54-10.85 across the probed cluster redshift
range (see Table~\ref{tab:sample}). The bulk of the analyses presented
here refer to the combined cluster galaxy sample across all five
clusters, down to a mass completeness limit of
log(M/M$_{\odot}$)=10.85.

\subsection{Field reference}
\label{sec:datafield}

In this work we use a field reference sample from the CANDELS/3D-HST
survey fields in GOODS-S, AEGIS, COSMOS, and UDS
\citep{grogin2011,koekemoer2011,brammer2012}\footnote{The GOODS-N
  field is not included because of the lack of F814W band
  imaging. This allows us to keep the photometric coverage for the
  field reference sample as close as possible to the one used for the
  cluster galaxy sample.}. We use specifically multiwavelength
photometry and derived photometric redshifts (photo-zs), stellar
masses and restframe colours from \citet{skelton2014}, together with
parametric morphology measurements from \citet{vanderwel2014},
analogous to those performed in the cluster fields as described in
Sect.~\ref{sec:datamorph}. Following \citet{vanderwel2014}, stellar
masses are corrected based on the ratio of the flux in the photometric
catalogue used for stellar mass estimation to the total flux as measured
by GALFIT in the \citet{vanderwel2014} morphological analysis. For the
galaxy sample relevant to this work (that is, within the photo-z and
stellar mass range of interest, and within morphological analysis flag
constraints as described below) this results in a median correction of
stellar masses by $\sim$6\%, with $<$1\% of galaxies having correction
factors larger than 2.

In the following analyses we focus on (various sub-samples of) galaxies
with photo-zs in the 1.3-1.8 range and stellar mass above the relevant
mass completeness limit for the corresponding cluster galaxy samples,
with uniform quality photometry \citep[i.e. use$\_$phot$>$0 as
  defined in][]{skelton2014}. With this selection, we start from
$\sim$900 galaxies in the combined GOODS-S, AEGIS, COSMOS and UDS
sample from the \citet{skelton2014} catalogues. We then remove sources
flagged as bad fits \citep[with bad fits identified by morphology flag
  $f\geq2$ as defined in][]{vanderwel2012} in the
\citet{vanderwel2014} morphology measurements, which leaves $\sim$770
galaxies. We further remove from this sample $\sim$1\% of sources with
estimated uncertainties on Sersic index or effective radius larger
than 20\%.  The resulting sample with available morphological
measurements thus contains 84\% of the photo-z and stellar-mass
selected sample at log(M/M$_{\odot}>10.85$). Finally, this sample
still contains potentially suspicious morphological measurements
\citep[morphology flag $f=1$, see definition details
  in][]{vanderwel2012,vanderwel2014}, which if removed reduce the
field reference sample with morphological measurements to $\sim$570
galaxies, or 63\% of the photo-z and mass selected sample. Apart from
lowering as discussed the statistics of the control field sample, we
verify that removing these suspicious measurements does not affect the
results presented in the following (very few, marginal exceptions are
mentioned where relevant). Results and figures in the following are
thus presented by default for the clean sample (excluding suspicious
measurements).

We classify galaxies from the field reference sample as quiescent
vs\ star-forming based on restframe UVJ colours from
\citet{skelton2014}. We use in particular the \citet{williams2009} UVJ
classification criterion. To limit the impact of the boundary region
between the quiescent and star-forming populations in the UVJ colour
plane, some results are also quoted excluding from the analysis a band
of $\pm$0.1 mag around the dividing line.

Finally, for the analyses below we need in some cases the F140W band
magnitude of the reference field galaxies. The depth of the CANDELS
F140W-band imaging is shallower than the depth of the F140W data on
the cluster fields ($\sim$800~s vs\ $\gtrsim$2400~s per pointing),
however in the magnitude range relevant to this work (m140$\leq$23.5)
the S/N of sources is still sufficiently high ($>$10 at the faintest
end) for all our purposes. However, as the CANDELS fields are only
partially covered with F140W imaging \citep{brammer2012}, for some
sources there is no F140W band magnitude available. In such cases
($\sim$40\% of the used field galaxy sample at $1.3<z<1.8$ and
log(M/M$_{\odot}$)>10.85 or m140$<$23.5), we substitute a m140 proxy
based on the (deeper) F160W band magnitude. We calibrate this proxy on
sources in the magnitude and redshift range of interest from the
F140W-imaged part of the survey, for quiescent and star-forming
galaxies separately, which yields a m140-m160 colour term of
$\sim$0.21~mag independent of magnitude for quiescent sources, and
between $\sim$0.2~mag and $\sim$0.1~mag across the magnitude range of
interest for star-forming galaxies. By comparison with sources in the
F140W-imaged part of the survey we estimate that, in the magnitude (or
mass) and redshift range relevant to this work, our estimated m140
magnitudes have negligible systematic offsets with respect to actually
measured ones (within 0.02~mag at most), and a scatter of at most
0.1~mag across the full magnitude range of interest. Over such
magnitude range, this scatter is dominated by scatter in the colour
term (rather than by F160W band magnitude uncertainty), and it is
indeed larger than the estimated uncertainty of the actually measured
F140 magnitude for sources where it is available. We thus use actually
measured (rather than estimated from F160W) F140W band magnitudes
where they are available.

\subsection{Morphological analysis of cluster galaxies}
\label{sec:datamorph}

We use GALFIT \citep{peng2002,peng2010} to fit PSF-convolved,
single-component Sersic profiles to the galaxy surface brightness
distribution in the {\it HST}/WFC3 F140W band images, for all sources
detected in the cluster fields down to m140=24~mag. At the clusters'
redshift, the F140W band images approximately probe the restframe V
band.  

GALFIT is run on the F140W images in counts units, with a non-zero
background \citep[see e.g. ][]{peng2002}, and with pixel scale of
0.03''. An independent PSF model for each of the five cluster fields
is created from non-saturated, high-S/N point-like sources selected
from the F140W-band source catalogues described in S19.  We assess the
impact of the adopted PSF model by comparing results obtained with PSF
models of the other clusters, or derived from other surveys, or
produced from our data but with different procedures. From these
comparisons we estimate at most a systematic offset of 5\% and 1\% for
Sersic indices and effective radii, respectively. This level of
systematics has negligible impact on the results presented in this
work - including concerning Sersic indices, which are only used for a
rather broad morphological classification, given the intrinsic scatter
and uncertainties in the relation between Sersic index and detailed
structural properties. Indeed, in the following we mostly
consider three broad morphology classes with Sersic index $n<1.5$
(disk-dominated galaxies), $n>2.5$ (bulge-dominated galaxies), and the
remaining intermediate sources with $1.5<n<2.5$.

The surface brightness profile of galaxies in the cluster fields is
modelled by means of an unsupervised fitting, to limit the introduction
of arbitrariness in the procedure. For each source in the F140W-band
catalogue satisfying the m140$<$24 magnitude limit, we model an image
area centred on the target and with size depending on the projected
dimensions of the target itself (10 times the size estimated from the
SExtractor detection, in both x and y directions).  The background
level is assumed to be constant over the cutout area and is modelled as
a free parameter.  All objects falling in the cutout area, and down to
a limit of 3 magnitudes fainter than the given target, are
simultaneously modelled, as well as bright objects outside of the
selected region but close and bright enough that their light could
produce asymmetric background features. A proper treatment of
neighbouring sources is indeed necessary also to allow GALFIT to
derive a robust background level, a critical point even for
single-Sersic profile fitting, especially for high Sersic index
sources having faint extended emission.

To quantify uncertainties on our estimates of morphological
parameters, both for individual sources and overall for the galaxy
(sub-)populations of interest in this work, we use simulations
(injecting artificial sources in the image, with known profile
parameters spanning the range of interest) as well as multiple fitting
results deriving from objects falling in more than one modelled cutout
area. Over the magnitude range relevant to this work (m140$<$23.5), we
estimate that typical statistical uncertainties on the measured Sersic
index and effective radius are within $\sim$10\% and 5\%,
respectively, with systematics of $<$2\% at most at all magnitudes of
interest.

With this surface brightness modelling, we obtain measurements of
structural parameters (and in particular of the effective radius along
the major axis of the galaxy, the Sersic index, and the axis ratio)
for 94\% (97\%) of the magnitude (stellar mass, respectively) limited
samples above the magnitude (mass completeness) limit adopted for the
individual clusters. For the overall mass-complete sample above
log(M/M$_{\odot}$)>10.85 used for the bulk of the analyses presented
here, structural measurements are available for 96 out of 99 galaxies.
For consistency with the constraints applied for the measurements in
the reference field sample \citep[][see below and
  Sect.~\ref{sec:datafield}]{vanderwel2014} we also exclude from the
analysis a further four sources with $n>8$ (specifically, $8<n<13$)
above the log(M/M$_{\odot}$)>10.85 limit\footnote{A total of 17
  sources with $n>8$, out of 321 with available structural
  measurements in the combined magnitude-limited samples across all
  clusters, are excluded from the analysis.}. This has no relevant
impact on the results presented here, which would remain well within
the quoted uncertainties if these four sources were included.

In many analyses presented in the following, we rely on the comparison
of structural properties of cluster galaxies to those of field
counterparts from a control field. As described in
Sect.~\ref{sec:datafield}, for this purpose we use morphological
measurements published in \citet{vanderwel2014}. Although the adopted
parametric modelling in this work and in \citet{vanderwel2014} is
similar, a number of details in the adopted procedures, as well as in
the reduction and analysis of the imaging data, may potentially
introduce biases in the measurements used in the following, especially
in the more sensitive parameter estimates, namely Sersic indices and
effective radii. As there is no overlap between the cluster and
control fields, in an attempt to quantify such potential biases and
assess their impact on our results, we reduce and analyse a limited
portion of the control field imaging data, and carry out the
morphological analysis with exactly the same procedures that we
adopted for the cluster fields. We then compare our results with the
\citet{vanderwel2014} measurements we use in the following, as
detailed in Appendix~\ref{sec:appendixmorphbias}. Based on that
comparison, our estimated effective radii for the overall galaxy
population in the magnitude range considered here, as well as split by
star-forming and quiescent sources, might be at face value slightly
biased towards lower values with respect to \citet{vanderwel2014}, by
$<$5\% for the overall population in the probed magnitude range.  As
it could be expected, such a bias might depend on the Sersic index of
the source (at face value, increasing from $\sim$3\% up to $\sim$8\%
for low- to high-Sersic index galaxies, see
Appendix~\ref{sec:appendixmorphbias}).  We specifically consider the
impact of this potential bias on our results on the environmental
dependence of galaxy sizes in Sect.~\ref{sec:galmasssize}.

Concerning Sersic indices, based on this comparison our estimates for
the whole population in the considered magnitude range might be biased
by $\sim$4\% towards lower values with respect to
\citet{vanderwel2014}. Also here there might be potential - but of
marginal significance with the estimated uncertainties - dependence of
the bias on the Sersic index (anyway at the $<$7\% level for all
sub-populations, see Appendix~\ref{sec:appendixmorphbias}). The impact
of such bias on the results presented in the following is always well
within the estimated statistical uncertainties, and it is in fact in
most cases unnoticeable. Therefore, we consider it negligible for the
purposes of this work, and all results presented in the following are
shown for our estimated Sersic indices with no corrections.

Based on the comparison discussed above, we estimate that any residual
systematics between cluster and field galaxy samples on both effective
radii and Sersic indices will not exceed the $\sim$5\% level. We
comment where relevant in the following on the impact of potential
systematics at such level on the presented results.

In the following, galaxy sizes are intended as effective radii, and we
adopt specifically the semi-major axis of the ellipse containing half
of the total flux of the best-fitting Sersic model. The average
restframe wavelength across the probed redshift range, corresponding
to the F140W band imaging where cluster galaxy sizes are determined,
is 5500~$\AA$. For the field galaxy sample, we adopt sizes from
\citet{vanderwel2014} measured in the F160W band, for which the
average restframe wavelength across the probed redshift range is
6270~$\AA$. To account for the wavelength dependence of galaxy sizes,
all sizes discussed in the following for both cluster and field
samples are scaled to a restframe wavelength of 5500~$\AA$.  For this
purpose, we proceed more specifically as follows: to scale the
measured size to 5500~$\AA$ restframe, we adopt the relevant scalings
of size as a function of wavelength as given in \citet{vanderwel2014}
for both quiescent and star-forming galaxies. In general, we use the
quiescent or star-forming scaling for galaxies that are classified as
quiescent or star-forming, respectively. For galaxies within $\pm$0.1
mag from the colour separation between quiescent and star-forming
sources, we adopt the scaling for quiescent (star-forming) galaxies
based on their Sersic index being larger (smaller, respectively) than
2. For a minority of sources ($\sim$10\% and 20\% of the
log(M/M$_{\odot}$)$>$10.85 field and cluster galaxy samples,
respectively) we adopt instead the alternative scaling with respect to
that determined with these criteria, based on their measured
sizes. More specifically: we first determine the ``reference size'' of
each source, as the size corresponding to its stellar mass and photo-z
according to the \citet{vanderwel2014} mass-size relation for
quiescent and star-forming galaxies, scaled to the observed wavelength
(F140W for cluster galaxies, F160W for field galaxies). Then, for
sources where the quiescent galaxy scaling should be adopted based on
the aforementioned criteria, but whose sizes are larger than a factor
2x their supposed reference size and closer to the star-forming
mass-size relation, we use the scaling for star-forming galaxies - and
conversely for star-forming sources smaller than 2x their reference
size.  We stress that, as also described in \citet{vanderwel2014}, we
apply these corrections with the aim of limiting the introduction of
systematic biases, but given the small redshift range of interest and
the chosen 5500~$\AA$ reference wavelength being close to the observed
wavelength, corrections are small. For the cluster galaxy sample, all
corrections for individual galaxies are within 3\%, and average
corrections across the five clusters go from $\sim$0.2\% to
$\sim$1.5\% depending on the cluster redshift. For field galaxies in
the relevant mass and redshift range, corrections range from $\sim$1\%
to $\sim$10\% with an average correction of 4\% and $>$90\% of the
galaxies having corrections $<$8\%.

\section{Structural properties of cluster galaxies }
\label{sec:galmorph}

In the following we combine the S19 measurements summarised in
Sect.~\ref{sec:galsamples} with the morphological analysis described
in Sect.~\ref{sec:datamorph}. We focus in particular on the
  relation between broad structural and stellar population properties,
  and environmental dependence of galaxy morphologies, in
  Sect.~\ref{sec:galmorphmix}, and on the stellar mass - size relation
  and environmental dependence of galaxy sizes in
  Sect.~\ref{sec:galmasssize}.

In the analyses described in this Section we often mark or exclude as
it is customary the cluster brightest central galaxies (BCGs), because
of their intrinsically peculiar nature that may be reflected in their
structural properties. As detailed in S19, BCGs are identified in each
cluster as the brightest red galaxy\footnote{In fact, the BCG
  identification is not affected by the specific choice of red colour
  selection (we formally considered m814-m140$>$1.5), because the
  adopted BCG is the brightest candidate member - irrespective of
  colour - in the considered $<$100~kpc region. The colour-magnitude
  diagrams in the whole field probed in these clusters, shown in
  Fig.~1 of S19, indeed show no blue candidate members brighter than
  the brightest red galaxies.  } within a distance of 100~kpc (proper)
from the projected number density peak of candidate members (Fig.~4 in
S19). The adopted BCG is thus the brightest red - and most massive,
with stellar masses in the range
2$\times$10$^{11}$-10$^{12}$~M$_{\odot}$ - galaxy at the centre of the
red galaxy concentration associated with the cluster (see S19 for more
details).  All BCGs in this cluster sample are classified as quiescent
galaxies. Their median Sersic index is $\sim$5 (with individual
estimates in the range $n\sim 3 - 6$). Estimated sizes are shown in
Fig.~\ref{fig:masssize}.  We note that, as discussed in S19, the
  adopted BCG is also the most massive galaxy of the whole candidate
  member sample in all clusters except SPT-CLJ0459 (where six formally
  more massive (by factors 5\% to 70\%) sources are spread along the
  galaxy overdensity).  Furthermore, in SPT-CLJ0446 and SPT-CLJ0459 a
similarly massive, bulge-dominated, quiescent galaxy identified as a
very likely cluster member is found within a distance of $<$7~kpc. In
such cases of ``double BCG'' the BCG definition is arbitrary to this
extent, but this does not affect any of the results presented here.

We further note that, for what is relevant to the purposes of this
analysis, we cannot identify significant deviations between the
cluster and field galaxy stellar mass distributions in the probed mass
range. Although there could be a mild overabundance of the most
massive galaxies in clusters with respect to the field population
(largely due to BCGs), its significance with the uncertainties and
sample statistics available in this work is marginal.

\subsection{Morphological properties of cluster galaxy populations}
\label{sec:galmorphmix}

Figure~\ref{fig:UVJncoded} shows the
m814-m140 vs m140-[3.6] colour-colour diagram (as a proxy of the
restframe UVJ diagram, as discussed in Sect.~\ref{sec:galsamples}) of
candidate cluster members. Colours of cluster members in this diagram,
as well as the separation between star-forming and quiescent galaxies,
follow S19 (their Figure~6, see Sect.~\ref{sec:galsamples}). In
Fig.~\ref{fig:UVJncoded}, candidate members are coded according to
their best-fit Sersic index, dividing the overall population into low,
intermediate and high-Sersic index sources. The background colour
shading shows the median Sersic index of field counterparts across the
diagram\footnote{In Fig.~\ref{fig:UVJncoded} we show field sources
  with \citet{vanderwel2014} morphology fit quality flags $f=0$ (see
  Sect. \ref{sec:datafield}). Including also sources with $f=1$ does
  not affect the figure appreciably. Furthermore, a small fraction of
  the overall reference field area is not covered by F814W imaging
  (and it is thus not included in Fig.~\ref{fig:UVJncoded}),
  corresponding to $\sim$10\% of the total photo-z and m140-selected
  sample relevant to this figure.}.

The distribution of galaxies in Figure~\ref{fig:UVJncoded} reflects
the known correlation between broad stellar population and structural
properties, with quiescence and active star formation preferentially
occurring in high-Sersic index (``bulge-dominated'') and
disk-dominated structures, respectively. As shown in
Fig.~\ref{fig:UVJncoded}, in general terms this correlation occurs
similarly in both cluster and field galaxy samples.

\begin{figure}
  \includegraphics[height=0.253\textheight,viewport= 369 375 558 687 , clip]{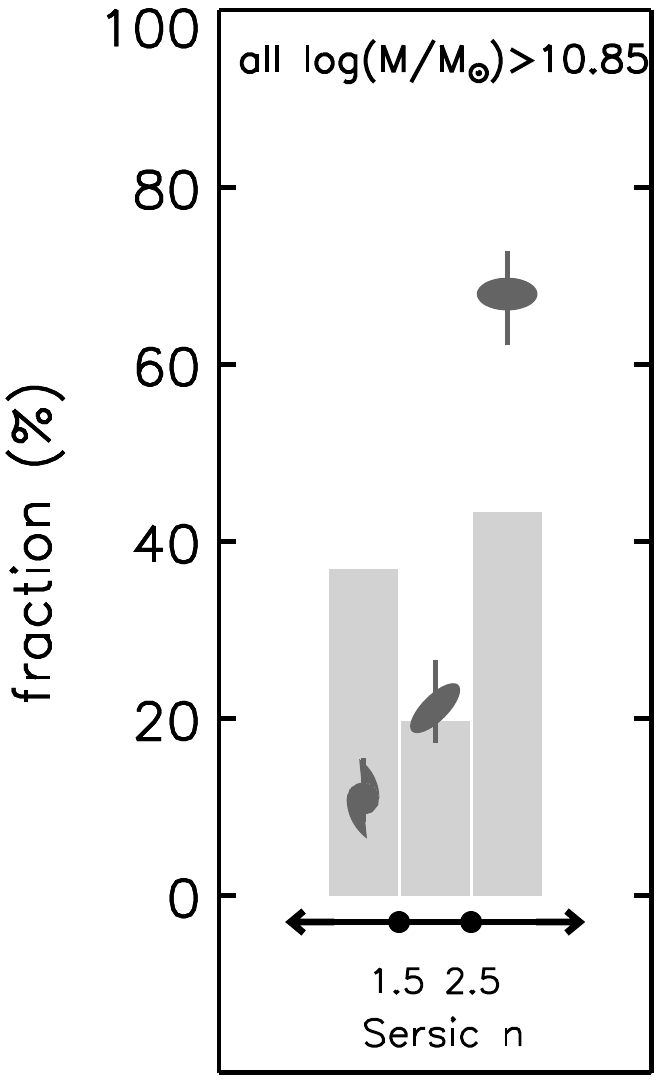}%
    \includegraphics[height=0.253\textheight,viewport= 111 375 368 687  , clip]{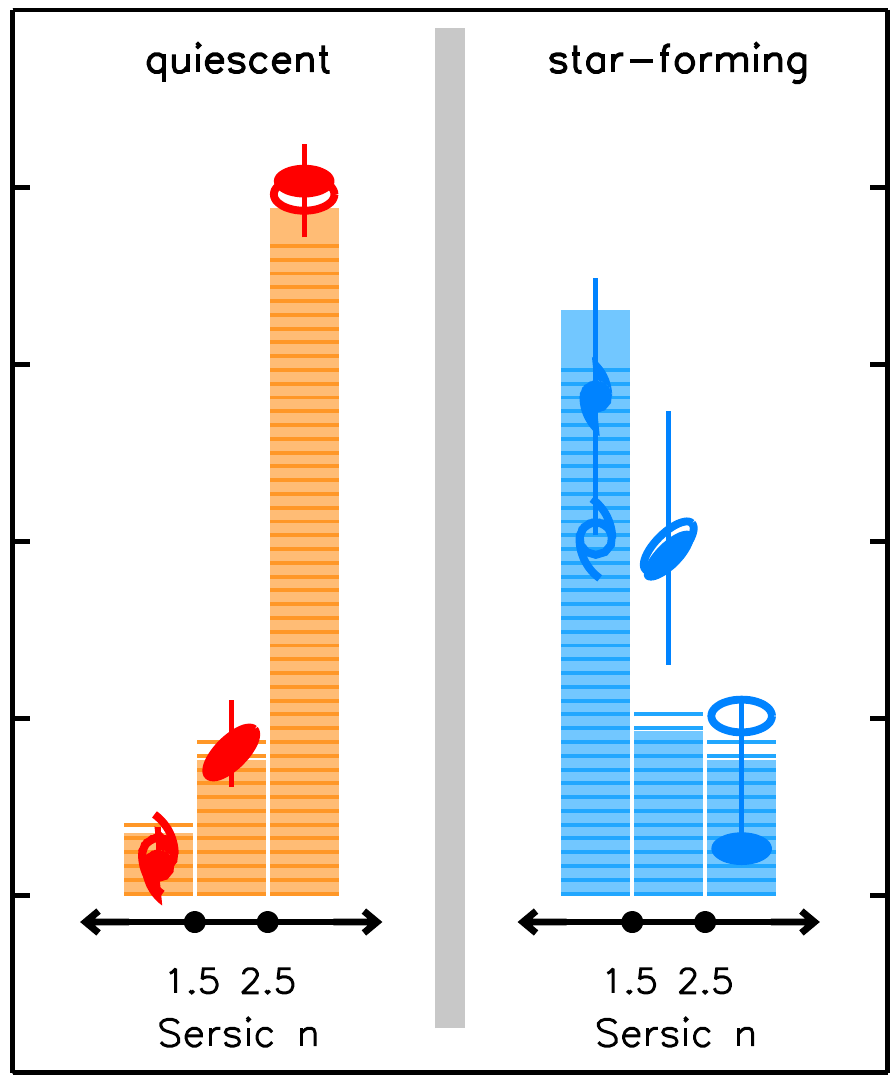}

 \caption{Morphology mix of massive galaxies in the probed cluster
   regions and in the reference field at $1.3<z<1.8$. {\it Left panel} shows the
   fraction of log(M/M$_{\odot}$) $>$ 10.85 cluster galaxies with Sersic
   indices $n\leq1.5$, $1.5<n<2.5$ and $n\geq2.5$ (dark grey symbols),
   and the corresponding fractions for the field reference sample
   (grey bars). For clarity, error bars
   \citep[binomial,][]{cameron2011p} are only shown for cluster
   samples. {\it Right panel} shows the analogous fractions for the quiescent
   and star-forming populations, as indicated. Solid (empty) symbols
   refer to cluster galaxy samples excluding (including) sources
   within $\pm$0.1 mag around the quiescent vs star-forming
   classification border. Solid (hatched) bars analogously refer to
   the corresponding field reference samples.
  \label{fig:morphmix}}
\end{figure}

We quantify this correlation by investigating the morphology mix of
different galaxy sub-populations in the cluster and reference
fields. For this purpose, we combine all five clusters and only
consider galaxies above the common mass completeness limit of
log(M/M$_{\odot}$)=10.85. We then select a field comparison sample at
$1.3<z<1.8$ and with the same stellar mass
limit. Figure~\ref{fig:morphmix} (left) shows the fraction of galaxies
with Sersic index $n\leq1.5$, $1.5<n<2.5$ and $n\geq2.5$ among the
full population of log(M/M$_{\odot}$)$>$10.85 cluster galaxies, as
compared to the analogous fractions in the field reference sample.

Figure~\ref{fig:morphmix} (left) shows that the morphology mix of the
massive galaxy population in the probed cluster regions is skewed
towards bulge-dominated systems with respect to the field, showing
that a morphology-density relation is already in place in these
environments.

On the other hand, the right-hand panel of Fig.~\ref{fig:morphmix}
shows that the morphology mix of the separate quiescent and
star-forming populations in cluster versus field environments is in fact
fully consistent. Therefore, the morphology-density relation in the
left-hand panel is largely connected to the higher quiescent fraction
in the cluster galaxy population with respect to the field.

Fig.~\ref{fig:morphmix} (right) also quantifies more specifically
  the correlation between broad structural and stellar population
  properties in the probed stellar mass range, with $\sim$80\% of
  quiescent galaxies being bulge-dominated ($n>2.5$) sources, whereas
  $\sim$60\% of star-forming galaxies are disk-dominated ($n<1.5$), in
  both cluster and field environments. Likewise, the vast majority of
  bulge-dominated galaxies is quiescent ($89^{+6}_{-3}$\% and
  $77\pm3$\% in the cluster and field samples,
  respectively\footnote{With this analysis and in the probed
    stellar mass range, we do not find an excess of star-forming
    early-type galaxies in clusters with respect to the field at this
    redshift \citep[as found in e.g. ][]{mei2022}, with our estimated
    star-forming fraction for cluster early-types being if anything
    somewhat lower than for field counterparts (we note however the
    differences in stellar mass range, definitions and methodologies
    between the two studies). }), while disk-dominated sources are
  predominantly ($\sim$80-90\%) star-forming.

Fractions for cluster galaxy samples in Fig.~\ref{fig:morphmix} are
based on all candidate members weighted as described in
Sect.~\ref{sec:galsamples} to account for residual background
contamination. Such contamination amounts to $\lesssim$10\% for all
quiescent samples (including Sersic index selected sub-samples), to
typically 40-50\% for star-forming samples, and to almost 20\% for the
overall full population, going from 13\% for n$>$2.5 sources to 35\%
for n$<$1.5.  As also shown in Fig.~\ref{fig:morphmix}, the impact of
excluding a region of $\pm$0.1 mag around the star-forming
vs\ quiescent classification border is typically marginal (and
largely within the estimated uncertainties) for both cluster and field
samples. The impact of matching the redshift distribution of the field
sample to that of the cluster galaxy sample (which is made of three
spikes at $z\sim1.4$, 1.5 and 1.7) is also marginal ($<$10\%
differences) and in any case within the statistical uncertainties. The
impact of residual systematics on the Sersic index at the level
described in Sect.~\ref{sec:datamorph}, and of potential residual
systematics on stellar masses as discussed in
Sect.~\ref{sec:galsamples}, is largely negligible and remains anyway
well within the uncertainties shown in Fig.~\ref{fig:morphmix}.

We note that with these observations our classification of quiescent
vs star-forming galaxies is necessarily broad in nature, and cannot
identify recently quenched, transition sources in any reliable nor
statistically meaningful way. Furthermore, focusing on the relatively
central cluster regions implies that our cluster galaxy samples are
dominated by sources accreted onto the cluster at earlier epochs. For
these reasons, a potential environmental signature more noticeable
only on recently quenched galaxies might involve a minority population
in our samples, with no clear identification, and might thus go
undetected in our analysis. Furthermore, our analysis focusses on
massive galaxies, while more significant signatures might be visible
at lower masses \citep[e.g.][]{chan2021}, possibly in relation to a
potential mass dependence of environmental quenching mechanisms in
clusters at this redshift, as discussed in
e.g. \citet{vanderburg2020,baxter2022}. However, according to these
studies the environmental quenching efficiency is higher for more
massive galaxies; therefore, the stellar mass regime probed here
should, in that respect, facilitate the detection of signatures
related to environmental quenching.  We further note that the Sersic
index classification adopted here to separate bulge- vs disk-dominated
galaxies has known limitations and might also possibly affect the
detection of specific signatures. Nonetheless, these observations show
no obvious environmental signatures in the correlation between broad
structural and stellar population properties as probed here,
suggesting a connection between environmental quenching and structural
evolution on timescales smaller than we can disentangle with this
analysis.

The average Sersic indices for quiescent versus star-forming cluster
galaxies above the log(M/M$_{\odot}$)$>$10.85 stellar mass limit, all
clusters combined and accounting for background contamination, are
n=3.8$\pm$0.2 and n=1.3$\pm$0.2, respectively (uncertainties are
determined by bootstrap, BCGs are excluded). Corresponding values for
the field reference sample above the same mass limit are
n=3.49$\pm$0.09 and 1.50$\pm$0.07 at $1.3<z<1.8$, with no significant
variation across this redshift range. We note though that including
suspicious sources (see Sect.~\ref{sec:datafield}) gives average
Sersic indices of, respectively, n=3.76$\pm$0.09 and 1.66$\pm$0.07.
Although there might be small differences in the average Sersic
indices of quiescent and star-forming cluster galaxies versus field
counterparts, they are minor and cannot be considered significant
given the estimated measurement uncertainties.

\begin{figure}
 \includegraphics[width=0.49\textwidth,viewport= 89 372 537 697, clip]{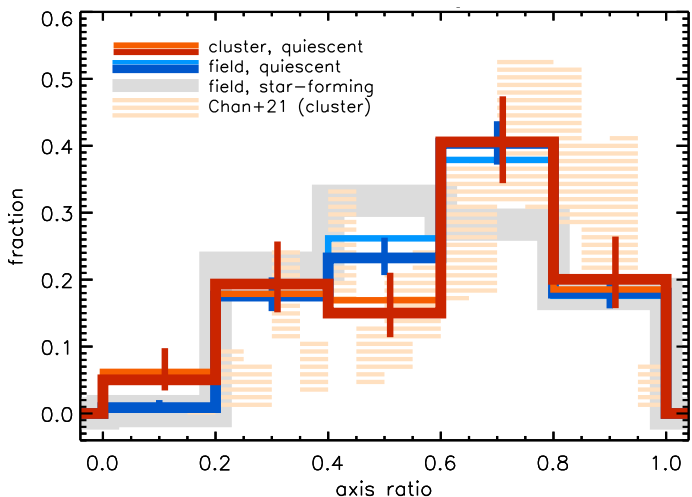}
 \caption{Distribution of axis ratios of log(M/M$_{\odot}$)$>$10.85
   quiescent cluster galaxies in the probed cluster regions across the
   whole cluster sample (red shades) and in the reference field at
   $1.3<z<1.8$ (blue shades, as indicated). For both cluster and field
   samples, darker (lighter) lines correspond to excluding (including)
   sources within $\pm$0.1 mag around the quiescent vs star-forming
   classification border. Error bars on the fractions over the whole
   population \citep[binomial,][]{cameron2011p} are slightly offset
   for clarity.  For comparison, the axis ratio distribution of
   star-forming field galaxies in the same mass and redshift range
   (grey line, errors within line width), and for $1<z<1.4$ cluster
   galaxies in the same stellar mass range from \citet[][hatched
     region, see text for details]{chan2021} are also shown.
  \label{fig:axisratio}}
\end{figure}

\begin{figure*}
  \hspace{1.5cm}
  \includegraphics[width=0.44\textwidth,viewport= 130 600 460 652 , clip]{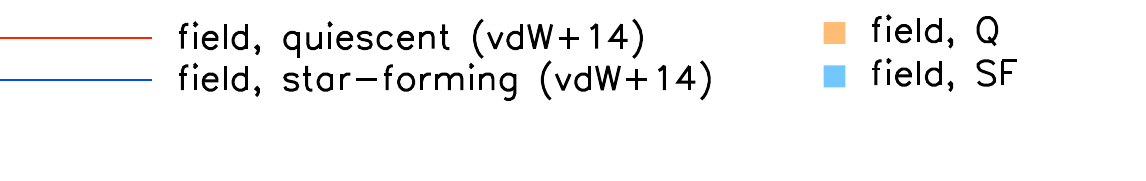}%
    \includegraphics[width=0.44\textwidth,viewport= 138 600 454 652 , clip]{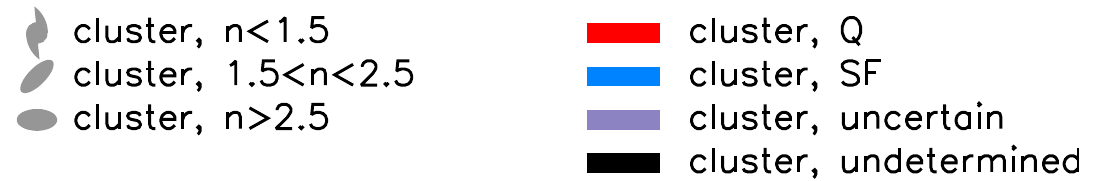}
  \includegraphics[height=0.348\textwidth,viewport=43 420 69 680, clip]{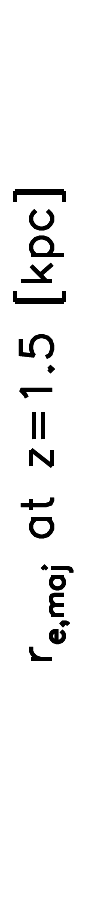}%
  \includegraphics[height=0.348\textwidth,viewport=67 372 397 706, clip]{fig4c.pdf}%
    \includegraphics[height=0.348\textwidth,viewport=101 372 397 706, clip]{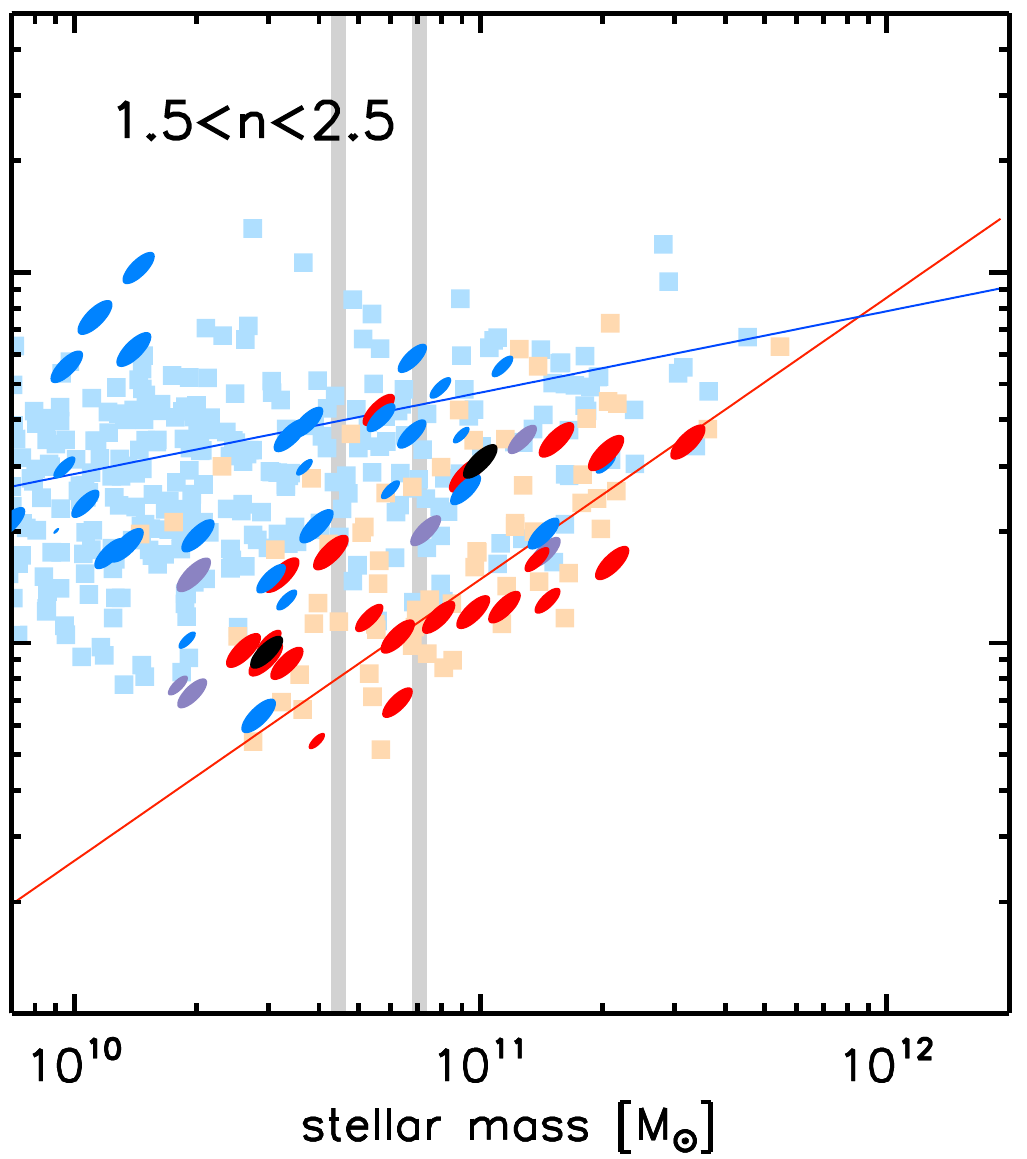}%
    \includegraphics[height=0.348\textwidth,viewport=101 372 397 706, clip]{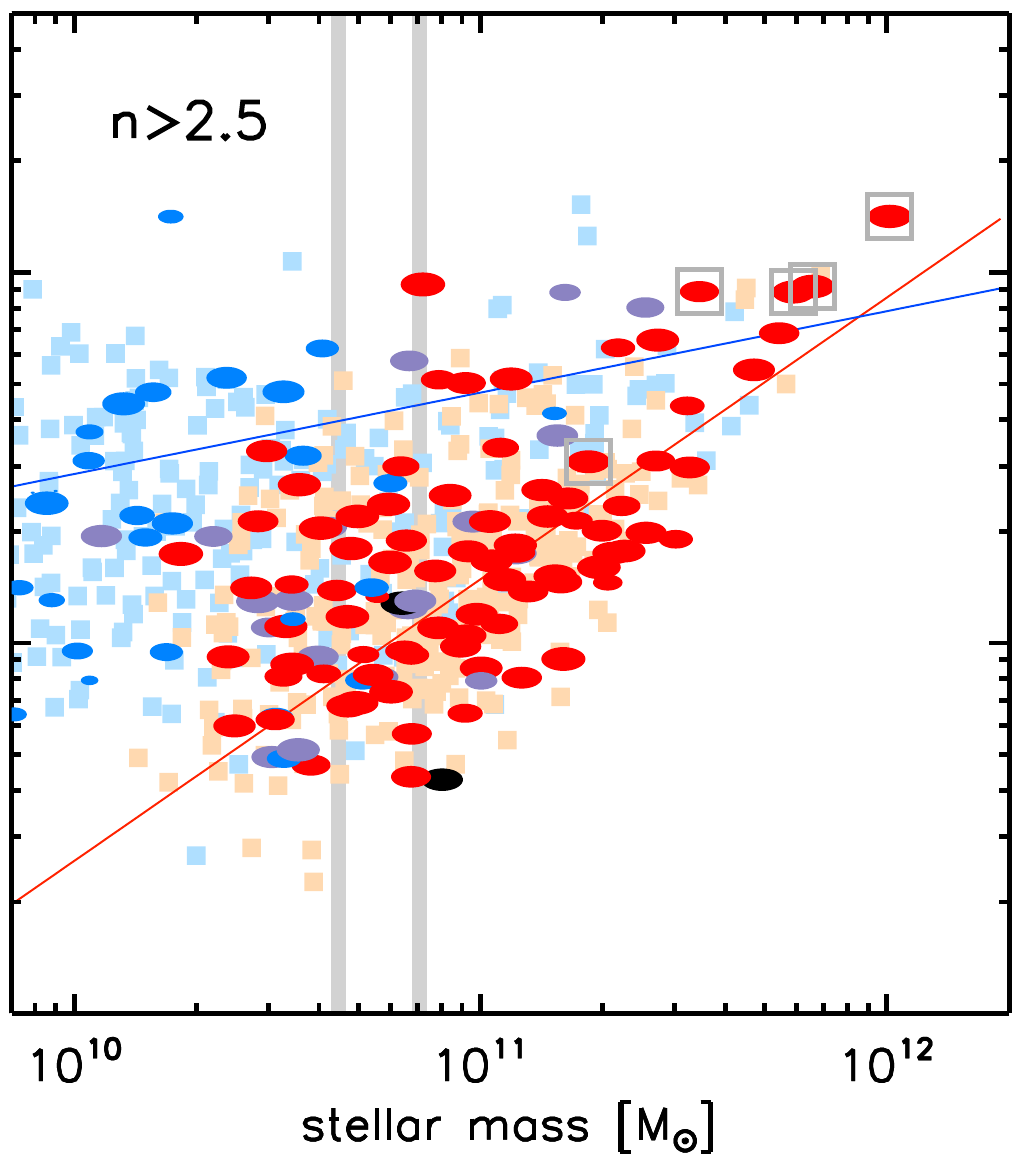}%

 \caption{Stellar mass - size relation of candidate cluster members
   and field counterparts, split by Sersic index as indicated.  All
   candidate members in the m140-limited samples across the five
   clusters are shown, with symbol size scaling with residual
   statistical background subtraction weights
   (Sect.~\ref{sec:galsamples}).  All sizes for both candidate members
   and field counterparts are scaled to a common redshift $z=1.5$ for
   the purpose of this figure (see text).  Candidate cluster members
   are shown with darker symbols, colour-coded by quiescent (red)
   vs star-forming (blue) classification (purple and black symbols
   show, respectively, galaxies within $\pm$0.1 mag from the
   classification border, and with no reliable classification). Large
   grey open squares indicate cluster BCGs (right hand panel). Light
   blue and light red squares in the background show counterparts from
   the control field sample at $1.35<z<1.75$. The red and blue lines
   show the best-fit mass-size relations for quiescent and
   star-forming galaxies from \citet[][also scaled to
     $z=1.5$]{vanderwel2014}. Vertical grey lines mark the minimum and
   maximum stellar mass completeness limits across the five clusters.
  \label{fig:masssize}}
\end{figure*}

As a further probe of potential environmental signatures on the
morphology of the quiescent galaxy population related to environmental
quenching, we compare in Figure~\ref{fig:axisratio} the axis ratio
distribution of massive (log(M/M$_{\odot}$)$>$10.85) quiescent
galaxies in the cluster and field ($1.3<z<1.8$) samples. As in
Fig.~\ref{fig:morphmix}, all candidate cluster members are shown,
weighted as described in Sect.~\ref{sec:galsamples} to account for
residual background contamination. Excluding sources within $\pm$0.1
mag around the quiescent vs star-forming classification border does
not affect the result, as shown.  Fig.~\ref{fig:axisratio} does not
show any significant difference between the axis ratio distributions
of quiescent galaxies in the cluster versus field samples (as also
confirmed by K-S test, both including and excluding sources around the
classification border). As mentioned in Sect.~\ref{sec:intro},
different studies found seemingly contrasting results on this aspect.
In particular at redshifts bracketing the range of clusters studied
here, \citet[][at stellar masses log(M/M$_{\odot})>10.8$, in the
  massive cluster JKCS~041 at $z=1.8$]{newman2014} highlighted a lack
of flattened, oblate quiescent galaxies in cluster versus field samples,
whereas \citet[][11 clusters at $1<z<1.4$ from the GOGREEN
  survey]{chan2021} found an excess (but most remarkably at lower
masses, $10.4<\textrm{log(M/M}_{\odot} <10.8$). We show in
Fig.~\ref{fig:axisratio} a rendition of the
\citet{chan2021}\footnote{A similar comparison with \citet{newman2014}
  results is heavily affected by their small sample size resulting in
  large statistical uncertainties that prevent the investigation of
  subtle differences. } axis ratio distribution of massive, quiescent
cluster galaxies (hatched region, corresponding to the $1\sigma$ range
around the weighted average\footnote{ \citet{chan2021} present axis
  ratio distributions in four stellar mass bins in the range
  log(M/M$_{\odot}) \sim 9.8 - 12$. Their two highest-mass bins
  overlap with our sample. We show a weighted average of their axis
  ratio distribution in their two most massive bins, weighting
  according to the actual contribution of galaxies in these mass
  ranges in our sample.}). Although this rendition suggests that
\citet{chan2021} results are largely consistent with our determination
in this stellar mass range, our axis ratio distribution might be
somewhat less skewed towards very high axis ratios, albeit the
significance of this difference is marginal given the estimated
uncertainties. We note though that \citet{chan2021} find that the most
noticeable environmental signatures on the axis ratio distribution of
quiescent galaxies is found at lower stellar masses than those probed
here.

\subsection{The stellar mass-size relation}
\label{sec:galmasssize}

As discussed in Sect.~\ref{sec:datamorph}, we find a (small) bias
between size measurements in the cluster and control field
samples. Given the direct impact of such bias on results presented
here, in the following we apply a correction to the measured sizes of
cluster galaxies to bring them on the same scale as the reference
field measurements. We specifically comment as needed on the impact of
this correction on the results.

Figure~\ref{fig:masssize} shows the stellar mass - size relation of
selected galaxy populations in both the combined cluster and field
samples. To combine all sources into one figure limiting the impact of
redshift-dependent effects, all galaxy sizes are scaled to a common
redshift of $z=1.5$ for the purpose of this figure, using the redshift
evolution\footnote{For each galaxy, we scale sizes to $z=1.5$ based on
  the \citet[][their Figure~6]{vanderwel2014} relations for quiescent
  or star-forming galaxies, adopting the star-forming or quiescent
  scaling based on the same criteria described in
  Sect.~\ref{sec:datamorph} to scale sizes to 5500~$\AA$ restframe.} of
the mass-size relation for star-forming and quiescent galaxies as
determined in \citet{vanderwel2014}.  Cluster and field samples are
split based on Sersic index into disk-dominated ($n\leq1.5$),
intermediate ($1.5<n<2.5$) and bulge-dominated ($n\geq2.5$) sources,
and coded according to their quiescent vs\ star-forming
classification.

Figure~\ref{fig:masssize} shows a substantial similarity in the
mass-size relation of cluster and field galaxies, and highlights again
the clear correlation between broad structural and stellar population
properties discussed in Sect.~\ref{sec:galmorphmix}. We note in
particular that star-forming and quiescent cluster galaxy populations
largely follow the respective \citet{vanderwel2014} mass-size
relations determined for field galaxies. Estimating a best-fit
mass-size relation for quiescent cluster galaxies from the data in
Fig.~\ref{fig:masssize}, assuming a linear form in log(size)
vs\ log(mass) in the probed mass range log(M/M$_{\odot}$)$>$10.85,
would give a best-fit relation within 5\% of the adopted determination
based on \citet{vanderwel2014}, and fully consistent within the
estimated uncertainties\footnote{Although there might be evidence of
  possible deviations at lower stellar masses, with sizes of quiescent
  cluster galaxies potentially lying at face value preferentially
  above the best-fit relation, due to the characteristics of the
  current observations (Sect.~\ref{sec:alldata}) and the possible
  related biases on the results we do not explore this regime in this
  work. We leave any related analysis to future investigations with
  data properly suited to investigate lower-mass populations.}. In the
following we thus use the \citet{vanderwel2014} relations as a
reference, in quantifying size differences between cluster galaxies
and field counterparts.

The distribution of the selected subpopulations in the mass-size plane
is remarkably similar in the cluster and field environments, the most
noticeable difference occurring for intermediate Sersic index sources,
for which larger, typically star-forming galaxies are underrepresented
in clusters with respect to the field.  Although this would be
potentially very interesting in relation to the investigation of the
correlation and timescales of star formation quenching and structural
evolution in cluster versus field environments, we note that at least
part of this effect might be due to the combination of uncertainties
in the estimated Sersic indices and the higher quiescent fraction in
the cluster versus field samples. This combination results in a larger
number of star-forming (thus largely disk-dominated, and larger in
size) galaxies being scattered into the intermediate-$n$ sub-sample
for the field, whereas a larger number of quiescent (thus largely
bulge-dominated, and smaller in size) galaxies are scattered into the
intermediate-$n$ sub-sample for the clusters. Indeed, as
  noticeable in Fig.~\ref{fig:masssize}, the quiescent galaxy fraction
  in the intermediate-$n$ sample is tendentially higher in clusters
  than in the field, though with marginal significance with the given
  uncertainties ($\sim$60$\pm$15\% vs 40$\pm$10\%). Although for the
  reason just discussed we refrain from elaborating on this further,
  we note that this is qualitatively similar (quantitative comparison is
  affected by the different stellar mass range probed) to what
  observed in $z\sim1$ clusters by \citet[][see also references
    therein]{matharu2019}.

\begin{figure*}[ht]
  \centering
 \includegraphics[width=0.95\textwidth,viewport=  59 532 550 689, clip]{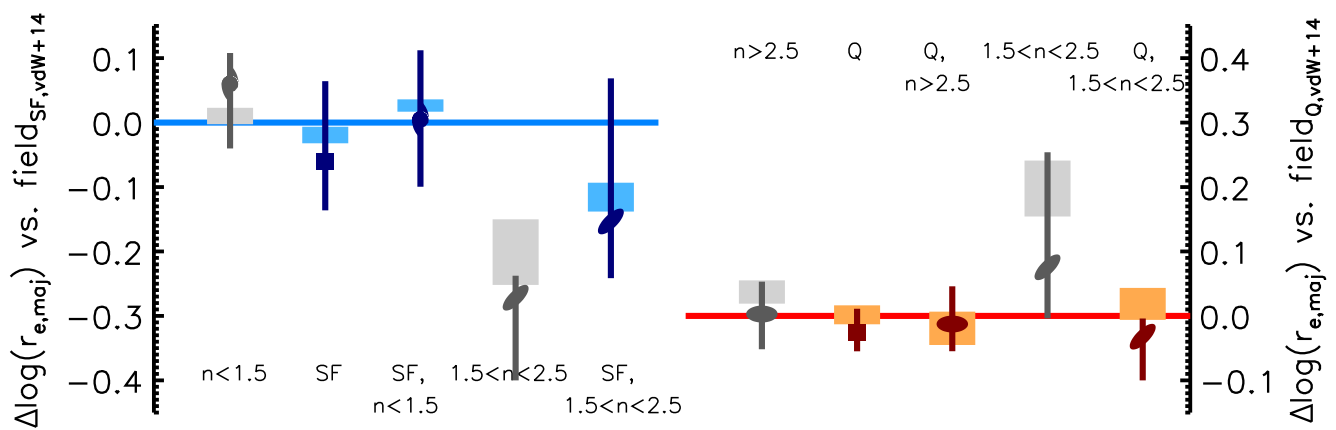} 
 \caption{Median log size difference of log(M/M$_{\odot}$)>10.85
   cluster galaxies (darker symbols with error bars, corrected for
   residual background subtraction, see text) and of field
   counterparts at $1.3<z<1.8$ (shaded rectangles showing the
   16$^{th}$-84$^{th}$ percentile range, see text) with respect to the
   expected size from the \citet{vanderwel2014} star-forming (left
   panel) and quiescent (right panel) relations. Median differences
   are shown for different sub-samples based on Sersic index (symbols
   as in Fig.~\ref{fig:masssize}, or squares for no Sersic index
   selection) and star-forming (blue colour shades) vs quiescent (red
   colour shades) classification (grey symbols show samples selected by
   Sersic index only, regardless of star-forming vs quiescent
   classification; see text).
  \label{fig:deltasize}}
\end{figure*}

In Figure~\ref{fig:deltasize} we show the median size
difference\footnote{As stated throughout, size measurements of cluster
  galaxies considered here are corrected by the estimated systematic
  offset with respect to the adopted control field measurements, as
  discussed in Sect.~\ref{sec:datamorph}. For the sake of
  completeness, we show for reference the corresponding measurements
  of the median size difference of cluster versus field galaxies without
  this correction in Fig.~\ref{fig:deltasizefullfig}.} of cluster and
field galaxies more massive than log(M/M$_{\odot}$)=10.85, adopting as
a reference the \citet{vanderwel2014} best-fit mass-size relations for
star-forming (left panel) and quiescent (right panel) galaxies. For
each galaxy in the combined cluster and in the control field samples,
we estimate the log difference of the measured size with respect to
the expected size from the \citet{vanderwel2014}
relations\footnote{The \citet{vanderwel2014} relations are scaled to
  sizes at 5500~\AA\ restframe, for consistency with the adopted
  measurements (see Sect.~\ref{sec:datamorph}).  } for a galaxy of the
given stellar mass and redshift (for cluster galaxies, the cluster
redshift is assumed).

Given the intrinsic variety within, and overlap between, colour-- and
morphology-- selected galaxy populations, in order to probe the
potential effect of galaxy sample selection on the identification of
environmental signatures on galaxy sizes, we consider for both cluster
and field samples several galaxy sub-samples (always excluding
BCGs). As shown in Fig.~\ref{fig:deltasize}, we divide the population
according to Sersic index ($n<1.5$, $1.5<n<2.5$, $n>2.5$) and
star-forming vs\ quiescent classification (to limit differential
contamination issues between the quiescent and star-forming
populations along the border of the UVJ-like classification in cluster
versus field samples, Fig.~\ref{fig:deltasize} shows results excluding
the $\pm$0.1 mag around the star-forming vs\ quiescent classification
border, but the impact of this removal remains well within the
$1\sigma$ uncertainties for all measurements shown, and has no impact
on the results).  For each sub-sample, we compute for both cluster and
field populations the median log difference of the measured galaxy
size with respect to the expected size.  To account for background
contamination of the cluster candidate member sample, for each
sub-sample we estimate the expected number of contaminants based on the
statistical weights described in Sect.~\ref{sec:galsamples}, and
randomly remove the estimated number of interlopers according to the
size distribution of the corresponding field sub-sample.  The
uncertainties shown in Fig.~\ref{fig:deltasize} for both cluster and
field median size offsets with respect to the \citet{vanderwel2014}
mass-size relations are estimated from the
16$^{\textrm{th}}$-84$^{\textrm{th}}$ percentile range of the
distribution of median size offsets obtained by bootstrapping 10000
times for each sub-sample. The quoted uncertainties thus reflect
Poisson noise given the sample size, and background contamination
removal for cluster galaxy samples.  We note that, given the lack of
evidence for a redshift dependence of the median size offset of the
control field sample (see Fig.~\ref{fig:deltasizefullfig}) we do not
match the redshift distribution of field galaxies to that of the
cluster galaxy sample, but use instead the whole selected field sample
in the $1.3<z<1.8$ range.

\begin{figure}[t]
  \centering
  \includegraphics[width=0.49\textwidth,viewport=  59 532 303 689, clip]{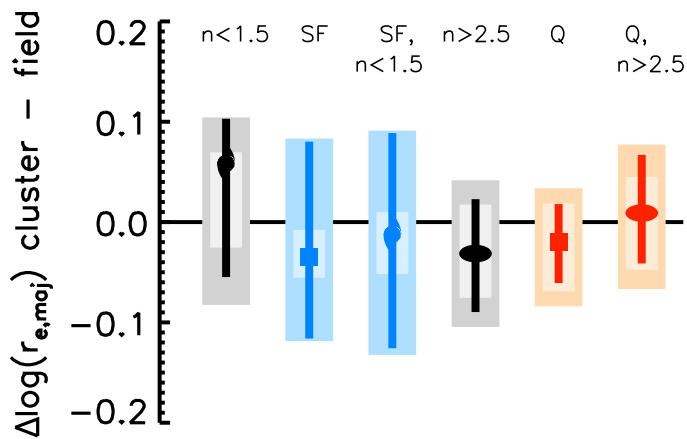} 
 \caption{Median log size difference between cluster and field
   galaxies as derived from Fig.~\ref{fig:deltasize}, for different
   population sub-samples (see text).  Symbols and colour coding follow
   Fig.~\ref{fig:deltasize}. Error bars show statistical uncertainties
   (including statistical background subtraction). Light shaded
   rectangles show the maximum impact on the median difference from
   potential residual uncorrected systematics. Dark shaded rectangles
   show the combined error including both statistical and systematic
   uncertainties (see text).
  \label{fig:deltasizecluvsfield}}
\end{figure}

Figure~\ref{fig:deltasize} shows that for all considered sub-samples
the median size offsets of cluster and field galaxies with respect to
the reference relations are generally consistent\footnote{A K-S
    test further confirms that the distributions of size offsets of
    cluster and field galaxies with respect to the reference relations
    are consistent. A two-sample K-S test is carried out for all
    sub-populations shown in Fig.~\ref{fig:deltasize}, with 10000
    realisations for each sub-population to statistically remove
    background contamination as discussed above. From these tests, the
    cluster and field samples are consistent with being drawn from the
    same distribution in 100\% of the 10000 realisations for all
    sub-populations except the $n<1.5$ (90\%), star-forming (97\%) and
    $1.5 < n < 2.5$ star-forming (83\% of the realisations) samples.
} according to the statistical uncertainties estimated as
above. Cluster versus\ field galaxy size median differences are within
$\sim$10\% (see Fig.~\ref{fig:deltasizecluvsfield}) for all considered
subpopulations with the exception of intermediate Sersic index
sources, where cluster galaxies might be at face value smaller than
field counterparts, but average sizes are still consistent given the
significant uncertainties.  Figures~\ref{fig:masssize} (middle) and
\ref{fig:deltasize} show that this effect, if real, would largely be
due to the different composition of the cluster and field
intermediate-$n$ samples (see earlier discussion on
Fig.~\ref{fig:masssize}), with a larger fraction of star-forming --
and thus typically larger -- sources in the field sample increasing
the average size of intermediate-$n$ field galaxies with respect to
cluster analogues. However, as discussed above in relation to
Fig.~\ref{fig:masssize}, this is at least partly due to the higher
quiescent fraction in the cluster versus field populations, producing a
differential contamination of the intermediate-$n$ samples. Also in
consideration of the significant statistical uncertainty that anyway
affects this estimated size difference, we thus do not comment further
on this potential difference in the following.

The general similarity of sizes of cluster and field galaxies in
Fig.~\ref{fig:deltasize} is reflected in
Fig.~\ref{fig:deltasizecluvsfield} that quantifies more explicitly the
median size difference of cluster versus field galaxies, as derived from
the results in Fig.~\ref{fig:deltasize}.  We focus here on six most
relevant sub-samples, concerning the bulk of star-forming
(disk-dominated galaxies, star-forming galaxies, and star-forming disk
galaxies) and quiescent populations (bulge-dominated galaxies,
quiescent galaxies, and quiescent bulge-dominated galaxies).  Thin
error bars show statistical errors (as derived from bootstrap as
discussed above for Fig.~\ref{fig:deltasize}). We estimate the
possible impact of potential residual systematics on stellar masses,
effective radii and Sersic indices (as discussed in
Sects.~\ref{sec:galsamples} and \ref{sec:datamorph}, respectively) by
applying random combinations of such systematics up to the estimated
maximum level discussed in the relevant Sections, and evaluating the
impact on the estimated median size difference between cluster and
field galaxies, which is shown as the lighter thick error bar in
Fig.~\ref{fig:deltasizecluvsfield}. The darker thick error bar shows
the estimated overall uncertainties (adding in quadrature the
statistical and systematic error contributions).

From Fig.~\ref{fig:deltasizecluvsfield} we conclude that we see no
evidence of a size difference between massive cluster and field
galaxies with these observations in the probed redshift and cluster
mass range, and in the probed cluster regions. On the other hand, we
note that potential differences of up to $\sim$15\% remain within our
1$\sigma$ uncertainties even for our better-constrained quiescent
samples (and up to 20-25\% if including our estimate of potential
unaccounted systematics; for star-forming galaxies our measurements
are even less constraining, as shown in
Fig.~\ref{fig:deltasizecluvsfield}).

\section{Summary and conclusions}

We investigate structural properties of galaxy populations in the
central regions of five of the most massive known clusters at
$1.4\lesssim z \lesssim 1.7$. This cluster sample is homogeneously
selected from the wide-area SPT-SZ Sunyaev Zel’dovich effect survey,
and it is deemed to be representative of the general population of very
massive (M$_{200}>4 \cdot 10^{14}$~M$_{\odot}$) clusters at this
cosmic time.

With a dedicated, homogeneous imaging follow-up from HST and Spitzer,
we probe the connection between broad structural and stellar
population properties. We find that a clear relation between the two
is already in place at this epoch in these cluster environments, which
is in line with several studies in the field up to a similar redshift
\citep[e.g.][see discussion in
  Sect.~\ref{sec:intro}]{wuyts2011,bell2012,kawinwanichakij2017,
  suess2021}. In the probed stellar mass range
(log(M/M$_{\odot}$)>10.85), quiescent galaxies in both cluster and
field environments are largely ($\sim$80\%) dominated by
bulge-dominated sources, while star-forming galaxies are typically
($\sim$60\%) disk-dominated.

Although the morphology mix of both the quiescent and the star-forming
populations is largely the same in clusters and in the field, the
larger quiescent galaxy fraction observed in these clusters (S19) is
reflected in a significant morphology-density relation, with
bulge-dominated galaxies already clearly dominating the massive galaxy
population in these clusters at $z\sim1.5$.

On the other hand, in the probed stellar mass range, we do not detect
any significant environmental dependence of the broad structural
properties of quiescent galaxies, which overall show a similar
fraction of bulge- versus disk-dominated sources, similar average
Sersic indices, as well as a similar axis ratio distribution in the
probed cluster and field environments.  As discussed in
Sect.~\ref{sec:galmorph}, the analysis presented here is affected by
intrinsic limitations that might impact the detection of potential
environmental signatures. Indeed, the adopted photometric
classification of quiescent versus star-forming galaxies is rather
broad in nature, and totally lacks the ability of robustly identifying
recently quenched galaxies transitioning towards the quiescent
population.  Similarly, the classification of bulge-
versus\ disk-dominated systems based on the Sersic index is also
inherently broad. Furthermore, this work focusses on the cluster
central regions, which are dominated by sources that are accreted onto
the cluster at earlier epochs. These limitations may thus affect our
ability to detect signatures that might be mostly visible in more
recently accreted, recently quenched populations.  Nonetheless, the
results from this analysis show no significant environmental
signatures in the correlation between broad structural and stellar
population properties, thus suggesting that a tight connection between
star formation quenching and structural evolution towards a
bulge-dominated morphology also holds in central cluster regions at
this redshift. Any differences in timescales between these two
processes in cluster versus field environments are smaller than we can
disentangle with this analysis and these observations.

We further probe the stellar mass - size relation of cluster galaxies,
finding that star-forming and quiescent cluster galaxy populations
largely follow the same relations as their field counterparts. When
dividing the population according to the estimated Sersic index, we
find that cluster galaxies populate the mass-size plane similarly to
field counterparts. A possible exception might be sources with
intermediate Sersic index, for which cluster galaxies might tend to be
smaller on average due to a larger contribution of quiescent - and
thus smaller - sources; see discussion in
Sect.~\ref{sec:galmasssize}. We estimate the size difference at fixed
stellar mass between analogously selected cluster and field galaxies
for different sub-populations. With our measurements, we do not find
any significant size difference for any of the probed populations. On
the other hand, in relation to the results from previous studies
\citep[e.g.][see more references and a discussion in
  Sect.~\ref{sec:intro}]{andreon2018,matharu2019}, we note that
potential differences of $\lesssim$15\% are within our estimated
statistical uncertainties (1$\sigma$; and up to 20-25\% if including
potential systematics, as discussed in Sect.~\ref{sec:galmasssize}),
even for the best constrained sub-populations.

This work benefits from a large sample of candidate cluster members,
which is unusual for studies of cluster galaxies at this redshift, and
it was made possible in this case by the small yet sizeable sample of
clusters and by their high mass, implying high richness. These
improved statistics result in a better constrained cluster
versus\ field comparison with respect to previous studies at a
similar redshift that rely on single and/or lower mass clusters.

Nonetheless, some of our results remain affected
by limitations, as discussed above, including -- in particular -- the
relatively coarse characterisation of both structural and stellar
population properties, the limited field of view of the observations
probing only relatively central cluster regions, and residual
uncertainties related to cluster membership especially for some of the
probed sub-populations, as discussed. We strive to obtain additional
follow-up observations to improve on these specific aspects, thus
better exploiting the characteristics of this cluster sample to
investigate the interplay between star formation quenching and
structural evolution in connection with environmental effects, which is a
crucial aspect of galaxy evolution within the most massive structures.

~\\

\begin{acknowledgements}
  MP and VS acknowledge support by the German Space Agency (DLR)
  through {\it Verbundforschung} project ID~50OR1603, and by the
  German Research Foundation (DFG) through research grant
  STR~1574/1-1. AS, MP and VS acknowledge support by the ERC-StG
  "ClustersXCosmo" grant agreement 716762. AS acknowledges support
  by the FARE-MIUR grant 'ClustersXEuclid' R165SBKTMA, and by
  the INFN InDark grant. CLR acknowledges support from the Australian
  Research Council’s Discovery Projects scheme (DP200101068). The
  South Pole Telescope is supported by the National Science Foundation
  through grants PLR-1248097, OPP-1852617, and OPP-2147371.  Based on
  observations made with the NASA/ESA {\it Hubble} Space Telescope
  under program GO-14252, obtained from the Data Archive at the Space
  Telescope Science Institute, which is operated by the Association of
  Universities for Research in Astronomy, Inc., under NASA contract
  NAS 5-26555.  Based on observations made with the {\it Spitzer}
  Space Telescope (program ID~12030) which is operated by the Jet
  Propulsion Laboratory, California Institute of Technology under NASA
  contract.

\end{acknowledgements}
%\newpage
\bibliographystyle{aa}
%\bibliography{ms}

\begin{thebibliography}{124}
\expandafter\ifx\csname natexlab\endcsname\relax\def\natexlab#1{#1}\fi

\bibitem[{{Afanasiev} {et~al.}(2022){Afanasiev}, {Mei}, {Fu}, {Shankar},
  {Amodeo}, {Stern}, {Cooke}, {Gonzalez}, {Noirot}, {Rettura}, {Wylezalek}, {De
  Breuck}, {Hatch}, {Stanford}, \& {Vernet}}]{afanasiev2022}
{Afanasiev}, A.~V., {Mei}, S., {Fu}, H., {et~al.} 2022, arXiv e-prints,
  arXiv:2212.00031

\bibitem[{{Alberts} {et~al.}(2021){Alberts}, {Lee}, {Pope}, {Brodwin},
  {Chiang}, {McKinney}, {Xue}, {Huang}, {Brown}, {Dey}, {Eisenhardt},
  {Jannuzi}, {Popescu}, {Ramakrishnan}, {Stanford}, \& {Weiner}}]{alberts2021}
{Alberts}, S., {Lee}, K.-S., {Pope}, A., {et~al.} 2021, \mnras, 501, 1970

\bibitem[{{Alberts} {et~al.}(2016){Alberts}, {Pope}, {Brodwin}, {Chung},
  {Cybulski}, {Dey}, {Eisenhardt}, {Galametz}, {Gonzalez}, {Jannuzi},
  {Stanford}, {Snyder}, {Stern}, \& {Zeimann}}]{alberts2016}
{Alberts}, S., {Pope}, A., {Brodwin}, M., {et~al.} 2016, \apj, 825, 72

\bibitem[{{Andersson} {et~al.}(2011){Andersson}, {Benson}, {Ade}, {Aird},
  {Armstrong}, {Bautz}, {Bleem}, {Brodwin}, {Carlstrom}, {Chang}, {Crawford},
  {Crites}, {de Haan}, {Desai}, {Dobbs}, {Dudley}, {Foley}, {Forman},
  {Garmire}, {George}, {Gladders}, {Halverson}, {High}, {Holder}, {Holzapfel},
  {Hrubes}, {Jones}, {Joy}, {Keisler}, {Knox}, {Lee}, {Leitch}, {Lueker},
  {Marrone}, {McMahon}, {Mehl}, {Meyer}, {Mohr}, {Montroy}, {Murray}, {Padin},
  {Plagge}, {Pryke}, {Reichardt}, {Rest}, {Ruel}, {Ruhl}, {Schaffer}, {Shaw},
  {Shirokoff}, {Song}, {Spieler}, {Stalder}, {Staniszewski}, {Stark}, {Stubbs},
  {Vanderlinde}, {Vieira}, {Vikhlinin}, {Williamson}, {Yang}, {Zahn}, \&
  {Zenteno}}]{andersson2011}
{Andersson}, K., {Benson}, B.~A., {Ade}, P.~A.~R., {et~al.} 2011, \apj, 738, 48

\bibitem[{{Andreon}(2018)}]{andreon2018}
{Andreon}, S. 2018, \aap, 617, A53

\bibitem[{{Andreon} {et~al.}(2009){Andreon}, {Maughan}, {Trinchieri}, \&
  {Kurk}}]{andreon2009}
{Andreon}, S., {Maughan}, B., {Trinchieri}, G., \& {Kurk}, J. 2009, \aap, 507,
  147

\bibitem[{{Andreon} {et~al.}(2014){Andreon}, {Newman}, {Trinchieri},
  {Raichoor}, {Ellis}, \& {Treu}}]{andreon2014}
{Andreon}, S., {Newman}, A.~B., {Trinchieri}, G., {et~al.} 2014, \aap, 565,
  A120

\bibitem[{{Baldry} {et~al.}(2006){Baldry}, {Balogh}, {Bower}, {Glazebrook},
  {Nichol}, {Bamford}, \& {Budavari}}]{baldry2006}
{Baldry}, I.~K., {Balogh}, M.~L., {Bower}, R.~G., {et~al.} 2006, \mnras, 373,
  469

\bibitem[{{Balogh} {et~al.}(2004){Balogh}, {Baldry}, {Nichol}, {Miller},
  {Bower}, \& {Glazebrook}}]{balogh2004}
{Balogh}, M.~L., {Baldry}, I.~K., {Nichol}, R., {et~al.} 2004, \apjl, 615, L101

\bibitem[{{Bassett} {et~al.}(2013){Bassett}, {Papovich}, {Lotz}, {Bell},
  {Finkelstein}, {Newman}, {Tran}, {Almaini}, {Lani}, {Cooper}, {Croton},
  {Dekel}, {Ferguson}, {Kocevski}, {Koekemoer}, {Koo}, {McGrath}, {McIntosh},
  \& {Wechsler}}]{bassett2013}
{Bassett}, R., {Papovich}, C., {Lotz}, J.~M., {et~al.} 2013, \apj, 770, 58

\bibitem[{{Baxter} {et~al.}(2022){Baxter}, {Cooper}, {Balogh}, {Carleton},
  {Cerulo}, {De Lucia}, {Demarco}, {McGee}, {Muzzin}, {Nantais},
  {Pintos-Castro}, {Reeves}, {Rudnick}, {Sarron}, {van der Burg}, {Vulcani},
  {Wilson}, \& {Zaritsky}}]{baxter2022}
{Baxter}, D.~C., {Cooper}, M.~C., {Balogh}, M.~L., {et~al.} 2022, \mnras, 515,
  5479

\bibitem[{{Bayliss} {et~al.}(2014){Bayliss}, {Ashby}, {Ruel}, {Brodwin},
  {Aird}, {Bautz}, {Benson}, {Bleem}, {Bocquet}, {Carlstrom}, {Chang}, {Cho},
  {Clocchiatti}, {Crawford}, {Crites}, {Desai}, {Dobbs}, {Dudley}, {Foley},
  {Forman}, {George}, {Gettings}, {Gladders}, {Gonzalez}, {de Haan},
  {Halverson}, {High}, {Holder}, {Holzapfel}, {Hoover}, {Hrubes}, {Jones},
  {Joy}, {Keisler}, {Knox}, {Lee}, {Leitch}, {Liu}, {Lueker}, {Luong-Van},
  {Mantz}, {Marrone}, {Mawatari}, {McDonald}, {McMahon}, {Mehl}, {Meyer},
  {Miller}, {Mocanu}, {Mohr}, {Montroy}, {Murray}, {Padin}, {Plagge}, {Pryke},
  {Reichardt}, {Rest}, {Ruhl}, {Saliwanchik}, {Saro}, {Sayre}, {Schaffer},
  {Shirokoff}, {Song}, {Stalder}, {{\v S}uhada}, {Spieler}, {Stanford},
  {Staniszewski}, {Stark}, {Story}, {Stubbs}, {van Engelen}, {Vanderlinde},
  {Vieira}, {Vikhlinin}, {Williamson}, {Zahn}, \& {Zenteno}}]{bayliss2014}
{Bayliss}, M.~B., {Ashby}, M.~L.~N., {Ruel}, J., {et~al.} 2014, \apj, 794, 12

\bibitem[{{Bell} {et~al.}(2012){Bell}, {van der Wel}, {Papovich}, {Kocevski},
  {Lotz}, {McIntosh}, {Kartaltepe}, {Faber}, {Ferguson}, {Koekemoer}, {Grogin},
  {Wuyts}, {Cheung}, {Conselice}, {Dekel}, {Dunlop}, {Giavalisco},
  {Herrington}, {Koo}, {McGrath}, {de Mello}, {Rix}, {Robaina}, \&
  {Williams}}]{bell2012}
{Bell}, E.~F., {van der Wel}, A., {Papovich}, C., {et~al.} 2012, \apj, 753, 167

\bibitem[{{Bertin} \& {Arnouts}(1996)}]{sextractor}
{Bertin}, E. \& {Arnouts}, S. 1996, \aaps, 117, 393

\bibitem[{{Bleem} {et~al.}(2015){Bleem}, {Stalder}, {de Haan}, {Aird}, {Allen},
  {Applegate}, {Ashby}, {Bautz}, {Bayliss}, {Benson}, {Bocquet}, {Brodwin},
  {Carlstrom}, {Chang}, {Chiu}, {Cho}, {Clocchiatti}, {Crawford}, {Crites},
  {Desai}, {Dietrich}, {Dobbs}, {Foley}, {Forman}, {George}, {Gladders},
  {Gonzalez}, {Halverson}, {Hennig}, {Hoekstra}, {Holder}, {Holzapfel},
  {Hrubes}, {Jones}, {Keisler}, {Knox}, {Lee}, {Leitch}, {Liu}, {Lueker},
  {Luong-Van}, {Mantz}, {Marrone}, {McDonald}, {McMahon}, {Meyer}, {Mocanu},
  {Mohr}, {Murray}, {Padin}, {Pryke}, {Reichardt}, {Rest}, {Ruel}, {Ruhl},
  {Saliwanchik}, {Saro}, {Sayre}, {Schaffer}, {Schrabback}, {Shirokoff},
  {Song}, {Spieler}, {Stanford}, {Staniszewski}, {Stark}, {Story}, {Stubbs},
  {Vanderlinde}, {Vieira}, {Vikhlinin}, {Williamson}, {Zahn}, \&
  {Zenteno}}]{bleem2015}
{Bleem}, L.~E., {Stalder}, B., {de Haan}, T., {et~al.} 2015, \apjs, 216, 27

\bibitem[{{Bocquet} {et~al.}(2019){Bocquet}, {Dietrich}, {Schrabback}, {Bleem},
  {Klein}, {Allen}, {Applegate}, {Ashby}, {Bautz}, {Bayliss}, {Benson},
  {Brodwin}, {Bulbul}, {Canning}, {Capasso}, {Carlstrom}, {Chang}, {Chiu},
  {Cho}, {Clocchiatti}, {Crawford}, {Crites}, {de Haan}, {Desai}, {Dobbs},
  {Foley}, {Forman}, {Garmire}, {George}, {Gladders}, {Gonzalez}, {Grandis},
  {Gupta}, {Halverson}, {Hlavacek-Larrondo}, {Hoekstra}, {Holder}, {Holzapfel},
  {Hou}, {Hrubes}, {Huang}, {Jones}, {Khullar}, {Knox}, {Kraft}, {Lee}, {von
  der Linden}, {Luong-Van}, {Mantz}, {Marrone}, {McDonald}, {McMahon}, {Meyer},
  {Mocanu}, {Mohr}, {Morris}, {Padin}, {Patil}, {Pryke}, {Rapetti},
  {Reichardt}, {Rest}, {Ruhl}, {Saliwanchik}, {Saro}, {Sayre}, {Schaffer},
  {Shirokoff}, {Stalder}, {Stanford}, {Staniszewski}, {Stark}, {Story},
  {Strazzullo}, {Stubbs}, {Vanderlinde}, {Vieira}, {Vikhlinin}, {Williamson},
  \& {Zenteno}}]{bocquet2019}
{Bocquet}, S., {Dietrich}, J.~P., {Schrabback}, T., {et~al.} 2019, \apj, 878,
  55

\bibitem[{{Bocquet} {et~al.}(2015){Bocquet}, {Saro}, {Mohr}, {Aird}, {Ashby},
  {Bautz}, {Bayliss}, {Bazin}, {Benson}, {Bleem}, {Brodwin}, {Carlstrom},
  {Chang}, {Chiu}, {Cho}, {Clocchiatti}, {Crawford}, {Crites}, {Desai}, {de
  Haan}, {Dietrich}, {Dobbs}, {Foley}, {Forman}, {Gangkofner}, {George},
  {Gladders}, {Gonzalez}, {Halverson}, {Hennig}, {Hlavacek-Larrondo}, {Holder},
  {Holzapfel}, {Hrubes}, {Jones}, {Keisler}, {Knox}, {Lee}, {Leitch}, {Liu},
  {Lueker}, {Luong-Van}, {Marrone}, {McDonald}, {McMahon}, {Meyer}, {Mocanu},
  {Murray}, {Padin}, {Pryke}, {Reichardt}, {Rest}, {Ruel}, {Ruhl},
  {Saliwanchik}, {Sayre}, {Schaffer}, {Shirokoff}, {Spieler}, {Stalder},
  {Stanford}, {Staniszewski}, {Stark}, {Story}, {Stubbs}, {Vanderlinde},
  {Vieira}, {Vikhlinin}, {Williamson}, {Zahn}, \& {Zenteno}}]{bocquet2015}
{Bocquet}, S., {Saro}, A., {Mohr}, J.~J., {et~al.} 2015, \apj, 799, 214

\bibitem[{{Boselli} {et~al.}(2016){Boselli}, {Roehlly}, {Fossati}, {Buat},
  {Boissier}, {Boquien}, {Burgarella}, {Ciesla}, {Gavazzi}, \&
  {Serra}}]{boselli2016}
{Boselli}, A., {Roehlly}, Y., {Fossati}, M., {et~al.} 2016, \aap, 596, A11

\bibitem[{{Brammer} {et~al.}(2012){Brammer}, {van Dokkum}, {Franx},
  {Fumagalli}, {Patel}, {Rix}, {Skelton}, {Kriek}, {Nelson}, {Schmidt},
  {Bezanson}, {da Cunha}, {Erb}, {Fan}, {F{\"o}rster Schreiber}, {Illingworth},
  {Labb{\'e}}, {Leja}, {Lundgren}, {Magee}, {Marchesini}, {McCarthy},
  {Momcheva}, {Muzzin}, {Quadri}, {Steidel}, {Tal}, {Wake}, {Whitaker}, \&
  {Williams}}]{brammer2012}
{Brammer}, G.~B., {van Dokkum}, P.~G., {Franx}, M., {et~al.} 2012, \apjs, 200,
  13

\bibitem[{{Brodwin} {et~al.}(2012){Brodwin}, {Gonzalez}, {Stanford}, {Plagge},
  {Marrone}, {Carlstrom}, {Dey}, {Eisenhardt}, {Fedeli}, {Gettings}, {Jannuzi},
  {Joy}, {Leitch}, {Mancone}, {Snyder}, {Stern}, \& {Zeimann}}]{brodwin2012}
{Brodwin}, M., {Gonzalez}, A.~H., {Stanford}, S.~A., {et~al.} 2012, \apj, 753,
  162

\bibitem[{{Brodwin} {et~al.}(2013){Brodwin}, {Stanford}, {Gonzalez}, {Zeimann},
  {Snyder}, {Mancone}, {Pope}, {Eisenhardt}, {Stern}, {Alberts}, {Ashby},
  {Brown}, {Chary}, {Dey}, {Galametz}, {Gettings}, {Jannuzi}, {Miller},
  {Moustakas}, \& {Moustakas}}]{brodwin2013}
{Brodwin}, M., {Stanford}, S.~A., {Gonzalez}, A.~H., {et~al.} 2013, \apj, 779,
  138

\bibitem[{{Bruzual} \& {Charlot}(2003)}]{BC03}
{Bruzual}, G. \& {Charlot}, S. 2003, \mnras, 344, 1000

\bibitem[{{Cameron}(2011)}]{cameron2011p}
{Cameron}, E. 2011, PASA, 28, 128

\bibitem[{{Capak} {et~al.}(2007){Capak}, {Abraham}, {Ellis}, {Mobasher},
  {Scoville}, {Sheth}, \& {Koekemoer}}]{capak2007b}
{Capak}, P., {Abraham}, R.~G., {Ellis}, R.~S., {et~al.} 2007, \apjs, 172, 284

\bibitem[{{Cappellari} {et~al.}(2011){Cappellari}, {Emsellem}, {Krajnovi{\'c}},
  {McDermid}, {Serra}, {Alatalo}, {Blitz}, {Bois}, {Bournaud}, {Bureau},
  {Davies}, {Davis}, {de Zeeuw}, {Khochfar}, {Kuntschner}, {Lablanche},
  {Morganti}, {Naab}, {Oosterloo}, {Sarzi}, {Scott}, {Weijmans}, \&
  {Young}}]{cappellari2011}
{Cappellari}, M., {Emsellem}, E., {Krajnovi{\'c}}, D., {et~al.} 2011, \mnras,
  416, 1680

\bibitem[{{Carlberg} {et~al.}(2001){Carlberg}, {Yee}, {Morris}, {Lin}, {Hall},
  {Patton}, {Sawicki}, \& {Shepherd}}]{carlberg2001}
{Carlberg}, R.~G., {Yee}, H.~K.~C., {Morris}, S.~L., {et~al.} 2001, \apj, 563,
  736

\bibitem[{{Carlstrom} {et~al.}(2011){Carlstrom}, {Ade}, {Aird}, {Benson},
  {Bleem}, {Busetti}, {Chang}, {Chauvin}, {Cho}, {Crawford}, {Crites}, {Dobbs},
  {Halverson}, {Heimsath}, {Holzapfel}, {Hrubes}, {Joy}, {Keisler}, {Lanting},
  {Lee}, {Leitch}, {Leong}, {Lu}, {Lueker}, {Luong-Van}, {McMahon}, {Mehl},
  {Meyer}, {Mohr}, {Montroy}, {Padin}, {Plagge}, {Pryke}, {Ruhl}, {Schaffer},
  {Schwan}, {Shirokoff}, {Spieler}, {Staniszewski}, {Stark}, {Tucker},
  {Vanderlinde}, {Vieira}, \& {Williamson}}]{carlstrom2011}
{Carlstrom}, J.~E., {Ade}, P.~A.~R., {Aird}, K.~A., {et~al.} 2011, \pasp, 123,
  568

\bibitem[{{Carollo} {et~al.}(2016){Carollo}, {Cibinel}, {Lilly}, {Pipino},
  {Bonoli}, {Finoguenov}, {Miniati}, {Norberg}, \& {Silverman}}]{carollo2016}
{Carollo}, C.~M., {Cibinel}, A., {Lilly}, S.~J., {et~al.} 2016, \apj, 818, 180

\bibitem[{{Chan} {et~al.}(2018){Chan}, {Beifiori}, {Saglia}, {Mendel}, {Stott},
  {Bender}, {Galametz}, {Wilman}, {Cappellari}, {Davies}, {Houghton},
  {Prichard}, {Lewis}, {Sharples}, \& {Wegner}}]{chan2018}
{Chan}, J. C.~C., {Beifiori}, A., {Saglia}, R.~P., {et~al.} 2018, \apj, 856, 8

\bibitem[{{Chan} {et~al.}(2021){Chan}, {Wilson}, {Balogh}, {Rudnick}, {van der
  Burg}, {Muzzin}, {Webb}, {Biviano}, {Cerulo}, {Cooper}, {De Lucia},
  {Demarco}, {Forrest}, {Jablonka}, {Lidman}, {McGee}, {Nantais}, {Old},
  {Pintos-Castro}, {Poggianti}, {Reeves}, {Vulcani}, {Yee}, \&
  {Zaritsky}}]{chan2021}
{Chan}, J. C.~C., {Wilson}, G., {Balogh}, M., {et~al.} 2021, \apj, 920, 32

\bibitem[{{Cooke} {et~al.}(2016){Cooke}, {Hatch}, {Stern}, {Rettura},
  {Brodwin}, {Galametz}, {Wylezalek}, {Bridge}, {Conselice}, {De Breuck},
  {Gonzalez}, \& {Jarvis}}]{cooke2016}
{Cooke}, E.~A., {Hatch}, N.~A., {Stern}, D., {et~al.} 2016, \apj, 816, 83

\bibitem[{de~Haan {et~al.}(2016)de~Haan, Benson, Bleem, Allen, Applegate,
  Ashby, Bautz, Bayliss, Bocquet, Brodwin, Carlstrom, Chang, Chiu, Cho,
  Clocchiatti, Crawford, Crites, Desai, Dietrich, Dobbs, Doucouliagos, Foley,
  Forman, Garmire, George, Gladders, Gonzalez, Gupta, Halverson,
  Hlavacek-Larrondo, Hoekstra, Holder, Holzapfel, Hou, Hrubes, Huang, Jones,
  Keisler, Knox, Lee, Leitch, von~der Linden, Luong-Van, Mantz, Marrone,
  McDonald, McMahon, Meyer, Mocanu, Mohr, Murray, Padin, Pryke, Rapetti,
  Reichardt, Rest, Ruel, Ruhl, Saliwanchik, Saro, Sayre, Schaffer, Schrabback,
  Shirokoff, Song, Spieler, Stalder, Stanford, Staniszewski, Stark, Story,
  Stubbs, Vanderlinde, Vieira, Vikhlinin, Williamson, \& Zenteno}]{dehaan2016}
de~Haan, T., Benson, B., Bleem, L., {et~al.} 2016, \apj, 832, 95

\bibitem[{{de Vaucouleurs}(1961)}]{devaucouleurs1961}
{de Vaucouleurs}, G. 1961, \apjs, 5, 233

\bibitem[{{Delaye} {et~al.}(2014){Delaye}, {Huertas-Company}, {Mei}, {Lidman},
  {Licitra}, {Newman}, {Raichoor}, {Shankar}, {Barrientos}, {Bernardi},
  {Cerulo}, {Couch}, {Demarco}, {Mu{\~n}oz}, {S{\'a}nchez-Janssen}, \&
  {Tanaka}}]{delaye2014}
{Delaye}, L., {Huertas-Company}, M., {Mei}, S., {et~al.} 2014, \mnras, 441, 203

\bibitem[{{Di Mascolo} {et~al.}(2020){Di Mascolo}, {Mroczkowski}, {Churazov},
  {Moravec}, {Brodwin}, {Gonzalez}, {Decker}, {Eisenhardt}, {Stanford},
  {Stern}, {Sunyaev}, \& {Wylezalek}}]{dimascolo2020}
{Di Mascolo}, L., {Mroczkowski}, T., {Churazov}, E., {et~al.} 2020, \aap, 638,
  A70

\bibitem[{{Dicker} {et~al.}(2020){Dicker}, {Romero}, {Di Mascolo},
  {Mroczkowski}, {Sievers}, {Moravec}, {Bhandarkar}, {Brodwin}, {Connor},
  {Decker}, {Devlin}, {Gonzalez}, {Lowe}, {Mason}, {Sarazin}, {Stanford},
  {Stern}, {Thongkham}, {Wylezalek}, \& {Zago}}]{dicker2020}
{Dicker}, S.~R., {Romero}, C.~E., {Di Mascolo}, L., {et~al.} 2020, \apj, 902,
  144

\bibitem[{{Dressler}(1980)}]{dressler1980}
{Dressler}, A. 1980, \apj, 236, 351

\bibitem[{{Finner} {et~al.}(2020){Finner}, {James Jee}, {Webb}, {Wilson},
  {Perlmutter}, {Muzzin}, \& {Hlavacek-Larrondo}}]{finner2020}
{Finner}, K., {James Jee}, M., {Webb}, T., {et~al.} 2020, \apj, 893, 10

\bibitem[{{Fioc} \& {Rocca-Volmerange}(1999)}]{fioceroccavolmerange1999}
{Fioc}, M. \& {Rocca-Volmerange}, B. 1999, \aap, 351, 869

\bibitem[{{Gobat} {et~al.}(2008){Gobat}, {Rosati}, {Strazzullo}, {Rettura},
  {Demarco}, \& {Nonino}}]{gobat2008}
{Gobat}, R., {Rosati}, P., {Strazzullo}, V., {et~al.} 2008, \aap, 488, 853

\bibitem[{{G{\'o}mez} {et~al.}(2003){G{\'o}mez}, {Nichol}, {Miller}, {Balogh},
  {Goto}, {Zabludoff}, {Romer}, {Bernardi}, {Sheth}, {Hopkins}, {Castander},
  {Connolly}, {Schneider}, {Brinkmann}, {Lamb}, {SubbaRao}, \&
  {York}}]{gomez2003}
{G{\'o}mez}, P.~L., {Nichol}, R.~C., {Miller}, C.~J., {et~al.} 2003, \apj, 584,
  210

\bibitem[{{Goto} {et~al.}(2003){Goto}, {Yamauchi}, {Fujita}, {Okamura},
  {Sekiguchi}, {Smail}, {Bernardi}, \& {Gomez}}]{goto2003}
{Goto}, T., {Yamauchi}, C., {Fujita}, Y., {et~al.} 2003, \mnras, 346, 601

\bibitem[{{Grogin} {et~al.}(2011){Grogin}, {Kocevski}, {Faber}, {Ferguson},
  {Koekemoer}, {Riess}, {Acquaviva}, {Alexander}, {Almaini}, {Ashby}, {Barden},
  {Bell}, {Bournaud}, {Brown}, {Caputi}, {Casertano}, {Cassata}, {Castellano},
  {Challis}, {Chary}, {Cheung}, {Cirasuolo}, {Conselice}, {Roshan Cooray},
  {Croton}, {Daddi}, {Dahlen}, {Dav{\'e}}, {de Mello}, {Dekel}, {Dickinson},
  {Dolch}, {Donley}, {Dunlop}, {Dutton}, {Elbaz}, {Fazio}, {Filippenko},
  {Finkelstein}, {Fontana}, {Gardner}, {Garnavich}, {Gawiser}, {Giavalisco},
  {Grazian}, {Guo}, {Hathi}, {H{\"a}ussler}, {Hopkins}, {Huang}, {Huang},
  {Jha}, {Kartaltepe}, {Kirshner}, {Koo}, {Lai}, {Lee}, {Li}, {Lotz}, {Lucas},
  {Madau}, {McCarthy}, {McGrath}, {McIntosh}, {McLure}, {Mobasher},
  {Moustakas}, {Mozena}, {Nandra}, {Newman}, {Niemi}, {Noeske}, {Papovich},
  {Pentericci}, {Pope}, {Primack}, {Rajan}, {Ravindranath}, {Reddy}, {Renzini},
  {Rix}, {Robaina}, {Rodney}, {Rosario}, {Rosati}, {Salimbeni}, {Scarlata},
  {Siana}, {Simard}, {Smidt}, {Somerville}, {Spinrad}, {Straughn}, {Strolger},
  {Telford}, {Teplitz}, {Trump}, {van der Wel}, {Villforth}, {Wechsler},
  {Weiner}, {Wiklind}, {Wild}, {Wilson}, {Wuyts}, {Yan}, \& {Yun}}]{grogin2011}
{Grogin}, N.~A., {Kocevski}, D.~D., {Faber}, S.~M., {et~al.} 2011, \apjs, 197,
  35

\bibitem[{{Hayashi} {et~al.}(2010){Hayashi}, {Kodama}, {Koyama}, {Tanaka},
  {Shimasaku}, \& {Okamura}}]{hayashi2010}
{Hayashi}, M., {Kodama}, T., {Koyama}, Y., {et~al.} 2010, \mnras, 402, 1980

\bibitem[{{Hill} {et~al.}(2019){Hill}, {van der Wel}, {Franx}, {Muzzin},
  {Skelton}, {Momcheva}, {van Dokkum}, \& {Whitaker}}]{hill2019}
{Hill}, A.~R., {van der Wel}, A., {Franx}, M., {et~al.} 2019, \apj, 871, 76

\bibitem[{{Holden} {et~al.}(2007){Holden}, {Illingworth}, {Franx}, {Blakeslee},
  {Postman}, {Kelson}, {van der Wel}, {Demarco}, {Magee}, {Tran}, {Zirm},
  {Ford}, {Rosati}, \& {Homeier}}]{holden2007}
{Holden}, B.~P., {Illingworth}, G.~D., {Franx}, M., {et~al.} 2007, \apj, 670,
  190

\bibitem[{{Humason}(1936)}]{humason1936}
{Humason}, M.~L. 1936, \apj, 83, 10

\bibitem[{{J{\o}rgensen} {et~al.}(2017){J{\o}rgensen}, {Chiboucas}, {Berkson},
  {Smith}, {Takamiya}, \& {Villaume}}]{jorgensen2017}
{J{\o}rgensen}, I., {Chiboucas}, K., {Berkson}, E., {et~al.} 2017, \aj, 154,
  251

\bibitem[{{Kauffmann} {et~al.}(2004){Kauffmann}, {White}, {Heckman},
  {M{\'e}nard}, {Brinchmann}, {Charlot}, {Tremonti}, {Brinkmann}, \&
  {whatever}}]{kauffmann2004}
{Kauffmann}, G., {White}, S.~D.~M., {Heckman}, T., {et~al.} 2004, \mnras, 353,
  713

\bibitem[{{Kawinwanichakij} {et~al.}(2017){Kawinwanichakij}, {Papovich},
  {Quadri}, {Glazebrook}, {Kacprzak}, {Allen}, {Bell}, {Croton}, {Dekel},
  {Ferguson}, {Forrest}, {Grogin}, {Guo}, {Kocevski}, {Koekemoer}, {Labb{\'e}},
  {Lucas}, {Nanayakkara}, {Spitler}, {Straatman}, {Tran}, {Tomczak}, \& {van
  Dokkum}}]{kawinwanichakij2017}
{Kawinwanichakij}, L., {Papovich}, C., {Quadri}, R.~F., {et~al.} 2017, \apj,
  847, 134

\bibitem[{{Khullar} {et~al.}(2022){Khullar}, {Bayliss}, {Gladders}, {Kim},
  {Calzadilla}, {Strazzullo}, {Bleem}, {Mahler}, {McDonald}, {Floyd},
  {Reichardt}, {Ruppin}, {Saro}, {Sharon}, {Somboonpanyakul}, {Stalder}, \&
  {Stark}}]{khullar2022}
{Khullar}, G., {Bayliss}, M.~B., {Gladders}, M.~D., {et~al.} 2022, \apj, 934,
  177

\bibitem[{{Khullar} {et~al.}(2019){Khullar}, {Bleem}, {Bayliss}, {Gladders},
  {Benson}, {McDonald}, {Allen}, {Applegate}, {Ashby}, {Bocquet}, {Brodwin},
  {Bulbul}, {Canning}, {Capasso}, {Chiu}, {Crawford}, {de Haan}, {Dietrich},
  {Gonzalez}, {Hlavacek-Larrondo}, {Hoekstra}, {Holzapfel}, {von der Linden},
  {Mantz}, {Patil}, {Reichardt}, {Saro}, {Sharon}, {Stalder}, {Stanford},
  {Stark}, \& {Strazzullo}}]{khullar2019}
{Khullar}, G., {Bleem}, L.~E., {Bayliss}, M.~B., {et~al.} 2019, \apj, 870, 7

\bibitem[{{Kim} {et~al.}(2018){Kim}, {Malhotra}, {Rhoads}, {Joshi}, {Fererras},
  \& {Pasquali}}]{kim2018}
{Kim}, K., {Malhotra}, S., {Rhoads}, J.~E., {et~al.} 2018, \apj, 867, 118

\bibitem[{{Kodama} {et~al.}(2007){Kodama}, {Tanaka}, {Kajisawa}, {Kurk},
  {Venemans}, {De Breuck}, {Vernet}, \& {Lidman}}]{kodama2007}
{Kodama}, T., {Tanaka}, I., {Kajisawa}, M., {et~al.} 2007, \mnras, 377, 1717

\bibitem[{{Koekemoer} {et~al.}(2011){Koekemoer}, {Faber}, {Ferguson}, {Grogin},
  {Kocevski}, {Koo}, {Lai}, {Lotz}, {Lucas}, {McGrath}, {Ogaz}, {Rajan},
  {Riess}, {Rodney}, {Strolger}, {Casertano}, {Castellano}, {Dahlen},
  {Dickinson}, {Dolch}, {Fontana}, {Giavalisco}, {Grazian}, {Guo}, {Hathi},
  {Huang}, {van der Wel}, {Yan}, {Acquaviva}, {Alexander}, {Almaini}, {Ashby},
  {Barden}, {Bell}, {Bournaud}, {Brown}, {Caputi}, {Cassata}, {Challis},
  {Chary}, {Cheung}, {Cirasuolo}, {Conselice}, {Roshan Cooray}, {Croton},
  {Daddi}, {Dav{\'e}}, {de Mello}, {de Ravel}, {Dekel}, {Donley}, {Dunlop},
  {Dutton}, {Elbaz}, {Fazio}, {Filippenko}, {Finkelstein}, {Frazer}, {Gardner},
  {Garnavich}, {Gawiser}, {Gruetzbauch}, {Hartley}, {H{\"a}ussler},
  {Herrington}, {Hopkins}, {Huang}, {Jha}, {Johnson}, {Kartaltepe},
  {Khostovan}, {Kirshner}, {Lani}, {Lee}, {Li}, {Madau}, {McCarthy},
  {McIntosh}, {McLure}, {McPartland}, {Mobasher}, {Moreira}, {Mortlock},
  {Moustakas}, {Mozena}, {Nandra}, {Newman}, {Nielsen}, {Niemi}, {Noeske},
  {Papovich}, {Pentericci}, {Pope}, {Primack}, {Ravindranath}, {Reddy},
  {Renzini}, {Rix}, {Robaina}, {Rosario}, {Rosati}, {Salimbeni}, {Scarlata},
  {Siana}, {Simard}, {Smidt}, {Snyder}, {Somerville}, {Spinrad}, {Straughn},
  {Telford}, {Teplitz}, {Trump}, {Vargas}, {Villforth}, {Wagner}, {Wandro},
  {Wechsler}, {Weiner}, {Wiklind}, {Wild}, {Wilson}, {Wuyts}, \&
  {Yun}}]{koekemoer2011}
{Koekemoer}, A.~M., {Faber}, S.~M., {Ferguson}, H.~C., {et~al.} 2011, \apjs,
  197, 36

\bibitem[{{Labb{\'e}} {et~al.}(2005){Labb{\'e}}, {Huang}, {Franx}, {Rudnick},
  {Barmby}, {Daddi}, {van Dokkum}, {Fazio}, {Schreiber}, {Moorwood}, {Rix},
  {R{\"o}ttgering}, {Trujillo}, \& {van der Werf}}]{labbe2005}
{Labb{\'e}}, I., {Huang}, J., {Franx}, M., {et~al.} 2005, \apjl, 624, L81

\bibitem[{{Lang} {et~al.}(2014){Lang}, {Wuyts}, {Somerville}, {F{\"o}rster
  Schreiber}, {Genzel}, {Bell}, {Brammer}, {Dekel}, {Faber}, {Ferguson},
  {Grogin}, {Kocevski}, {Koekemoer}, {Lutz}, {McGrath}, {Momcheva}, {Nelson},
  {Primack}, {Rosario}, {Skelton}, {Tacconi}, {van Dokkum}, \&
  {Whitaker}}]{lang2014}
{Lang}, P., {Wuyts}, S., {Somerville}, R.~S., {et~al.} 2014, \apj, 788, 11

\bibitem[{{Lani} {et~al.}(2013){Lani}, {Almaini}, {Hartley}, {Mortlock},
  {H{\"a}u{\ss}ler}, {Chuter}, {Simpson}, {van der Wel}, {Gr{\"u}tzbauch},
  {Conselice}, {Bradshaw}, {Cooper}, {Faber}, {Grogin}, {Kocevski},
  {Koekemoer}, \& {Lai}}]{lani2013}
{Lani}, C., {Almaini}, O., {Hartley}, W.~G., {et~al.} 2013, \mnras, 435, 207

\bibitem[{{Lee} {et~al.}(2013){Lee}, {Giavalisco}, {Williams}, {Guo}, {Lotz},
  {Van der Wel}, {Ferguson}, {Faber}, {Koekemoer}, {Grogin}, {Kocevski},
  {Conselice}, {Wuyts}, {Dekel}, {Kartaltepe}, \& {Bell}}]{lee2013}
{Lee}, B., {Giavalisco}, M., {Williams}, C.~C., {et~al.} 2013, \apj, 774, 47

\bibitem[{{Lemaux} {et~al.}(2019){Lemaux}, {Tomczak}, {Lubin}, {Gal}, {Shen},
  {Pelliccia}, {Wu}, {Hung}, {Mei}, {Le F{\`e}vre}, {Rumbaugh}, {Kocevski}, \&
  {Squires}}]{lemaux2019}
{Lemaux}, B.~C., {Tomczak}, A.~R., {Lubin}, L.~M., {et~al.} 2019, \mnras, 490,
  1231

\bibitem[{{Lustig} {et~al.}(2021){Lustig}, {Strazzullo}, {D'Eugenio}, {Daddi},
  {Pannella}, {Renzini}, {Cimatti}, {Gobat}, {Jin}, {Mohr}, \&
  {Onodera}}]{lustig2021}
{Lustig}, P., {Strazzullo}, V., {D'Eugenio}, C., {et~al.} 2021, \mnras, 501,
  2659

\bibitem[{{Mancone} {et~al.}(2010){Mancone}, {Gonzalez}, {Brodwin}, {Stanford},
  {Eisenhardt}, {Stern}, \& {Jones}}]{mancone2010}
{Mancone}, C.~L., {Gonzalez}, A.~H., {Brodwin}, M., {et~al.} 2010, \apj, 720,
  284

\bibitem[{{Mantz} {et~al.}(2020){Mantz}, {Allen}, {Morris}, {Canning},
  {Bayliss}, {Bleem}, {Floyd}, \& {McDonald}}]{mantz2020}
{Mantz}, A.~B., {Allen}, S.~W., {Morris}, R.~G., {et~al.} 2020, \mnras, 496,
  1554

\bibitem[{{Matharu} {et~al.}(2019){Matharu}, {Muzzin}, {Brammer}, {van der
  Burg}, {Auger}, {Hewett}, {van der Wel}, {van Dokkum}, {Balogh}, {Chan},
  {Demarco}, {Marchesini}, {Nelson}, {Noble}, {Wilson}, \& {Yee}}]{matharu2019}
{Matharu}, J., {Muzzin}, A., {Brammer}, G.~B., {et~al.} 2019, \mnras, 484, 595

\bibitem[{{McConachie} {et~al.}(2022){McConachie}, {Wilson}, {Forrest},
  {Marsan}, {Muzzin}, {Cooper}, {Annunziatella}, {Marchesini}, {Chan}, {Gomez},
  {Abdullah}, {Saracco}, \& {Nantais}}]{mcconachie2022}
{McConachie}, I., {Wilson}, G., {Forrest}, B., {et~al.} 2022, \apj, 926, 37

\bibitem[{{McLure} {et~al.}(2013){McLure}, {Pearce}, {Dunlop}, {Cirasuolo},
  {Curtis-Lake}, {Bruce}, {Caputi}, {Almaini}, {Bonfield}, {Bradshaw},
  {Buitrago}, {Chuter}, {Foucaud}, {Hartley}, \& {Jarvis}}]{mclure2013}
{McLure}, R.~J., {Pearce}, H.~J., {Dunlop}, J.~S., {et~al.} 2013, \mnras, 428,
  1088

\bibitem[{{Mei} {et~al.}(2022){Mei}, {Hatch}, {Amodeo}, {Afanasiev}, {De
  Breuck}, {Stern}, {Cooke}, {Gonzalez}, {Noirot}, {Rettura}, {Seymour},
  {Stanford}, {Vernet}, \& {Wylezalek}}]{mei2022}
{Mei}, S., {Hatch}, N.~A., {Amodeo}, S., {et~al.} 2022, arXiv e-prints,
  arXiv:2209.02078

\bibitem[{{Mei} {et~al.}(2009){Mei}, {Holden}, {Blakeslee}, {Ford}, {Franx},
  {Homeier}, {Illingworth}, {Jee}, {Overzier}, {Postman}, {Rosati}, {Van der
  Wel}, \& {Bartlett}}]{mei2009}
{Mei}, S., {Holden}, B.~P., {Blakeslee}, J.~P., {et~al.} 2009, \apj, 690, 42

\bibitem[{{Mei} {et~al.}(2012){Mei}, {Stanford}, {Holden}, {Raichoor},
  {Postman}, {Nakata}, {Finoguenov}, {Ford}, {Illingworth}, {Kodama}, {Rosati},
  {Tanaka}, {Huertas-Company}, {Rettura}, {Shankar}, {Carrasco}, {Demarco},
  {Eisenhardt}, {Jee}, {Koyama}, \& {White}}]{mei2012}
{Mei}, S., {Stanford}, S.~A., {Holden}, B.~P., {et~al.} 2012, \apj, 754, 141

\bibitem[{{Miller} {et~al.}(2018){Miller}, {Chapman}, {Aravena}, {Ashby},
  {Hayward}, {Vieira}, {Wei{\ss}}, {Babul}, {B{\'e}thermin}, {Bradford},
  {Brodwin}, {Carlstrom}, {Chen}, {Cunningham}, {De Breuck}, {Gonzalez},
  {Greve}, {Harnett}, {Hezaveh}, {Lacaille}, {Litke}, {Ma}, {Malkan},
  {Marrone}, {Morningstar}, {Murphy}, {Narayanan}, {Pass}, {Perry}, {Phadke},
  {Rennehan}, {Rotermund}, {Simpson}, {Spilker}, {Sreevani}, {Stark},
  {Strandet}, \& {Strom}}]{miller2018}
{Miller}, T.~B., {Chapman}, S.~C., {Aravena}, M., {et~al.} 2018, \nat, 556, 469

\bibitem[{{Morgan} \& {Mayall}(1957)}]{morganemayall1957}
{Morgan}, W.~W. \& {Mayall}, N.~U. 1957, \pasp, 69, 291

\bibitem[{{Mowla} {et~al.}(2019){Mowla}, {van Dokkum}, {Brammer}, {Momcheva},
  {van der Wel}, {Whitaker}, {Nelson}, {Bezanson}, {Muzzin}, {Franx},
  {MacKenty}, {Leja}, {Kriek}, \& {Marchesini}}]{mowla2019}
{Mowla}, L.~A., {van Dokkum}, P., {Brammer}, G.~B., {et~al.} 2019, \apj, 880,
  57

\bibitem[{{Mullis} {et~al.}(2005){Mullis}, {Rosati}, {Lamer}, {B{\" o}hringer},
  {Schwope}, {Schuecker}, \& {Fassbender}}]{mullis2005}
{Mullis}, C.~R., {Rosati}, P., {Lamer}, G., {et~al.} 2005, \apjl, 623, L85

\bibitem[{{Nantais} {et~al.}(2017){Nantais}, {Muzzin}, {van der Burg},
  {Wilson}, {Lidman}, {Foltz}, {DeGroot}, {Noble}, {Cooper}, \&
  {Demarco}}]{nantais2017}
{Nantais}, J.~B., {Muzzin}, A., {van der Burg}, R.~F.~J., {et~al.} 2017,
  \mnras, 465, L104

\bibitem[{{Newman} {et~al.}(2014){Newman}, {Ellis}, {Andreon}, {Treu},
  {Raichoor}, \& {Trinchieri}}]{newman2014}
{Newman}, A.~B., {Ellis}, R.~S., {Andreon}, S., {et~al.} 2014, \apj, 788, 51

\bibitem[{{Noordeh} {et~al.}(2021){Noordeh}, {Canning}, {Willis}, {Allen},
  {Mantz}, {Stanford}, \& {Brammer}}]{noordeh2021}
{Noordeh}, E., {Canning}, R.~E.~A., {Willis}, J.~P., {et~al.} 2021, \mnras,
  507, 5272

\bibitem[{{Osborne} {et~al.}(2020){Osborne}, {Salim}, {Damjanov}, {Faber},
  {Huertas-Company}, {Koo}, {Mantha}, {McIntosh}, {Primack}, \&
  {Tacchella}}]{osborne2020}
{Osborne}, C., {Salim}, S., {Damjanov}, I., {et~al.} 2020, \apj, 902, 77

\bibitem[{{Oteo} {et~al.}(2018){Oteo}, {Ivison}, {Dunne}, {Manilla-Robles},
  {Maddox}, {Lewis}, {de Zotti}, {Bremer}, {Clements}, {Cooray}, {Dannerbauer},
  {Eales}, {Greenslade}, {Omont}, {Perez{\textendash}Fourn{\'o}n}, {Riechers},
  {Scott}, {van der Werf}, {Weiss}, \& {Zhang}}]{oteo2018}
{Oteo}, I., {Ivison}, R.~J., {Dunne}, L., {et~al.} 2018, \apj, 856, 72

\bibitem[{{Papovich} {et~al.}(2012){Papovich}, {Bassett}, {Lotz}, {van der
  Wel}, {Tran}, {Finkelstein}, {Bell}, {Conselice}, {Dekel}, {Dunlop}, {Guo},
  {Faber}, {Farrah}, {Ferguson}, {Finkelstein}, {H{\"a}ussler}, {Kocevski},
  {Koekemoer}, {Koo}, {McGrath}, {McLure}, {McIntosh}, {Momcheva}, {Newman},
  {Rudnick}, {Weiner}, {Willmer}, \& {Wuyts}}]{papovich2012}
{Papovich}, C., {Bassett}, R., {Lotz}, J.~M., {et~al.} 2012, \apj, 750, 93

\bibitem[{{Papovich} {et~al.}(2010){Papovich}, {Momcheva}, {Willmer},
  {Finkelstein}, {Finkelstein}, {Tran}, {Brodwin}, {Dunlop}, {Farrah}, {Khan},
  {Lotz}, {McCarthy}, {McLure}, {Rieke}, {Rudnick}, {Sivanandam}, {Pacaud}, \&
  {Pierre}}]{papovich2010}
{Papovich}, C., {Momcheva}, I., {Willmer}, C.~N.~A., {et~al.} 2010, \apj, 716,
  1503

\bibitem[{{Paulino-Afonso} {et~al.}(2019){Paulino-Afonso}, {Sobral}, {Darvish},
  {Ribeiro}, {van der Wel}, {Stott}, {Buitrago}, {Best}, {Stroe}, \&
  {Craig}}]{paulinoafonso2019}
{Paulino-Afonso}, A., {Sobral}, D., {Darvish}, B., {et~al.} 2019, \aap, 630,
  A57

\bibitem[{{Peng} {et~al.}(2002){Peng}, {Ho}, {Impey}, \& {Rix}}]{peng2002}
{Peng}, C.~Y., {Ho}, L.~C., {Impey}, C.~D., \& {Rix}, H. 2002, \aj, 124, 266

\bibitem[{{Peng} {et~al.}(2010){Peng}, {Lilly}, {Kova{\v c}}, {Bolzonella},
  {Pozzetti}, {Renzini}, {Zamorani}, {Ilbert}, {Knobel}, {Iovino}, {Maier},
  {Cucciati}, {Tasca}, {Carollo}, {Silverman}, {Kampczyk}, {de Ravel},
  {Sanders}, {Scoville}, {Contini}, {Mainieri}, {Scodeggio}, {Kneib}, {Le
  F{\`e}vre}, {Bardelli}, {Bongiorno}, {Caputi}, {Coppa}, {de la Torre},
  {Franzetti}, {Garilli}, {Lamareille}, {Le Borgne}, {Le Brun}, {Mignoli},
  {Perez Montero}, {Pello}, {Ricciardelli}, {Tanaka}, {Tresse}, {Vergani},
  {Welikala}, {Zucca}, {Oesch}, {Abbas}, {Barnes}, {Bordoloi}, {Bottini},
  {Cappi}, {Cassata}, {Cimatti}, {Fumana}, {Hasinger}, {Koekemoer},
  {Leauthaud}, {Maccagni}, {Marinoni}, {McCracken}, {Memeo}, {Meneux}, {Nair},
  {Porciani}, {Presotto}, \& {Scaramella}}]{peng2010}
{Peng}, Y.-j., {Lilly}, S.~J., {Kova{\v c}}, K., {et~al.} 2010, \apj, 721, 193

\bibitem[{{Postman} {et~al.}(2005){Postman}, {Franx}, {Cross}, {Holden},
  {Ford}, {Illingworth}, {Goto}, {Demarco}, {Rosati}, {Blakeslee}, {Tran},
  {Ben{\'{\i}}tez}, {Clampin}, {Hartig}, {Homeier}, {Ardila}, {Bartko},
  {Bouwens}, {Bradley}, {Broadhurst}, {Brown}, {Burrows}, {Cheng}, {Feldman},
  {Golimowski}, {Gronwall}, {Infante}, {Kimble}, {Krist}, {Lesser}, {Martel},
  {Mei}, {Menanteau}, {Meurer}, {Miley}, {Motta}, {Sirianni}, {Sparks}, {Tran},
  {Tsvetanov}, {White}, \& {Zheng}}]{postman2005}
{Postman}, M., {Franx}, M., {Cross}, N.~J.~G., {et~al.} 2005, \apj, 623, 721

\bibitem[{{Raichoor} {et~al.}(2012){Raichoor}, {Mei}, {Stanford}, {Holden},
  {Nakata}, {Rosati}, {Shankar}, {Tanaka}, {Ford}, {Huertas-Company},
  {Illingworth}, {Kodama}, {Postman}, {Rettura}, {Blakeslee}, {Demarco}, {Jee},
  \& {White}}]{raichoor2012b}
{Raichoor}, A., {Mei}, S., {Stanford}, S.~A., {et~al.} 2012, \apj, 745, 130

\bibitem[{{Rosati} {et~al.}(2009){Rosati}, {Tozzi}, {Gobat}, {Santos},
  {Nonino}, {Demarco}, {Lidman}, {Mullis}, {Strazzullo}, {B{\"o}hringer},
  {Fassbender}, {Dawson}, {Tanaka}, {Jee}, {Ford}, {Lamer}, \&
  {Schwope}}]{rosati2009}
{Rosati}, P., {Tozzi}, P., {Gobat}, R., {et~al.} 2009, \aap, 508, 583

\bibitem[{{Salpeter}(1955)}]{salpeter1955}
{Salpeter}, E.~E. 1955, \apj, 121, 161

\bibitem[{{Santos} {et~al.}(2015){Santos}, {Altieri}, {Valtchanov}, {Nastasi},
  {B{\"o}hringer}, {Cresci}, {Elbaz}, {Fassbender}, {Rosati}, {Tozzi}, \&
  {Verdugo}}]{santos2015}
{Santos}, J.~S., {Altieri}, B., {Valtchanov}, I., {et~al.} 2015, \mnras, 447,
  L65

\bibitem[{{Saracco} {et~al.}(2017){Saracco}, {Gargiulo}, {Ciocca}, \&
  {Marchesini}}]{saracco2017}
{Saracco}, P., {Gargiulo}, A., {Ciocca}, F., \& {Marchesini}, D. 2017, \aap,
  597, A122

\bibitem[{{Sazonova} {et~al.}(2020){Sazonova}, {Alatalo}, {Lotz}, {Rowlands},
  {Snyder}, {Boone}, {Brodwin}, {Hayden}, {Lanz}, {Perlmutter}, \&
  {Rodriguez-Gomez}}]{sazonova2020}
{Sazonova}, E., {Alatalo}, K., {Lotz}, J., {et~al.} 2020, \apj, 899, 85

\bibitem[{{Sersic}(1968)}]{sersic1968}
{Sersic}, J.~L. 1968, {Atlas de galaxias australes} (Cordoba, Argentina:
  Observatorio Astronomico)

\bibitem[{{Skelton} {et~al.}(2014){Skelton}, {Whitaker}, {Momcheva}, {Brammer},
  {van Dokkum}, {Labb{\'e}}, {Franx}, {van der Wel}, {Bezanson}, {Da Cunha},
  {Fumagalli}, {F{\"o}rster Schreiber}, {Kriek}, {Leja}, {Lundgren}, {Magee},
  {Marchesini}, {Maseda}, {Nelson}, {Oesch}, {Pacifici}, {Patel}, {Price},
  {Rix}, {Tal}, {Wake}, \& {Wuyts}}]{skelton2014}
{Skelton}, R.~E., {Whitaker}, K.~E., {Momcheva}, I.~G., {et~al.} 2014, \apjs,
  214, 24

\bibitem[{{Spitler} {et~al.}(2012){Spitler}, {Labb{\'e}}, {Glazebrook},
  {Persson}, {Monson}, {Papovich}, {Tran}, {Poole}, {Quadri}, {van Dokkum},
  {Kelson}, {Kacprzak}, {McCarthy}, {Murphy}, {Straatman}, \&
  {Tilvi}}]{spitler2012}
{Spitler}, L.~R., {Labb{\'e}}, I., {Glazebrook}, K., {et~al.} 2012, \apjl, 748,
  L21

\bibitem[{{Stalder} {et~al.}(2013){Stalder}, {Ruel}, {{\v{S}}uhada}, {Brodwin},
  {Aird}, {Andersson}, {Armstrong}, {Ashby}, {Bautz}, {Bayliss}, {Bazin},
  {Benson}, {Bleem}, {Carlstrom}, {Chang}, {Cho}, {Clocchiatti}, {Crawford},
  {Crites}, {de Haan}, {Desai}, {Dobbs}, {Dudley}, {Foley}, {Forman}, {George},
  {Gettings}, {Gladders}, {Gonzalez}, {Halverson}, {Harrington}, {High},
  {Holder}, {Holzapfel}, {Hoover}, {Hrubes}, {Jones}, {Joy}, {Keisler}, {Knox},
  {Lee}, {Leitch}, {Liu}, {Lueker}, {Luong-Van}, {Mantz}, {Marrone},
  {McDonald}, {McMahon}, {Mehl}, {Meyer}, {Mocanu}, {Mohr}, {Montroy},
  {Murray}, {Natoli}, {Nurgaliev}, {Padin}, {Plagge}, {Pryke}, {Reichardt},
  {Rest}, {Ruhl}, {Saliwanchik}, {Saro}, {Sayre}, {Schaffer}, {Shaw},
  {Shirokoff}, {Song}, {Spieler}, {Stanford}, {Staniszewski}, {Stark}, {Story},
  {Stubbs}, {van Engelen}, {Vanderlinde}, {Vieira}, {Vikhlinin}, {Williamson},
  {Zahn}, \& {Zenteno}}]{stalder2013}
{Stalder}, B., {Ruel}, J., {{\v{S}}uhada}, R., {et~al.} 2013, \apj, 763, 93

\bibitem[{{Stanford} {et~al.}(2012){Stanford}, {Brodwin}, {Gonzalez},
  {Zeimann}, {Stern}, {Dey}, {Eisenhardt}, {Snyder}, \&
  {Mancone}}]{stanford2012}
{Stanford}, S.~A., {Brodwin}, M., {Gonzalez}, A.~H., {et~al.} 2012, \apj, 753,
  164

\bibitem[{{Stockmann} {et~al.}(2020){Stockmann}, {Toft}, {Gallazzi}, {Zibetti},
  {Conselice}, {Margalef-Bentabol}, {Zabl}, {J{\o}rgensen}, {Magdis},
  {G{\'o}mez-Guijarro}, {Valentino}, {Brammer}, {Ceverino}, {Cortzen},
  {Davidzon}, {Demarco}, {Faisst}, {Hirschmann}, {Krogager}, {Lagos}, {Man},
  {Mundy}, {Peng}, {Selsing}, {Steinhardt}, \& {Whitaker}}]{stockmann2020}
{Stockmann}, M., {Toft}, S., {Gallazzi}, A., {et~al.} 2020, \apj, 888, 4

\bibitem[{{Stockton} {et~al.}(2008){Stockton}, {McGrath}, {Canalizo}, {Iye}, \&
  {Maihara}}]{stockton2008}
{Stockton}, A., {McGrath}, E., {Canalizo}, G., {Iye}, M., \& {Maihara}, T.
  2008, \apj, 672, 146

\bibitem[{{Strateva} {et~al.}(2001){Strateva}, {Ivezi{\'c}}, {Knapp},
  {Narayanan}, {Strauss}, {Gunn}, {Lupton}, {Schlegel}, {Bahcall}, {Brinkmann},
  {Brunner}, {Budav{\'a}ri}, {Csabai}, {Castander}, {Doi}, {Fukugita},
  {Gy{\H{o}}ry}, {Hamabe}, {Hennessy}, {Ichikawa}, {Kunszt}, {Lamb}, {McKay},
  {Okamura}, {Racusin}, {Sekiguchi}, {Schneider}, {Shimasaku}, \&
  {York}}]{strateva2001}
{Strateva}, I., {Ivezi{\'c}}, {\v{Z}}., {Knapp}, G.~R., {et~al.} 2001, \aj,
  122, 1861

\bibitem[{{Strazzullo} {et~al.}(2018){Strazzullo}, {Coogan}, {Daddi},
  {Sargent}, {Gobat}, {Valentino}, {Bethermin}, {Pannella}, {Dickinson},
  {Renzini}, {Arimoto}, {Cimatti}, {Dannerbauer}, {Finoguenov}, {Liu}, \&
  {Onodera}}]{strazzullo2018}
{Strazzullo}, V., {Coogan}, R.~T., {Daddi}, E., {et~al.} 2018, \apj, 862, 64

\bibitem[{{Strazzullo} {et~al.}(2016){Strazzullo}, {Daddi}, {Gobat},
  {Valentino}, {Pannella}, {Dickinson}, {Renzini}, {Brammer}, {Onodera},
  {Finoguenov}, {Cimatti}, {Carollo}, \& {Arimoto}}]{strazzullo2016}
{Strazzullo}, V., {Daddi}, E., {Gobat}, R., {et~al.} 2016, \apjl, 833, L20

\bibitem[{{Strazzullo} {et~al.}(2013){Strazzullo}, {Gobat}, {Daddi}, {Onodera},
  {Carollo}, {Dickinson}, {Renzini}, {Arimoto}, {Cimatti}, {Finoguenov}, \&
  {Chary}}]{strazzullo2013}
{Strazzullo}, V., {Gobat}, R., {Daddi}, E., {et~al.} 2013, \apj, 772, 118

\bibitem[{{Strazzullo} {et~al.}(2019){Strazzullo}, {Pannella}, {Mohr}, {Saro},
  {Ashby}, {Bayliss}, {Bocquet}, {Bulbul}, {Khullar}, {Mantz}, {Stanford},
  {Benson}, {Bleem}, {Brodwin}, {Canning}, {Capasso}, {Chiu}, {Gonzalez},
  {Gupta}, {Hlavacek-Larrondo}, {Klein}, {McDonald}, {Noordeh}, {Rapetti},
  {Reichardt}, {Schrabback}, {Sharon}, \& {Stalder}}]{strazzullo2019}
{Strazzullo}, V., {Pannella}, M., {Mohr}, J.~J., {et~al.} 2019, \aap, 622, A117

\bibitem[{{Strazzullo} {et~al.}(2010){Strazzullo}, {Rosati}, {Pannella},
  {Gobat}, {Santos}, {Nonino}, {Demarco}, {Lidman}, {Tanaka}, {Mullis},
  {Nu{\~n}ez}, {Rettura}, {Jee}, {B{\"o}hringer}, {Bender}, {Bouwens},
  {Dawson}, {Fassbender}, {Franx}, {Perlmutter}, \&
  {Postman}}]{strazzullo2010b}
{Strazzullo}, V., {Rosati}, P., {Pannella}, M., {et~al.} 2010, \aap, 524, A17

\bibitem[{{Suess} {et~al.}(2021){Suess}, {Kriek}, {Price}, \&
  {Barro}}]{suess2021}
{Suess}, K.~A., {Kriek}, M., {Price}, S.~H., \& {Barro}, G. 2021, \apj, 915, 87

\bibitem[{{Sunyaev} \& {Zeldovich}(1972)}]{sunyaevzeldovich1972}
{Sunyaev}, R.~A. \& {Zeldovich}, Y.~B. 1972, Comments on Astrophysics and Space
  Physics, 4, 173

\bibitem[{{Tanaka} {et~al.}(2013){Tanaka}, {Toft}, {Marchesini}, {Zirm}, {De
  Breuck}, {Kodama}, {Koyama}, {Kurk}, \& {Tanaka}}]{tanaka2013}
{Tanaka}, M., {Toft}, S., {Marchesini}, D., {et~al.} 2013, \apj, 772, 113

\bibitem[{{Tozzi} {et~al.}(2015){Tozzi}, {Santos}, {Jee}, {Fassbender},
  {Rosati}, {Nastasi}, {Forman}, {Sartoris}, {Borgani}, {Boehringer},
  {Altieri}, {Pratt}, {Nonino}, \& {Jones}}]{tozzi2015}
{Tozzi}, P., {Santos}, J.~S., {Jee}, M.~J., {et~al.} 2015, \apj, 799, 93

\bibitem[{{Tran} {et~al.}(2010){Tran}, {Papovich}, {Saintonge}, {Brodwin},
  {Dunlop}, {Farrah}, {Finkelstein}, {Finkelstein}, {Lotz}, {McLure},
  {Momcheva}, \& {Willmer}}]{tran2010}
{Tran}, K., {Papovich}, C., {Saintonge}, A., {et~al.} 2010, \apjl, 719, L126

\bibitem[{{Tran} {et~al.}(2015){Tran}, {Nanayakkara}, {Yuan}, {Kacprzak},
  {Glazebrook}, {Kewley}, {Momcheva}, {Papovich}, {Quadri}, {Rudnick},
  {Saintonge}, {Spitler}, {Straatman}, \& {Tomczak}}]{tran2015}
{Tran}, K.-V.~H., {Nanayakkara}, T., {Yuan}, T., {et~al.} 2015, \apj, 811, 28

\bibitem[{{Trudeau} {et~al.}(2022){Trudeau}, {Willis}, {Rennehan}, {Canning},
  {Carnall}, {Poggianti}, {Noordeh}, \& {Pierre}}]{trudeau2022}
{Trudeau}, A., {Willis}, J.~P., {Rennehan}, D., {et~al.} 2022, \mnras, 515,
  2529

\bibitem[{{van der Burg} {et~al.}(2020){van der Burg}, {Rudnick}, {Balogh},
  {Muzzin}, {Lidman}, {Old}, {Shipley}, {Gilbank}, {McGee}, {Biviano},
  {Cerulo}, {Chan}, {Cooper}, {De Lucia}, {Demarco}, {Forrest}, {Gwyn},
  {Jablonka}, {Kukstas}, {Marchesini}, {Nantais}, {Noble}, {Pintos-Castro},
  {Poggianti}, {Reeves}, {Stefanon}, {Vulcani}, {Webb}, {Wilson}, {Yee}, \&
  {Zaritsky}}]{vanderburg2020}
{van der Burg}, R. F.~J., {Rudnick}, G., {Balogh}, M.~L., {et~al.} 2020, \aap,
  638, A112

\bibitem[{{van der Wel} {et~al.}(2012){van der Wel}, {Bell}, {H{\"a}ussler},
  {McGrath}, {Chang}, {Guo}, {McIntosh}, {Rix}, {Barden}, {Cheung}, {Faber},
  {Ferguson}, {Galametz}, {Grogin}, {Hartley}, {Kartaltepe}, {Kocevski},
  {Koekemoer}, {Lotz}, {Mozena}, {Peth}, \& {Peng}}]{vanderwel2012}
{van der Wel}, A., {Bell}, E.~F., {H{\"a}ussler}, B., {et~al.} 2012, \apjs,
  203, 24

\bibitem[{{van der Wel} {et~al.}(2014){van der Wel}, {Franx}, {van Dokkum},
  {Skelton}, {Momcheva}, {Whitaker}, {Brammer}, {Bell}, {Rix}, {Wuyts},
  {Ferguson}, {Holden}, {Barro}, {Koekemoer}, {Chang}, {McGrath},
  {H{\"a}ussler}, {Dekel}, {Behroozi}, {Fumagalli}, {Leja}, {Lundgren},
  {Maseda}, {Nelson}, {Wake}, {Patel}, {Labb{\'e}}, {Faber}, {Grogin}, \&
  {Kocevski}}]{vanderwel2014}
{van der Wel}, A., {Franx}, M., {van Dokkum}, P.~G., {et~al.} 2014, \apj, 788,
  28

\bibitem[{{van der Wel} {et~al.}(2007){van der Wel}, {Holden}, {Franx},
  {Illingworth}, {Postman}, {Kelson}, {Labb{\'e}}, {Wuyts}, {Blakeslee}, \&
  {Ford}}]{vanderwel2007}
{van der Wel}, A., {Holden}, B.~P., {Franx}, M., {et~al.} 2007, \apj, 670, 206

\bibitem[{{van der Wel} {et~al.}(2008){van der Wel}, {Holden}, {Zirm}, {Franx},
  {Rettura}, {Illingworth}, \& {Ford}}]{vanderwel2008}
{van der Wel}, A., {Holden}, B.~P., {Zirm}, A.~W., {et~al.} 2008, \apj, 688, 48

\bibitem[{{van der Wel} {et~al.}(2011){van der Wel}, {Rix}, {Wuyts}, {McGrath},
  {Koekemoer}, {Bell}, {Holden}, {Robaina}, \& {McIntosh}}]{vanderwel2011}
{van der Wel}, A., {Rix}, H.-W., {Wuyts}, S., {et~al.} 2011, \apj, 730, 38

\bibitem[{{Wang} {et~al.}(2016){Wang}, {Elbaz}, {Daddi}, {Finoguenov}, {Liu},
  {Schreiber}, {Mart{\'{\i}}n}, {Strazzullo}, {Valentino}, {van der Burg},
  {Zanella}, {Ciesla}, {Gobat}, {Le Brun}, {Pannella}, {Sargent}, {Shu}, {Tan},
  {Cappelluti}, \& {Li}}]{wang2016}
{Wang}, T., {Elbaz}, D., {Daddi}, E., {et~al.} 2016, \apj, 828, 56

\bibitem[{{Webb} {et~al.}(2020){Webb}, {Balogh}, {Leja}, {van der Burg},
  {Rudnick}, {Muzzin}, {Boak}, {Cerulo}, {Gilbank}, {Lidman}, {Old},
  {Pintos-Castro}, {McGee}, {Shipley}, {Biviano}, {Chan}, {Cooper}, {De Lucia},
  {Demarco}, {Forrest}, {Jablonka}, {Kukstas}, {McCarthy}, {McNab}, {Nantais},
  {Noble}, {Poggianti}, {Reeves}, {Vulcani}, {Wilson}, {Yee}, \&
  {Zaritsky}}]{webb2020}
{Webb}, K., {Balogh}, M.~L., {Leja}, J., {et~al.} 2020, \mnras, 498, 5317

\bibitem[{{Wijesinghe} {et~al.}(2012){Wijesinghe}, {Hopkins}, {Brough},
  {Taylor}, {Norberg}, {Bauer}, {Brown}, {Cameron}, {Conselice}, {Croom},
  {Driver}, {Grootes}, {Jones}, {Kelvin}, {Loveday}, {Pimbblet}, {Popescu},
  {Prescott}, {Sharp}, {Baldry}, {Sadler}, {Liske}, {Robotham}, {Bamford},
  {Bland-Hawthorn}, {Gunawardhana}, {Meyer}, {Parkinson}, {Drinkwater},
  {Peacock}, \& {Tuffs}}]{wijesinghe2012}
{Wijesinghe}, D.~B., {Hopkins}, A.~M., {Brough}, S., {et~al.} 2012, \mnras,
  423, 3679

\bibitem[{{Williams} {et~al.}(2009){Williams}, {Quadri}, {Franx}, {van Dokkum},
  {Labb{\'e}}, \& {whatever}}]{williams2009}
{Williams}, R.~J., {Quadri}, R.~F., {Franx}, M., {et~al.} 2009, \apj, 691, 1879

\bibitem[{{Willis} {et~al.}(2020){Willis}, {Canning}, {Noordeh}, {Allen},
  {King}, {Mantz}, {Morris}, {Stanford}, \& {Brammer}}]{willis2020}
{Willis}, J.~P., {Canning}, R.~E.~A., {Noordeh}, E.~S., {et~al.} 2020, \nat,
  577, 39

\bibitem[{{Wuyts} {et~al.}(2011){Wuyts}, {F{\"o}rster Schreiber}, {van der
  Wel}, {Magnelli}, {Guo}, {Genzel}, {Lutz}, {Aussel}, {Barro}, {Berta},
  {Cava}, {Graci{\'a}-Carpio}, {Hathi}, {Huang}, {Kocevski}, {Koekemoer},
  {Lee}, {Le Floc'h}, {McGrath}, {Nordon}, {Popesso}, {Pozzi}, {Riguccini},
  {Rodighiero}, {Saintonge}, \& {Tacconi}}]{wuyts2011}
{Wuyts}, S., {F{\"o}rster Schreiber}, N.~M., {van der Wel}, A., {et~al.} 2011,
  \apj, 742, 96

\bibitem[{{Zavala} {et~al.}(2019){Zavala}, {Casey}, {Scoville}, {Champagne},
  {Chiang}, {Dannerbauer}, {Drew}, {Fu}, {Spilker}, {Spitler}, {Tran},
  {Treister}, \& {Toft}}]{zavala2019}
{Zavala}, J.~A., {Casey}, C.~M., {Scoville}, N., {et~al.} 2019, \apj, 887, 183

\bibitem[{{Zirm} {et~al.}(2008){Zirm}, {Stanford}, {Postman}, {Overzier},
  {Blakeslee}, {Rosati}, {Kurk}, {Pentericci}, {Venemans}, {Miley},
  {R{\"o}ttgering}, {Franx}, {van der Wel}, {Demarco}, \& {van
  Breugel}}]{zirm2008}
{Zirm}, A.~W., {Stanford}, S.~A., {Postman}, M., {et~al.} 2008, \apj, 680, 224

\end{thebibliography}

\begin{appendix}

  \section{Estimation of potential bias in the measurements of  morphological parameters of cluster galaxies vs the adopted field reference measurements }
\label{sec:appendixmorphbias}

\begin{figure}[hb!]
    \centering
    \includegraphics[width=0.46\textwidth,viewport= 85 412 536 696, clip]{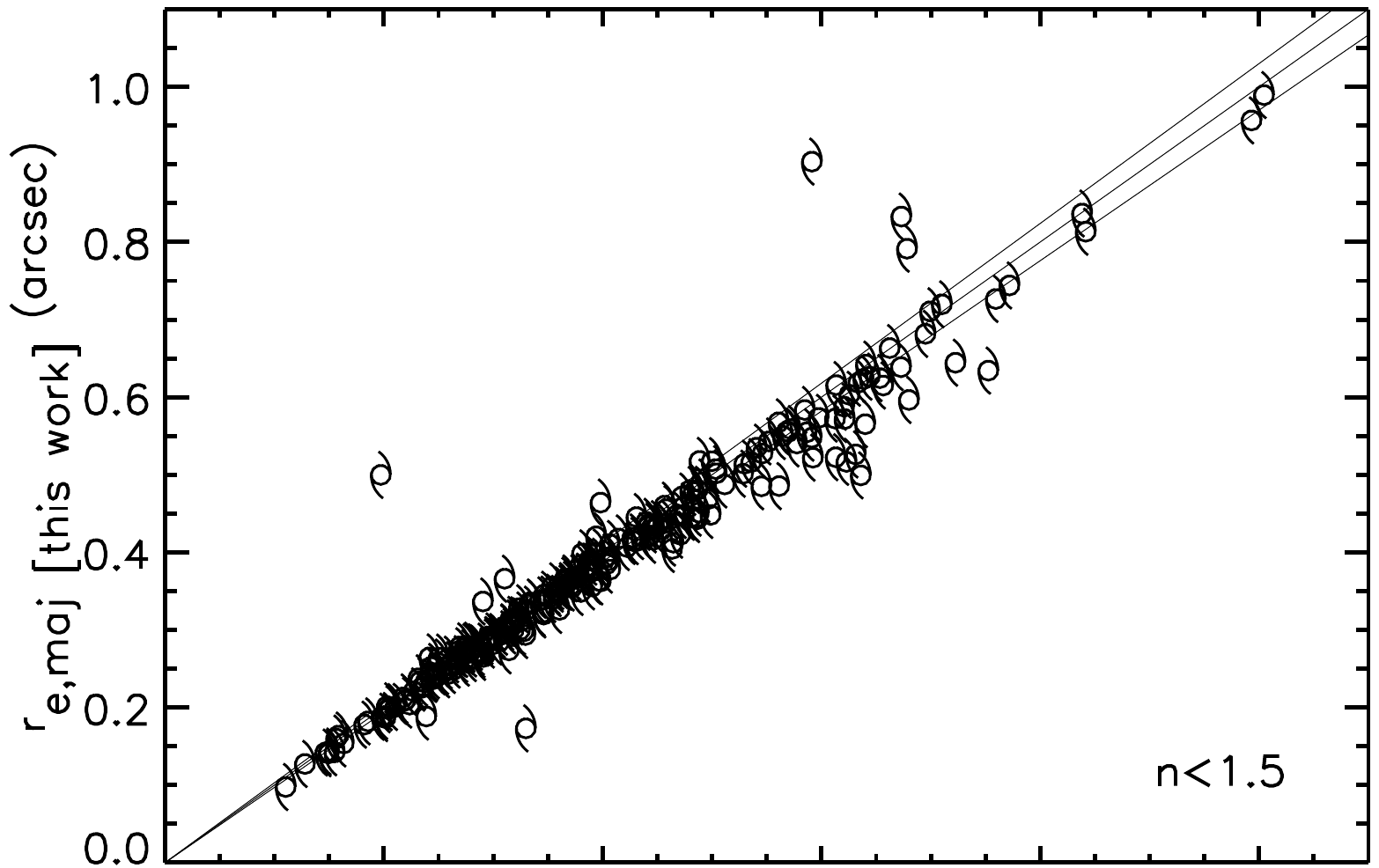}
    \includegraphics[width=0.46\textwidth,viewport= 85 412 536 696, clip]{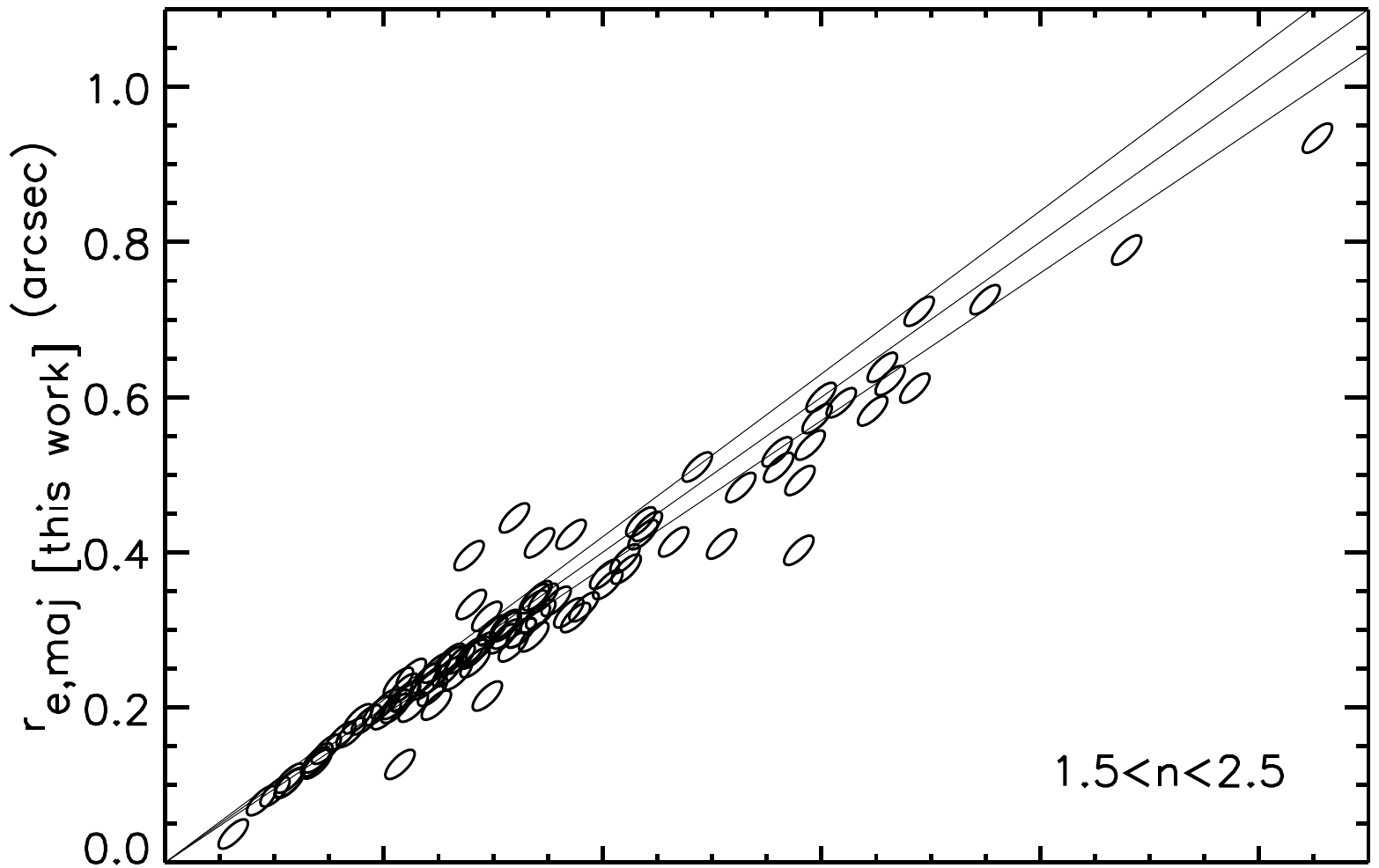}
    \includegraphics[width=0.46\textwidth,viewport= 85 362 536 696, clip]{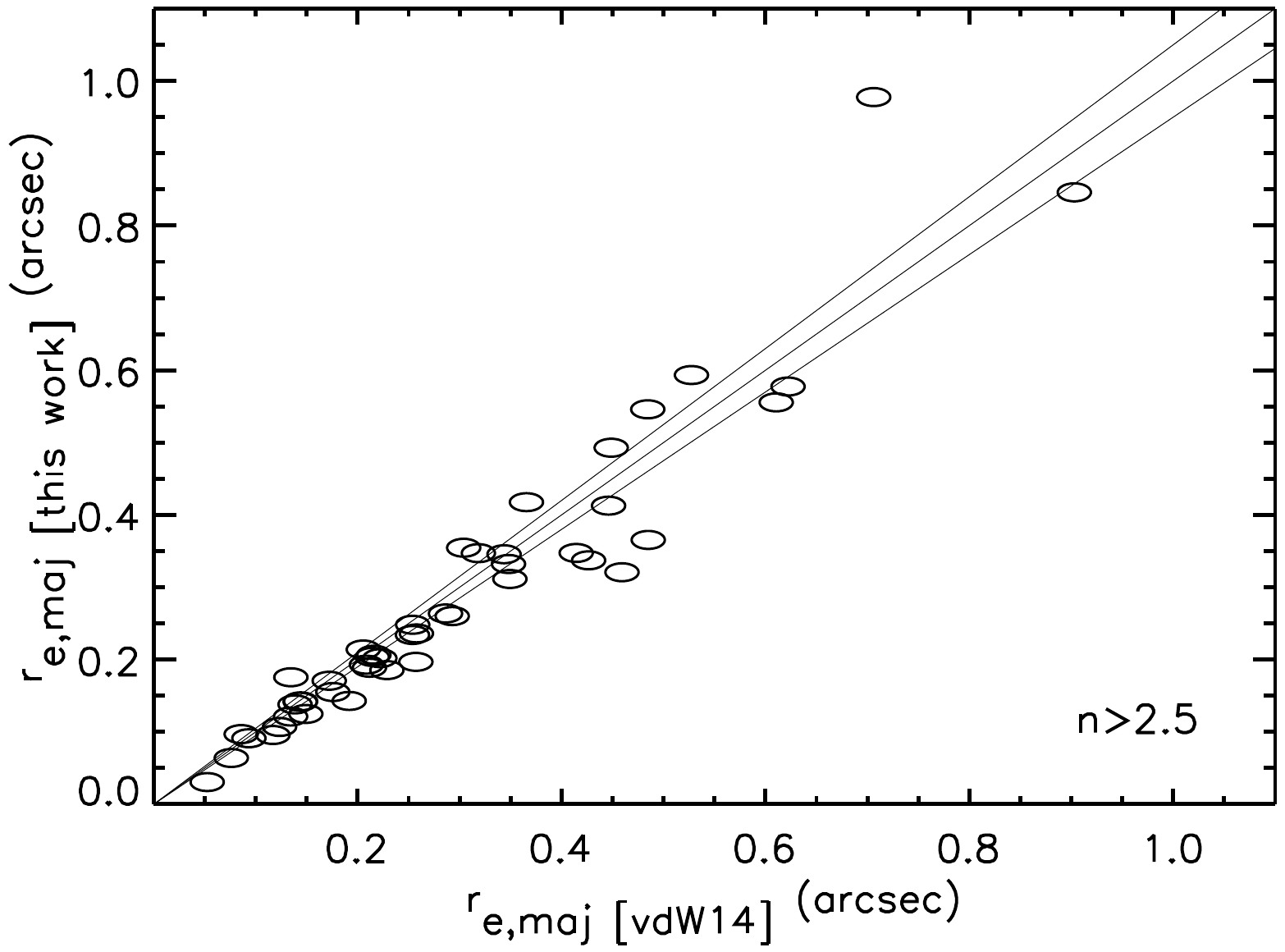}
 \caption{Comparison of effective radii as measured with this work
   procedures vs\ \citet{vanderwel2014} estimates adopted for the
   field reference sample. Measurements are compared for a random portion of the
   control field galaxy sample (see text), split by Sersic index as
   indicated. The solid lines show the bisector (middle), and the
   expected average uncertainty (1~$\sigma$) on effective radii from
   this work, as estimated with simulations (see
   Sect.~\ref{sec:datamorph}) for sources at the faint end of the
   probed magnitude range.
 \label{fig:morphbiasre1}}
\end{figure}

We describe here the assessment of the potential bias between the
measured morphological parameters - in particular Sersic index and
effective radius - for cluster galaxies, and the corresponding
measurements adopted for the control field sample from
\citet{vanderwel2014}. As discussed in Sect.~\ref{sec:datamorph},
given the lack of overlap between the cluster and control fields, a
small portion of the control field, corresponding to 10 WFC3 fields,
is reduced and analysed with the same procedures we adopted for the
cluster fields (also including PSF modelling), in order to enable a
direct comparison of the measurements for the same sources.

\begin{figure}[b]
      \centering
    \includegraphics[width=0.47\textwidth,viewport= 71 369 540 705, clip]{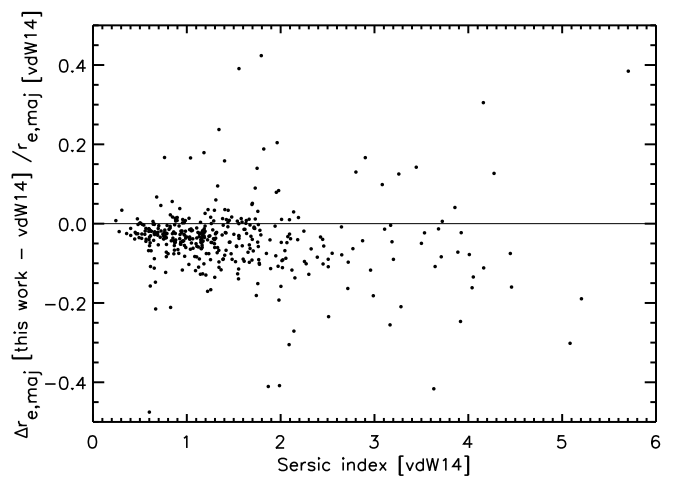}
   \caption{Comparison of effective radii as measured with this
     work procedures vs\ \citet{vanderwel2014} estimates adopted for
     the field reference sample. For the same sources as in
     Fig.~\ref{fig:morphbiasre1}, the fractional difference between
     the two measurements is shown as a function of Sersic index (two sources
     (<1\%) falling outside of the plotted range are not shown).
  \label{fig:morphbiasre2}}
\end{figure}

In Figures~\ref{fig:morphbiasre1} and \ref{fig:morphbiasre2} we
compare our measurements of effective radii with those from
\citet{vanderwel2014} for galaxies in these testing fields. Our
measurements show a systematic offset towards lower values with
respect to \citet{vanderwel2014}, possibly dependent on Sersic
index. At face value, the offset is $\sim$3\% (5\%, 8\%) for $n<1.5$
sources ($1.5<n<2.5$, $n>2.5$, respectively), in the magnitude range
relevant to this work. Besides the systematic offset, the dispersion
(rms) between the two measurements is $\sim 4\%$ (7\%, 15\%,
respectively) for the same sub-populations. For reference, our
expected average uncertainty on effective radii, as estimated from
simulations (Sect.~\ref{sec:datamorph}) for sources towards the faint
end of our probed range, is within $\sim$5\%. As discussed in
Sects.~\ref{sec:datamorph} and \ref{sec:galmasssize}, we correct our
estimates of effective radii to bring them on the same scale as the
\citet{vanderwel2014} measurements for the purpose of quantitative
comparison with field counterparts.

\begin{figure*}
    \centering
    \includegraphics[height=0.35\textwidth,viewport= 100 372 535 696, clip]{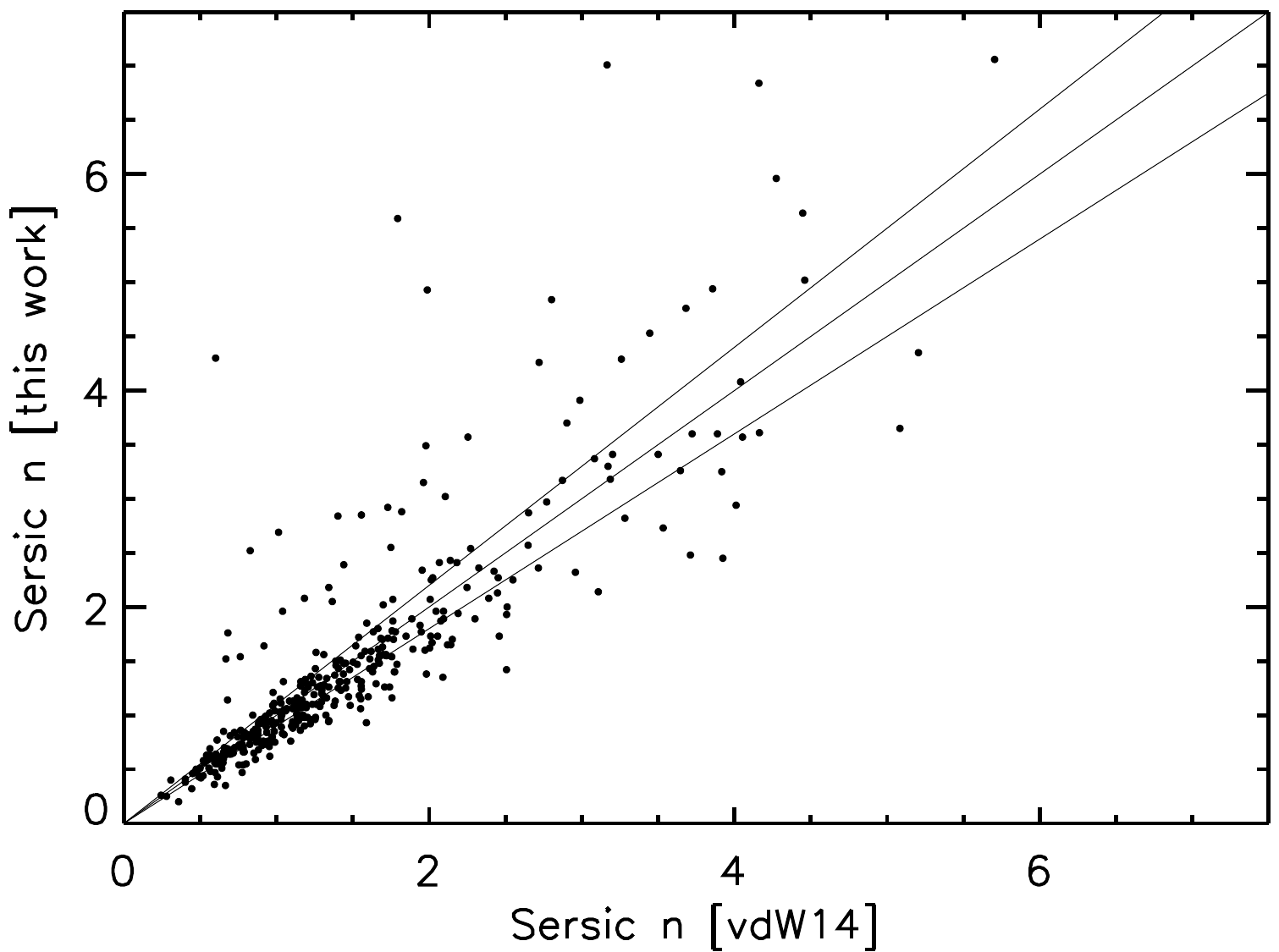}%
        \includegraphics[height=0.35\textwidth,viewport= 75 372 535 696, clip]{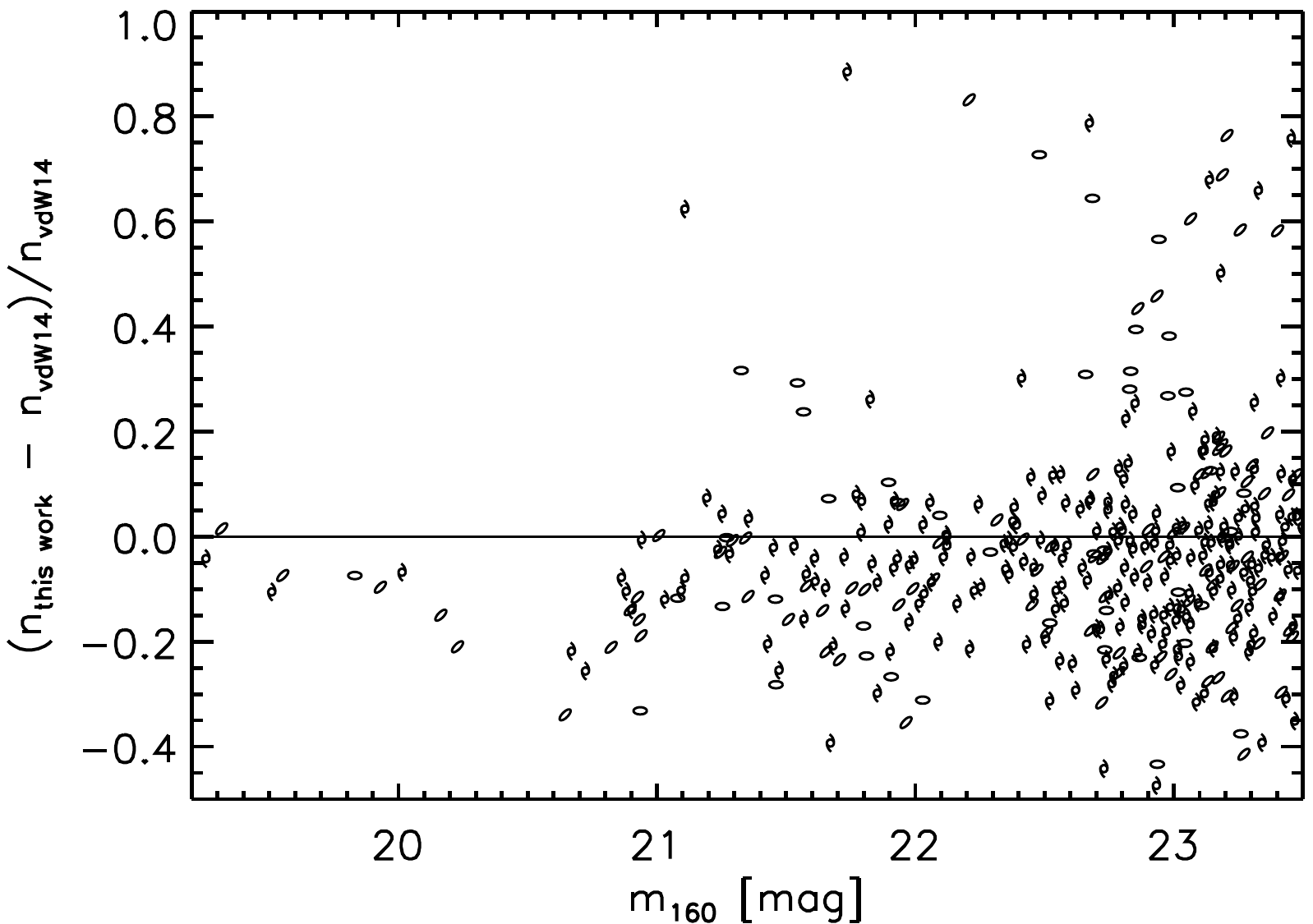}
 \caption{Comparison of Sersic indices as measured with this work
   procedures vs\ \citet{vanderwel2014} estimates adopted for the
   field sample, for the same sources as in Fig.~\ref{fig:morphbiasre1}
   (see text). {\it Left:} The direct comparison of the two
   measurements. The solid lines show the bisector (middle), and the
   expected average uncertainty (1~$\sigma$) on Sersic indices from
   this work, as estimated with simulations (see
   Sect.~\ref{sec:datamorph}) for sources at the faint end of the
   probed magnitude range. {\it Right:} The fractional difference
   between the two measurements as a function of $m_{160}$ magnitude
   (3\% of the sources out of the plotted range are not
   shown). Symbols are coded by Sersic index, as in
   Fig.~\ref{fig:morphbiasre1}.
  \label{fig:morphbiasnser}}
\end{figure*}

In Fig.~\ref{fig:morphbiasnser}, we similarly compare our Sersic index
measurements versus \citet{vanderwel2014} for the same sources.  We find
a systematic offset towards lower values than those in
\citet{vanderwel2014}, by $\sim$4\% for the overall population in the
 magnitude range of interest. At face value, the offset for sub-samples
split by Sersic index is $\sim$4\% and 7\% for $n<1.5$ and $1.5<n<2.5$
sources, respectively, and consistent with no systematic offset for $n>2.5$ sources,
but possible variation with Sersic index is not significant with the
given statistics. Indeed, Fig.~\ref{fig:morphbiasnser} clearly shows
the dispersion between the two measurements, of order 15\% for
disk-dominated galaxies and 20-25\% for intermediate and
bulge-dominated systems.  For reference, the expected average
uncertainty on Sersic indices, as estimated from simulations
(Sect.~\ref{sec:datamorph}) for sources towards the faint end of our
probed range, is $\sim$10\%. Based on this comparison, with respect to
the broad Sersic index classes considered in this work, and taking
\citet{vanderwel2014} measurements as the reference for the purpose of
this assessment, 6\% of $n<1.5$ sources would be misclassified as
$n>1.5$, out of which 2\% with $n>2.5$; at the same time, 20\% of
$n>2.5$ sources would be misclassified as $n<2.5$, out of which 2\% as
$n<1.5$. As may be expected, misclassification may be more severe
for intermediate Sersic sources, with almost 40\% of $1.5<n<2.5$
sources in \citet{vanderwel2014} being ``misclassified'' with our
measurements, mostly (25\%) as disky $n<1.5$ systems.

Given the intrinsically broad nature of morphology classes used in the
analysis in Sect.~\ref{sec:galmorph}, the impact of potential
systematics at the level discussed here above on the results presented
in this work turns out to be very marginal. Correcting Sersic indices
of cluster galaxies by the amounts discussed above gives differences
on the results and figures presented here that are always well within
the quoted uncertainties, and are actually not noticeable in most
cases. Therefore, as discussed in Sect.~\ref{sec:galmorph}, Sersic
indices presented throughout and all related results are shown with no
corrections applied.

\begin{figure*}
  \centering
 \includegraphics[width=0.97\textwidth,viewport= 59 532 550 689, clip]{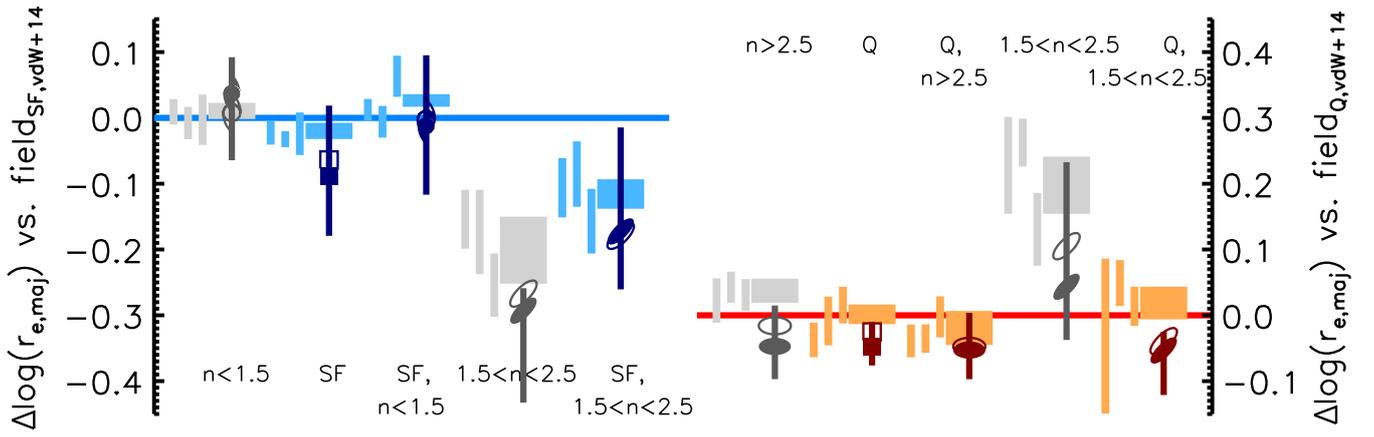}
 \caption{As Fig.~\ref{fig:deltasize}, but showing sizes for cluster
   galaxies as measured (not corrected for estimated size bias with
   respect to control field sample, see text for details). Empty
   symbols also show the measurement for the full relevant candidate
   cluster member samples, not accounting for residual background
   contamination (solid symbols account instead for residual
   background contamination as in Fig.~\ref{fig:deltasize}).  The
   three vertical lines on the left of the control field measurements
   show the field sample split by redshift ($1.3<z<1.5$, $1.4<z<1.6$,
   $1.6<z<1.8$), showing no redshift dependence in the field
   measurements across the redshift range of interest.
   \label{fig:deltasizefullfig}}
\end{figure*}

\section{Additional measurements related to size difference of cluster versus field galaxies }
\label{sec:appendixdeltasize}

Figure~\ref{fig:deltasizefullfig} shows additional measurements
related to the investigation of size differences between cluster and
field galaxies for different sub-populations, as discussed in
Sect.~\ref{sec:galmasssize}. In particular,
Fig.~\ref{fig:deltasizefullfig} shows 1) analogous measurements as in
Fig.~\ref{fig:deltasize}, but with average size differences as
measured (not corrected for estimated size bias with respect to
control field sample, see Sects.~\ref{sec:datamorph} and
\ref{sec:galmasssize}); 2) the impact of residual background
contamination correction (see Sects.~ \ref{sec:galsamples} and
\ref{sec:galmasssize}); 3) the lack of redshift dependence of average
size measurements for the reference field sample (see discussion in
Sect.~\ref{sec:galmasssize}).

\end{appendix}

\end{document}